%% file: TOPQ-2015-05.tex
\newcommand*{\ATLASLATEXPATH}{latex/}
\documentclass[cernpreprint,txfonts,UKenglish,texlive=2016]{\ATLASLATEXPATH
atlasdoc}

\usepackage[backend=biber]{\ATLASLATEXPATH atlaspackage}
\pdfoutput=1
\usepackage{commath}
\usepackage{longtable}
\usepackage{rotating}
\usepackage{epstopdf}
\newcommand{\onlyacomment}[1]{}
\onlyacomment{%
\usepackage[%
backend=biber
]{biblatex}
}
\usepackage{\ATLASLATEXPATH atlasbiblatex}

\usepackage[unit=false]{\ATLASLATEXPATH atlasphysics}

\usepackage{epstopdf}
\graphicspath{{logos/}{figs/}}

\usepackage{TOPQ-2015-05-defs}

\newif\ifAppSysTable
\AppSysTabletrue 

\input{TOPQ-2015-05-metadata}
\hypersetup{pdftitle={ATLAS document},pdfauthor={The ATLAS Collaboration}}

\begin{document}

\maketitle

\tableofcontents

\clearpage

\input{introduction}

\input{atlas_detector}

\input{samples}

\input{objects}

\input{selection}

\input{xsect_def}

\input{nn}

\input{sys}

\input{results_inclusive_xs}

\input{xsect_diff}

\FloatBarrier

\input{conclusion}

\section*{Acknowledgements}

\input{Acknowledgements}

\clearpage
\printbibliography[heading=bibintoc]

\newpage \input{atlas_authlist}

\end{document}

%% file: TOPQ-2015-05-metadata.tex
\AtlasTitle{Fiducial, total and differential cross-section
measurements of $t$-channel single top-quark production in $pp$ collisions at
\SI{8}{\TeV} using data collected by the ATLAS detector}

\author{The ATLAS Collaboration}

\AtlasRefCode{TOPQ-2015-05}

\PreprintIdNumber{CERN-EP-2016-223}

\arXivId{1702.02859}

\AtlasJournalRef{Eur. Phys. J. C 77 (2017) 531}
\AtlasDOI{10.1140/epjc/s10052-017-5061-9}

\AtlasAbstract{%
  Detailed measurements of $t$-channel single top-quark production are presented.
  They use \SI{20.2}{\per\fb} of data collected by the ATLAS experiment in proton--proton collisions
  at a centre-of-mass energy of \SI{8}{\TeV} at the LHC.
  Total, fiducial and differential cross-sections are measured for both
  top-quark and top-antiquark production.
  The fiducial cross-section is measured with a precision of \SI{5.8}{\%}
  (top quark) and \SI{7.8}{\%} (top antiquark), respectively.
  The total cross-sections are measured to be 
  $\sigtot(tq) = 56.7^{+4.3}_{-3.8}\;\si{\pb}$ for top-quark production and
  $\sigtot(\tbar q) = 32.9^{+3.0}_{-2.7}\;\si{\pb}$ for top-antiquark
  production, in agreement with the Standard Model prediction.
  In addition, the ratio of top-quark to top-antiquark
  production cross-sections is determined to be $R_t=1.72 \pm 0.09$.
  The differential cross-sections as a function of the transverse momentum and
  rapidity of both the top quark and the top antiquark are
  measured at both the parton and particle levels.
  The transverse momentum and rapidity differential cross-sections of the accompanying jet from the $t$-channel scattering are measured at particle level.
  All measurements are compared to various Monte Carlo predictions as well as
  to fixed-order QCD calculations where available.
}

%% file: introduction.tex
\section{Introduction}
\label{sec:intro}

\begin{figure}[htpb]
  \centering
  \subfloat[][]{
    \includegraphics[width=0.3\linewidth]{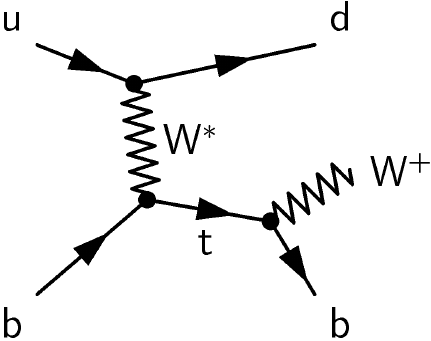}
    \label{subfig:top_quark}
  }
  \qquad
  \subfloat[][]{%
    \includegraphics[width=0.3\linewidth]{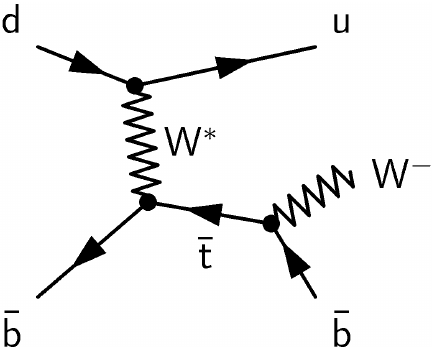}
    \label{subfig:top_antiquark}
  }
  \caption{Representative leading-order Feynman diagrams for
    \protect\subref{subfig:top_quark} single top-quark production and 
    \protect\subref{subfig:top_antiquark} single top-antiquark 
    production via the $t$-channel exchange of a virtual $W^*$ boson, including
    the decay of the top quark and top antiquark, respectively.}
  \label{fig:Feynman_tchan}
\end{figure}

Top quarks are produced singly in proton--proton (\pp) collisions via electroweak
charged-current interactions.
In leading-order (LO) perturbation theory, single top-quark production is described by three 
subprocesses that are distinguished by the virtuality of the exchanged $W$ boson. 
The dominant process is the $t$-channel exchange depicted in 
\Fig{\ref{fig:Feynman_tchan}}, where a light quark from one of the colliding
protons interacts with a $b$-quark from another proton by exchanging a virtual $W$ boson ($W^*$).
Since the valence $u$-quark density of the proton is about twice as high as the
valence $d$-quark density, the production cross-section of single top quarks,
$\sigma(tq)$,  
is expected to be about twice as high as the cross-section of top-antiquark
production, $\sigma(\bar{t}q)$.
At LO, subdominant single-top-quark processes are the associated production of a
$W$ boson and a top quark~($Wt$) and the $s$-channel production of $t\bar{b}$.
The $t$-channel and $s$-channel processes do not interfere even at
next-to-leading order (NLO) in perturbation theory and are thus well defined
with that precision.

This paper presents measurements of $\sigma(tq)$ and $\sigma(\bar{t}q)$ in
\pp collisions at a centre-of-mass energy of $\sqrt{s} =
\SI{8}{\TeV}$ at the Large Hadron Collider (LHC).
The analysis is based on the full ATLAS dataset collected in 2012, corresponding
to an integrated luminosity of \SI{20.2}{\per\fb}. 
Separate measurements of 
$tq$ and $\bar{t}q$ production provide sensitivity to the parton distribution 
functions (PDFs) of the $u$-quark and the $d$-quark, exploiting the different 
initial states of the two processes as shown in \Fig{\ref{fig:Feynman_tchan}}.
In addition, the cross-section ratio $R_t \equiv \sigma(\tq)/\sigma(\tbarq)$
is measured, which has smaller systematic uncertainties than the
individual cross-sections, because of partial cancellations of common
uncertainties.
Investigating $R_t$ also provides a way of searching for new-physics contributions in single top-quark (top-antiquark) 
production~\cite{AguilarSaavedra:2008gt} and of elucidating the nature of physics 
beyond the Standard Model (SM) if it were to be observed~\cite{Gao:2011fx}.

In general, measurements of single top-quark production provide insights into 
the properties of the $Wtb$ interaction.
The cross-sections are proportional to the
square of the coupling at the $Wtb$ production vertex.
In the SM, the
coupling is given by the Cabibbo--Kobayashi--Maskawa (CKM) matrix element 
$V_{tb}$~\cite{CKM1,CKM2} multiplied by the universal electroweak coupling constant.
All measurements presented in this paper are based on the assumption
that the production and the decay of top quarks via $Wts$ and $Wtd$ 
vertices are suppressed due to the fact that the CKM matrix elements 
$V_{ts}$ and $V_{td}$ are much smaller than $V_{tb}$.
Potential new-physics contributions to the $Wtb$ vertex are parameterised by
an additional left-handed form factor \fl~\cite{AguilarSaavedra:2008zc}, assumed to
be real. In this approach the Lorentz structure is assumed to be the same as in the SM, that is vector--axial-vector ($\mathrm{V}-\mathrm{A}$).
The inclusive cross-section $\sigma(\tq+\tbarq)$ is determined as the sum of $\sigma(\tq)$ 
and $\sigma(\tbarq)$ and used to determine \flvtb. 
Alternatively, the measurement of $\sigma(\tq+\tbarq)$ can be used to constrain
the $b$-quark PDF.
The measurement of $\sigma(\tq+\tbarq)$ is also sensitive to various models
of new-physics phenomena~\cite{Tait:2000sh}, such as extra heavy quarks, gauge
bosons, or scalar bosons.
Studies of differential cross-sections allow the modelling of the process to be probed in more detail
and provide a more sensitive search for effects of new physics.

Single top-quark production in the $t$-channel was first established 
in $p\bar{p}$ collisions at $\sqrt{s} = \SI{1.96}{\TeV}$ at the
Tevatron~\cite{Abazov:2009pa,Aaltonen:2014ura}.
Measurements of $t$-channel single top-quark production at the LHC at $\sqrt{s} = \SI{7}{\TeV}$ 
were performed by the ATLAS Collaboration~\cite{TOPQ-2011-14,TOPQ-2012-21} and the CMS 
Collaboration~\cite{CMS-TOP-10-008,CMS-TOP-11-021}.
At $\sqrt{s} = \SI{8}{\TeV}$ the CMS Collaboration measured the $t$-channel
cross-sections and the cross-section ratio, $R_t$~\cite{CMS-TOP-12-038}.

The total inclusive cross-sections of top-quark and top-antiquark production in
the $t$-channel in $pp$ collisions at $\sqrt{s} = \SI{8}{\TeV}$
are predicted to be
\begin{subequations}
\begin{align}
  \sigma(tq)          & = \SI[parse-numbers=false]{54.9 ^{+2.3}_{-1.9}}{\pb}\,,
  \label{eq:sigma_pred_tq} \\
  \sigma(\bar{t}q)    & = \SI[parse-numbers=false]{29.7 ^{+1.7}_{-1.5}}{\pb}\,,
  \label{eq:sigma_pred_tbarq} \\
  \sigma(tq+\bar{t}q) & = \SI[parse-numbers=false]{84.6 ^{+3.9}_{-3.4}}{\pb}\,,
  \label{eq:predicted_combined_xs}
\end{align}
\end{subequations}
at NLO accuracy in QCD. The cross-sections are
calculated with the \HATHOR~v2.1~\cite{Kant:2014oha} tool, which is based on
work documented in \Ref{\cite{Campbell:2009ss}}.
The top-quark mass $\mtop$ is assumed to be \SI{172.5}{\GeV}, the same value
which is used for the samples of simulated events in this analysis.
The central values quoted in \Eqnrange{\eqref{eq:sigma_pred_tq}}{\eqref{eq:predicted_combined_xs}} are determined
following the PDF4LHC prescription~\cite{Botje:2011sn},
which defines the central value as the midpoint
of the uncertainty envelope of three PDF sets:
\mstw~\cite{Martin:2009iq,Martin:2009bu}, \ct~NLO~\cite{Lai:2010vv} and
\nnpdfthree~\cite{Ball:2012cx}.
The uncertainty due to the PDFs and their $\alphas$ dependence is given by half
of the width of the envelope defined by these PDFs and is added in
quadrature to the scale uncertainty to obtain the total uncertainties quoted in \Eqnrange{\eqref{eq:sigma_pred_tq}}{\eqref{eq:predicted_combined_xs}}. 
The sensitivity of $\sigma(tq)$ and $\sigma(\bar{t}q)$ to the PDFs has recently
gained attention in the literature~\cite{Alekhin:2015cza}.
The scale uncertainties in the predictions are determined following a
prescription referred to as independent restricted scale variations, in which the
renormalisation scale (\mur) and the factorisation scale (\muf) are varied 
independently, considering the default choices \murdef and \mufdef, half the
default scales and two times the default scales. The combinations
($0.5 \murdef$, $2.0 \mufdef$) and ($2.0 \murdef$, $0.5 \mufdef$) 
are excluded, thus \enquote{restricted variations}.
The maximum deviations in the predicted cross-sections for the six probed
variations define the uncertainty.

Predictions of $\sigma(tq)$ and $\sigma(\bar{t}q)$ have recently been
calculated at next-to-next-to-leading order (NNLO)~\cite{Brucherseifer:2014ama}.
The calculation uses $\mtop=\SI{173.2}{\GeV}$ and $\mur=\muf=\mtop$, and 
results in a cross-section which is \SI{1.5}{\%} lower than the NLO value
calculated with the same settings. Only a limited number of scale variations 
are presented in \Ref{\cite{Brucherseifer:2014ama}}; however, they do indicate a
reduction in the scale uncertainties compared to the NLO result. 
Since the NLO computation implemented in \HATHOR allows a complete
treatment of the scale and PDF uncertainties, which is not currently available 
for the NNLO calculation, the NLO computation is used when extracting
\flvtb and for comparing the \rt measurement to different PDF sets.
The NLO results have been augmented by including the resummation of soft-gluon
terms at next-to-next-to-leading logarithmic (NNLL) 
accuracy~\cite{Kidonakis:2011wy, Kidonakis:2012db, Kidonakis:2015nna}, leading
to fixed-order predictions at the so-called NLO+NNLL level.

Cross-sections are measured in two ways:
over the full kinematic range and within a fiducial phase space,
defined to be as close as possible to the experimental measurement range.
The definition of the fiducial phase space is based on stable particles
output by Monte Carlo (MC) generators, with which reconstructed objects, such as
primary leptons, jets and missing transverse momentum, are defined.
The advantage of the fiducial cross-section measurements is a substantial
reduction of the size of the applied acceptance corrections,
leading to reduced systematic uncertainties.

Differential cross-sections are measured as
a function of the transverse momentum of the top (anti)quark, \pTt,
and as a function of the absolute value of its rapidity, \absyt.
The measured cross-sections are unfolded to both parton level and particle level.
Parton-level measurements can be directly compared to theory
predictions that use stable top quarks.
Particle-level measurements make use of a top-quark proxy which is constructed
with the objects used in the fiducial cross-section measurements.
At particle level, it is also possible to measure differential cross-sections 
as a function of the \pT and rapidity of the jet formed by the scattered light quark
in the $t$-channel exchange of a $W$ boson.

Events are selected targeting the $t \to \ell \nu b$ decay mode of the top quark
where the lepton can be either an electron or a muon originating from a $W$-boson decay.\footnote{%
  Events involving $W\rightarrow \tau\nu$ decays with a subsequent decay of the $\tau$ lepton
  to either $e\nu_e\nu_\tau$ or $\mu\nu_\mu\nu_\tau$ are included in the signal.}
The experimental signature of candidate events is thus given by one charged lepton (electron or 
muon), large values of the magnitude of the missing transverse momentum, \MET, and two hadronic jets with high transverse momentum.
Exactly one of the two hadronic jets is required to be identified as a jet
containing $b$-hadrons ($b$-jet).
The other hadronic jet is referred to as the \emph{untagged} jet
and is assumed to be the accompanying jet in the $t$-channel exchange.

Several other processes feature the same signature as single-top-quark events;
the main backgrounds being \wjets production and top-quark--top-antiquark (\ttbar) pair production.
Since a typical signature-based event selection yields only a relatively low signal purity,
a dedicated analysis  strategy is developed to separate signal and background events.
Several observables discriminating between signal and 
background events are combined by an artificial neural network (NN) into one
discriminant, \NNout, with improved signal-to-background separation.
The cross-section measurements are based on a maximum-likelihood fit to the
\NNout distribution.
In addition, a cut on \NNout is applied to obtain a sample of
events enriched in $t$-channel single-top-quark events.
These events are used to extract differential cross-sections as a function of
both the top-quark and untagged-jet variables.

This paper is organised as follows. The ATLAS detector is introduced in \Sect{\ref{sec:detector}};
details of both the data set and simulated event samples are given in
\Sect{\ref{sec:data}}.
The objects used to select events are introduced in \Sect{\ref{sec:object}},
while \Sect{\ref{sec:selection}} discusses the event selection criteria. In
\Sect{\ref{sec:background_estimate}} the background estimation is described. The
measured cross-sections are defined in detail in \Sect{\ref{sec:xsect_def}} before turning to the separation of signal from background using a neural network in \Sect{\ref{sec:nn}}.
The sources of systematic uncertainty considered in the analyses are covered
in \Sect{\ref{sec:sys}}.
The fiducial and inclusive cross-section measurements are the subject of \Sect{\ref{sec:xsect_tot}},
including the measurement of $R_{t}$ and \flvtb.
This is followed by the differential cross-section measurements in \Sect{\ref{sec:xsect_diff}},
which also explains the method used to unfold the cross-sections.
Finally, the conclusion is given in \Sect{\ref{sec:conclusion}}.

%% file: atlas_detector.tex
\newpage
\section{ATLAS detector}
\label{sec:detector}

The ATLAS experiment~\cite{PERF-2007-01} at the LHC is a multi-purpose particle detector
with a forward-backward symmetric cylindrical geometry and a near $4\pi$ coverage in 
solid angle.\footnote{%
ATLAS uses a right-handed coordinate system with its origin at the nominal interaction point (IP)
in the centre of the detector and the $z$-axis along the beam pipe.
The $x$-axis points from the IP to the centre of the LHC ring,
and the $y$-axis points upwards.
Cylindrical coordinates $(r,\phi)$ are used in the transverse plane, 
$\phi$ being the azimuthal angle around the $z$-axis.
The pseudorapidity is defined in terms of the polar angle $\theta$ as $\eta = -\ln \tan(\theta/2)$.
Angular distance is measured in units of $\Delta R \equiv \sqrt{(\Delta\eta)^{2} + (\Delta\phi)^{2}}$.}
It consists of an inner tracking detector (ID) surrounded by a thin superconducting solenoid
providing a \SI{2}{\tesla} axial magnetic field, electromagnetic and hadron calorimeters, and a muon spectrometer.
The ID covers the pseudorapidity range $|\eta| < 2.5$.
It consists of silicon pixel, silicon microstrip, and transition-radiation
tracking detectors.
Lead/liquid-argon (LAr) sampling calorimeters provide electromagnetic (EM) energy measurements
with high granularity.
A hadron (steel/scintillator-tile) calorimeter covers the central pseudorapidity
range ($|\eta| < 1.7$).
The endcap ($1.5 < |\eta| < 3.2$) and forward regions ($3.1 < |\eta| < 4.9$) are
instrumented with LAr calorimeters for both the EM and hadronic energy
measurements.
The muon spectrometer (MS) surrounds the calorimeters and is based on
three large air-core toroid superconducting magnets with eight coils each.
Its bending power ranges from \num{2.0} to
\SI{7.5}{Tm}.
It includes a system of precision tracking chambers and fast detectors for triggering.
A three-level trigger system is used to select events.
The first-level trigger is implemented in hardware and uses a subset of the detector information
to reduce the accepted rate to at most \SI{75}{\kilo\hertz}.
This is followed by two software-based trigger levels that
together reduce the accepted event rate to \SI{400}{\hertz} on average,
depending on the data-taking conditions during 2012.

%% file: samples.tex
\section{Data sample and simulation}
\label{sec:data}

This analysis is performed using \pp collision data 
recorded at a centre-of-mass energy of $\sqrt{s}=\SI{8}{\TeV}$ with the ATLAS
detector at the LHC.
Only the data-taking periods in which all the subdetectors were operational are considered.
The data sets used in this analysis are defined by high-\pT 
single-electron or single-muon triggers~\cite{PERF-2011-02, TRIG-2012-03}, resulting in a
data sample with an integrated luminosity of $L_{\text{int}} =
\SI{20.2}{\per\fb}$~\cite{newlumi}.

In the first-level trigger, electron-channel events are triggered by
a cluster of energy depositions in the electromagnetic calorimeter.
In the software-based triggers, a cluster of energy depositions in the calorimeter
needs to be matched to a track and the trigger electron candidate is required to 
have transverse energy $\ET > \SI{60}{\GeV}$, or $\ET > \SI{24}{\GeV}$ with
additional isolation requirements.

The single-muon trigger is based on muon candidates reconstructed in the muon spectrometer.
Muon-channel events are accepted by the trigger if they have either a muon with
transverse momentum $\pT > \SI{36}{\GeV}$ or an isolated muon with $\pT > \SI{24}{\GeV}$. 

Simulated signal and background samples were generated with an MC
technique. Detector and trigger simulations are performed 
within the dedicated ATLAS simulation software infrastructure
utilizing the 
\textsc{GEANT4} framework~\cite{SOFT-2010-01,Agostinelli:2002hh}.
The same offline reconstruction methods 
used with data events are applied to the samples of simulated events. 
Multiple inelastic $pp$ collisions (referred to as pile-up) are
simulated with \PYTHIAV{8}~\cite{Sjostrand:2007gs}, and are overlaid on
each MC event. Weights are assigned to the simulated events such that
the distribution of the number of pile-up interactions in the simulation 
matches the corresponding distribution in the data, which has an
average of 21~\cite{newlumi}.

Single-top-quark events from $t$-channel production are generated using the 
\POWHEGBOX~(r2556)~\cite{Frederix:2012dh} generator.
This generator uses the four-flavour scheme (4FS) for the NLO matrix
element (ME) calculations, since the 4FS leads to a more precise description of
the event kinematics compared to the five-flavour scheme (5FS).
Events are generated with the fixed four-flavour PDF 
set \ctf~\cite{Lai:2010vv} and the renormalisation and factorisation scales
are set to the recommendation given in Ref.~\cite{Frederix:2012dh}.
Top quarks are decayed at LO using \MADSPIN~\cite{Artoisenet:2012st}, preserving
all spin correlations.
The parton shower, hadronisation and the underlying event are 
modelled using the \PYTHIAV{6}~(v6.428)~\cite{Sjostrand:2006za} generator and 
a set of tuned parameters called the Perugia2012
tune (P2012)~\cite{Skands:2010ak}.

For the generation of single top-quarks in the $Wt$ and the $s$-channel
the \POWHEGBOX (r2819) generator~\cite{Re:2010bp,Alioli:2009je} with the \ct PDF set is used.
Samples of \ttbar events are generated with the \POWHEGBOX
(r3026)~\cite{Frixione:2007nw} and the \ct PDF set. 
In the event generation of \ttbar, the $\hdamp$ parameter, which controls the
\pT spectrum of the first additional emission beyond the Born configuration, is
set to the mass of the top quark.
The main effect of this is to regulate the high-\pT emission against 
which the \ttbar system recoils.
The parton shower, hadronisation and the underlying event are added
using \PYTHIAV{6} and the P2011C set of tuned parameters~\cite{Skands:2010ak}.

All top-quark processes are generated assuming a top-quark mass of \SI{172.5}{\GeV}. 
The decay of the top quark is assumed to be exclusively $t \to Wb$.

For studies of systematic uncertainties in all processes involving top quarks,
either alternative generators or parameter variations in the \POWHEGBOX + \PYTHIAV{6} setup are used.
To study the hadronisation modelling, the \POWHEGBOX generator interfaced 
to \HERWIGV{(v6.5.20)}~\cite{Corcella:2000bw} is used. 
The underlying event is simulated using the
\JIMMY~(v4.31)~\cite{Butterworth:1996zw} model with the ATLAS
AUET2~\cite{ATL-PHYS-PUB-2011-008} set of tuned parameters.
For studies of the NLO matching method,
\MGMCatNLO~(v2.2.2)~\cite{Alwall:2014hca} interfaced to \HERWIG is used.
Samples are generated using the \ctf PDF set in the ME calculations and the
renormalisation and factorisation scales are set to be the same as those implemented in \POWHEGBOX.
Again, the top quarks produced in the ME are decayed using \MADSPIN, preserving
all spin correlations.
Variations of the amount of additional radiation are studied by generating samples using
\POWHEGBOX + \PYTHIAV{6} after changing the hard-scatter scales and the
scales in the parton shower simultaneously.
In these samples, a variation of the factorisation and renormalisation 
scales by a factor of 2.0 is combined with the Perugia2012radLo parameters
and a variation of both parameters by a factor of 0.5 is combined with the
Perugia2012radHi parameters~\cite{Skands:2010ak}.
In the case of the up-variation,  the $\hdamp$ parameter is also changed
and set to two times the top-quark mass~\cite{ATL-PHYS-PUB-2015-002}.

Vector-boson production in association with jets, \vjets, is simulated using the 
multi-leg LO generator \SHERPAV{(v1.4.1)}~\cite{Gleisberg:2008ta} with its
own parameter tune and the \ct PDF set.
\SHERPA is used not only to generate the hard process, but also for the parton shower and
the modelling of the underlying event.
Samples of \wjets and \zjets events with up to four additional partons are generated.
The CKKW method~\cite{Hoeche:2009rj} is used to remove overlap between 
partonic configurations generated by the matrix element and by parton shower evolution.
Double counting between the inclusive $V$+$n$ parton samples and samples with 
associated heavy-quark pair production
is avoided consistently by applying the CKKW method 
also to heavy quarks~\cite{Hoeche:2009rj}. In \SHERPA, massive $c$- and
$b$-quarks are used in the ME as well as in the shower.

Diboson events, denoted $VV$, are also simulated using the \SHERPAV{(v1.4.1)}
generator. The matrix elements contain all diagrams with four electroweak
vertices.
They are calculated for zero additional partons at NLO and up to three additional partons at LO using the
same methodology as for \vjets production.
Only decay modes where one boson decays leptonically and the other boson decays hadronically are considered.
The \ct PDF set is used in conjunction with a dedicated set of parton-shower
parameters developed by the \SHERPA authors.

%% file: objects.tex
\section{Object definitions}
\label{sec:object}

Electron candidates are selected from energy deposits (clusters) in the LAr EM calorimeter 
associated with a well-measured track fulfilling strict quality requirements~\cite{PERF-2010-04, PERF-2013-03}.
Electron candidates are required to satisfy $\pT > \SI{25}{\GeV}$ and 
$|\eta_{\text{clus}}| < 2.47$, where $\eta_{\text{clus}}$ 
denotes the pseudorapidity of the cluster.  Clusters in the calorimeter barrel--endcap 
transition region, corresponding to $1.37<|\eta_{\text{clus}}|<1.52$, are ignored.
High-\pT~electrons associated with the $W$-boson decay can be mimicked by hadronic jets reconstructed 
as electrons, electrons from the decay of heavy quarks, and photon conversions.  
Since electrons from the $W$-boson decay are typically isolated from hadronic jet activity, backgrounds are
suppressed by isolation criteria, which require minimal calorimeter activity 
and only allow low-$\pT$ tracks in an $\eta$--$\phi$ cone around the electron candidate.
Isolation criteria are optimised to achieve a uniform selection efficiency of \SI{90}{\%} as a function of $\eta_{\text{clus}}$ 
and transverse energy, $\ET$.
The direction of the electron candidate is taken as that of the associated track.
Electron candidates are isolated by imposing thresholds on 
the scalar sum of the transverse momenta of calorimeter energy deposits
within a surrounding cone of size $\Delta R = 0.2$.
In addition, the scalar sum of all track transverse momenta within a cone of size $\Delta R = 0.3$ 
around the electron direction is required to be below a \pT-dependent threshold in the range between \SI{0.9}{\GeV} and \SI{2.5}{\GeV}.
The track belonging to the electron candidate is excluded from the sum. 

Muon candidates are reconstructed by matching track segments or complete tracks in the MS
with tracks found in the ID~\cite{PERF-2014-05}. The candidates are required to have 
$\pT > \SI{25}{\GeV}$ and to be in the pseudorapidity region $|\eta|<2.5$.
Isolation criteria are applied to reduce background events in which a high-\pT muon is produced
in the decay of a heavy-flavour quark.
An isolation variable is defined as the scalar sum of the transverse momenta of
all tracks with \pT above \SI{1}{\GeV}, excluding the one matched to the muon, within a cone of size $\Delta R_{\text{iso}} = \SI{10}{\GeV}/\pT(\mu)$.
The definition of $\Delta R_{\text{iso}}$ is inspired by the one used in
\Ref{\cite{Rehermann:2010vq}}. 
Muon candidates are accepted if they have an
isolation to $\pT(\mu)$ ratio of less than 0.05.
Events are rejected if the selected electron and the muon candidate share the
same ID track.

Jets are reconstructed using the \antikt algorithm~\cite{Cacciari:2008gp} 
with a radius parameter of $R=0.4$, 
using topological clusters~\cite{PERF-2014-07} as inputs to the jet finding. 
The clusters are calibrated with a local cluster weighting
method~\cite{PERF-2014-07}.
The jet energy is further corrected for the effect of multiple $pp$ interactions, both in 
data and in simulated events.
Calibrated jets~\cite{PERF-2012-01} using a transverse momentum- and $\eta$-dependent simulation-based
calibration scheme, with \emph{in situ} corrections based on data,
are required to have $\pT > \SI{30}{\GeV}$ and $|\eta|<4.5$. 
The minimum jet $\pT$ is raised to \SI{35}{\GeV} within the
transition region from the endcap to the forward calorimeter,
corresponding to $2.7<|\eta|<3.5$.

If any jet is within $\Delta R = 0.2$ of an electron, the closest jet is removed, since in these cases the 
jet and the electron are very likely to correspond to the same object. Remaining electron candidates 
overlapping with jets within a distance $\Delta R=0.4$ are subsequently rejected.
To reject jets from pile-up events, a so-called jet-vertex-fraction
criterion~\cite{PERF-2014-03} is applied for jets with $\pT < \SI{50}{\GeV}$ and $|\eta| <2.4$:
at least \SI{50}{\%} of the scalar sum of the \pT of tracks within a jet 
is required to be from tracks compatible with the 
primary vertex\footnote{The primary vertex is defined as the vertex with the largest $\sum \pT^2$ of the associated
tracks.} associated with the hard-scattering collision. 

Since $W+c$ production is a major background, a $b$-tagging algorithm
optimised to improve the rejection of $c$-quark jets is used. 
A neural-network-based algorithm is employed, which combines three different algorithms exploiting 
the properties of a $b$-hadron decay in a jet~\cite{PERF-2012-04}.
The resulting NN discriminant ranges from zero to one and is
required to be larger than 0.8349 for a jet to be considered $b$-tagged.
This requirement corresponds to a $b$-tagging efficiency of \SI{50}{\%}
and a $c$-quark jet and light-parton jet mistag acceptance of \SI{3.9}{\%} and
\SI{0.07}{\%}, respectively. 
These efficiencies are determined in simulated \ttbar
events.

The missing transverse momentum (with magnitude \MET) is calculated 
based on the vector sum of energy deposits in the calorimeter
projected onto the transverse plane~\cite{PERF-2011-07}.
All cluster energies are corrected using the local cluster weighting method.
Clusters associated with a high-\pT jet or electron are further calibrated using 
their respective energy corrections. In addition, the \pT
of muons with $\pT > \SI{5}{\GeV}$ is included in the calculation of \MET. The
muon energy deposited in the calorimeter is taken into account to avoid double counting.

%% file: selection.tex
\section{Event selection}
\label{sec:selection}

The event selection requires exactly one charged lepton ($\ell$), $e$ or $\mu$, 
exactly two jets, and $\MET > \SI{30}{\GeV}$.
Exactly one of the jets must be $b$-tagged. 
The selected lepton must be within $\Delta R=0.15$ of the lepton selected by the trigger. 
Candidate events are selected if they contain at least one good primary vertex candidate with at least five associated tracks, each of which has $\pT>\SI{400}{\MeV}$.
Events containing misreconstructed jets are rejected.
Misreconstructed jets are jets with $\pT > \SI{20}{\GeV}$ 
failing to satisfy quality criteria defined in \Ref{\cite{PERF-2011-03}}.

Multijet events produced in hard QCD processes may be selected, even though
there is no primary lepton from a weak-boson decay. This may happen if a jet is
misidentified as an isolated lepton, leading to a so-called fake lepton, or if the event has a
non-prompt lepton from a hadron decay which appears to be isolated.
The misidentification of jets as leptons is difficult to model in the detector
simulation, which is why two specific requirements are included in the event
selection to reduce the multijet background
without significantly reducing the signal efficiency. The first such requirement uses
the transverse mass of the lepton--\MET system,
\begin{equation}
  m_{\text{T}}\left(\ell\MET\right) = \sqrt{2 p_\mathrm{T}(\ell) \cdot \MET 
  \left[1-\cos \left( \Delta \phi \left(\ell, \MET \right) \right) \right]}\,, 
\label{eq:mTW}
\end{equation} 
and requires it to be larger than \SI{50}{\GeV}.
Further reduction of the multijet background is achieved by 
placing an additional requirement on events with a charged lepton 
that is back-to-back with the highest-\pT (leading) jet. 
This is realised by the following requirement between the lepton $\pT(\ell)$ and 
$\Delta \phi \left(j_1, \ell \right)$:
\begin{equation}
\pT\left(\ell\right) > \max \left(\SI{25}{\GeV}, \SI{40}{\GeV} \cdot \left(1 -  \frac{\pi - |\Delta \phi\left(j_1, \ell
\right)|}{\pi -1} \right) \right)\,,
\end{equation}
where $j_1$ denotes the leading jet.

Events with an additional lepton are vetoed to suppress \zjets and \ttbar
dilepton backgrounds. Only leptons with opposite charge to the
primary lepton are considered for this purpose.
These additional leptons are identified with less
stringent quality criteria than the primary lepton.
Additional leptons are not required to be isolated and must have $\pT >
\SI{10}{\GeV}$.
The pseudorapidity region in which additional electrons are identified
includes $|\eta(e)|<4.9$, and for additional muons $|\eta(\mu)|<2.5$.
Beyond the acceptance of the ID, forward electrons are identified within the
pseudorapidity range of $2.5<|\eta|<4.9$ based on calorimeter measurements
only~\cite{PERF-2010-04}.

Two separate vetoes are applied, depending on the flavour of the additional
lepton with respect to the primary lepton.
If the additional lepton has the same flavour as the primary lepton and the invariant mass of the
lepton pair is between 80 and \SI{100}{\GeV}, the event is rejected. 
If the additional lepton has a different flavour than the primary lepton,
the event is rejected unless the additional lepton is within $\Delta R=0.4$ to
the selected $b$-jet.

A requirement of $m(\ell b) < \SI{160}\GeV$, where $m(\ell b)$ is the invariant 
mass of the lepton and the $b$-tagged jet, is imposed, in order to exclude the 
off-shell region of top-quark decay beyond the kinematic limit of 
$m(\ell b)^2=\mtop^2-m_W^2$.
The off-shell region is not modelled well by the
currently available MC generators since off-shell effects are not
included in the underlying matrix-element calculation.

Selected events are divided into two different signal regions (SRs) according 
to the sign of the lepton charge. These two regions are denoted $\ell^+$ SR and
$\ell^-$ SR.

In addition, two validation regions (VRs) are defined to be orthogonal to the
SRs in the same kinematic phase space to validate the modelling of the main backgrounds, \wjets and \ttbar.
Events in the \wcr pass the same requirements as events in the SR except for the
$b$-tagging. Exactly one $b$-tagged jet is required, which is identified with a
less stringent $b$-tagging criterion than used to define the SR.
The NN-$b$-tagging discriminant must be in the interval (0.4051, 0.8349),
thereby excluding the SR beyond the higher threshold.
The \tcr is defined by requiring both jets to pass the same $b$-tagging
requirement that is used for the SR.

\section{Background estimation}
\label{sec:background_estimate}

For all background processes, except the multijet background, the normalisations are initially estimated by using 
MC simulation scaled to the theoretical cross-section predictions.
The associated production of an on-shell $W$ boson and a top quark ($Wt$) has a 
predicted production cross-section of \SI{22.3}{pb}~\cite{Kidonakis:2010ux}, 
calculated at NLO+NNLL accuracy.
The uncertainty in this cross-section is \SI{7.6}{\%}.
Predictions of the $s$-channel production are calculated at NLO using the same
methodology as for the $t$-channel production based on \Ref{\cite{Campbell:2004ch}} and yield a predicted cross-section
of \SI{5.2}{pb} with a total uncertainty of \SI{4.2}{\%}.

The predicted \ttbar cross-section is \SI{253}{pb}.
It is calculated with \texttt{Top++}
(v2.0)~\cite{Cacciari:2011hy,Baernreuther:2012ws,Czakon:2012zr,Czakon:2012pz,Czakon:2013goa,Czakon:2011xx}
at NNLO in QCD, including the resummation of NNLL soft-gluon terms.
The uncertainties due to the PDFs and $\alphas$ are calculated using the PDF4LHC
prescription~\cite{Botje:2011sn} with the \mstw \SI{68}{\%} CL NNLO, \ct NNLO
and \nnpdfthree PDF sets and are added in quadrature to the scale uncertainty,
leading to a total uncertainty in the cross-section of \SI{6}{\%}.

The cross-sections for inclusive $W$- and $Z$-boson production are predicted
with NNLO accuracy using the \textsc{FEWZ}
program~\cite{Gavin:2010az,Gavin:2012kw} to be \SI{37.0}{nb} and \SI{3.83}{nb}, respectively.
The uncertainty is \SI{4}{\%} and comprises the PDF and
scale uncertainties.

$VV$ events are normalised to the NLO cross-section of \SI{26.9}{pb} provided by 
\MCFM~\cite{Campbell:2011bn}.
The uncertainty in the inclusive cross-section for these processes is
\SI{5}{\%}.

The normalisation of the multijet background is obtained from a fit to the
observed \MET~distribution, performed independently in the signal and in the 
validation regions.
In order to select a pure sample of multijet background events, 
different methods are adopted for the electron and muon channels.
The \enquote{jet-lepton} model is used in the electron channel
while the \enquote{anti-muon} model is used in the muon channel~\cite{ATLAS-CONF-2014-058}.
In case of the \enquote{jet-lepton} model, a dedicated selection is imposed on MC simulated dijet events,
in order to enrich events with jets that are likely to resemble a lepton in the detector.
The jet candidates are treated as a lepton henceforth.
The \enquote{anti-muon} model imposes a dedicated selection on data to enrich events that contain fake muons.
To determine the normalisation of the multijet background, 
a binned maximum-likelihood fit is performed on the \MET~distribution using the observed data,
after applying all selection criteria except for the cut on \MET. 
Fits are performed separately in two $\eta$ regions for electrons: in the barrel ($|\eta| < 1.37$)
and endcap ($|\eta| > 1.52$)  region of the electromagnetic calorimeter, 
i.e.\ the transition region is excluded. 
For muons, the complete $\eta$ region is used.
For the purpose of this fit, the contributions from
\wjets, the contributions from \ttbar and single top-quark production, and the
contributions from \zjets and $VV$ production, are combined into one
template.
The normalisation of \zjets and $VV$ backgrounds is fixed during the fit, 
as their contribution is small.

\begin{figure}[htbp]
  \centering
  \subfloat[][]{%
    \includegraphics[width=0.46\textwidth]{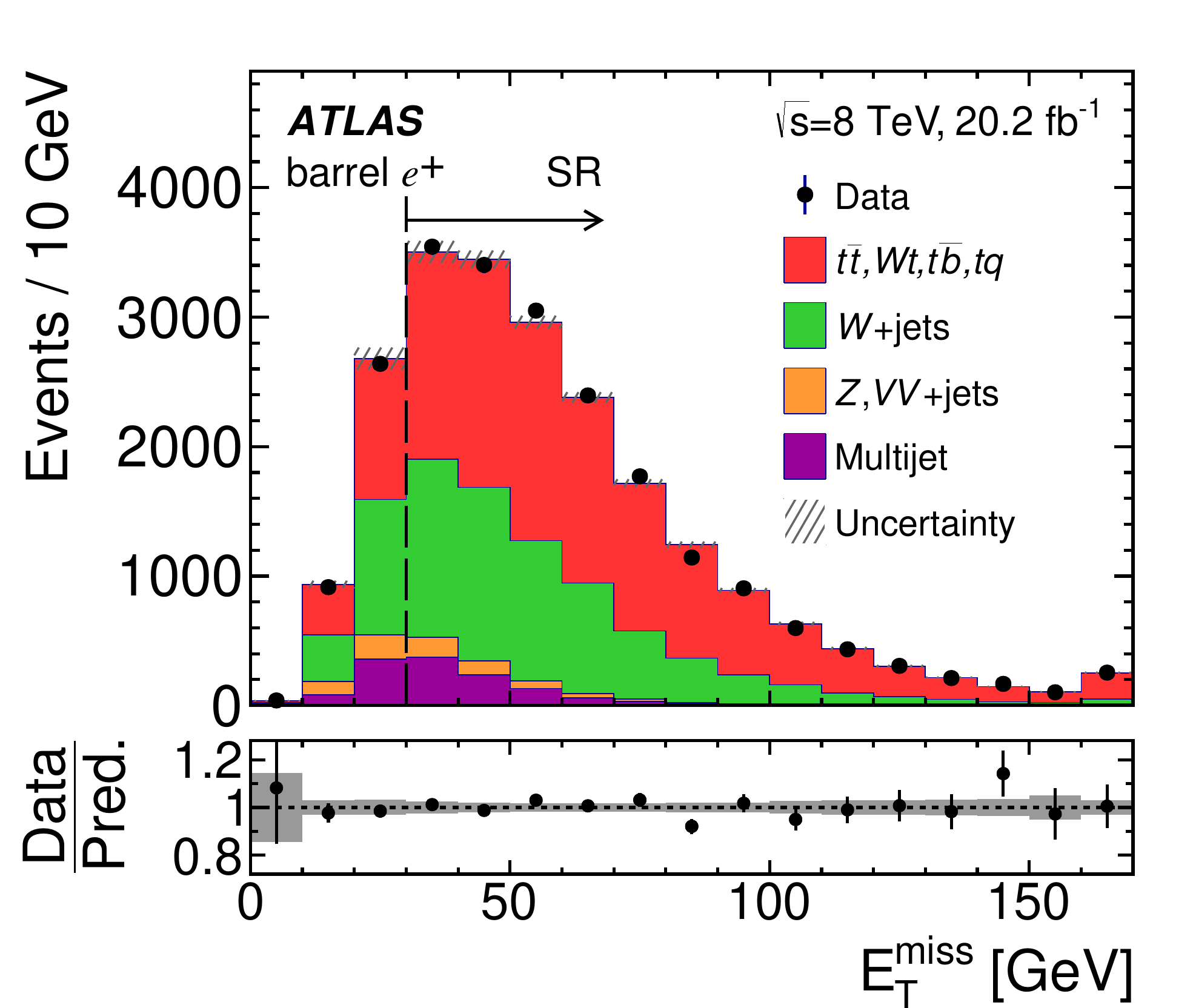}
    \label{fig:missetfit_ele}
  }
  \quad
  \subfloat[][]{%
    \includegraphics[width=0.46\textwidth]{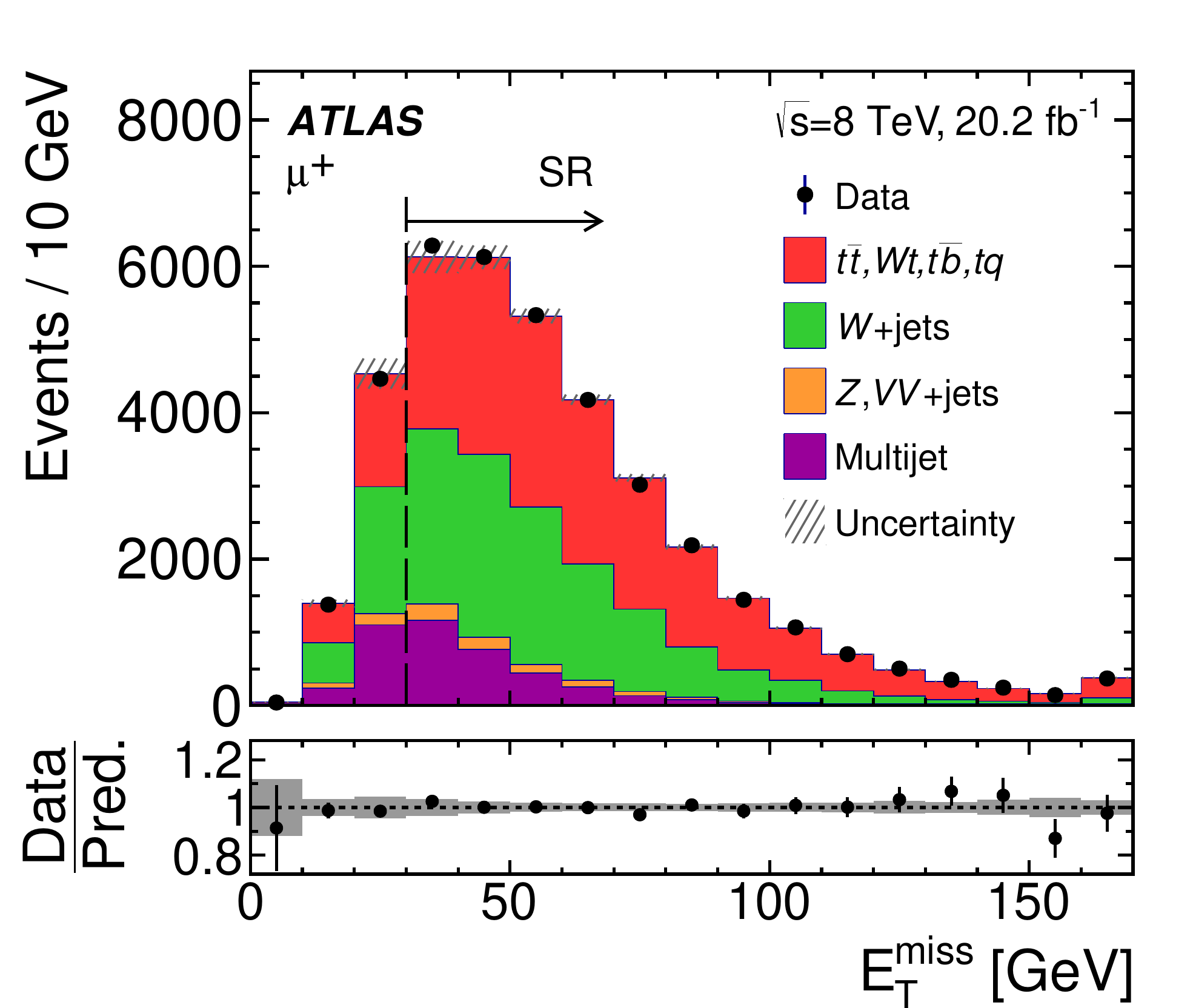}
    \label{fig:missetfit_muo}
  }
  \caption{\label{fig:missetfit}
    Observed distributions of the missing transverse momentum, \MET, in the
    signal region (SR), including events with $\MET<\SI{30}{\GeV}$, for \protect\subref{fig:missetfit_ele} events in the \epc with an electron in
    the barrel region and for \protect\subref{fig:missetfit_muo} events in the
    \mpc , compared to the model obtained from simulated events.
    The normalisation is obtained from the 
    binned maximum-likelihood fit to the full \MET distributions, and applied to the SR.
    The hatched uncertainty band represents the MC statistical uncertainty and the normalisation of the multijet background.
    The ratio of observed~(Data) to predicted~(Pred.) number of events in
    each bin is shown in the lower panel.
    Events beyond the $x$-axis range are included in the last bin.}
\end{figure}

The \MET~distributions, after rescaling the different backgrounds and the multijets template to their 
respective fit results, are shown in \Fig{\ref{fig:missetfit}} for both the \epc and \mpc.
The estimated event rates obtained from the binned maximum-likelihood fit for the combined contributions 
of \wjets, \ttbar and single top-quark production are not used in the later analysis
and are only applied to scale the respective backgrounds in order to check the modelling of the
kinematic distributions. For the later NN training, as well as for the final
statistical analysis, the normalisation for all but the multijets background is taken solely from MC simulations scaled 
to their respective cross-section predictions.
Based on comparisons of the rates using an alternative method, 
namely the matrix method~\cite{ATLAS-CONF-2014-058},
a systematic uncertainty of ~\SI{15}{\%} 
is assigned to the estimated multijet yields.

Table~\ref{tab:yields} summarises the event yields in the signal region for each of the
background processes considered, together with the event yields for the signal
process.
The quoted uncertainties are statistical uncertainties and the uncertainty in
the number of multijet events.
The yields are calculated using the acceptance from MC samples
normalised to their respective theoretical cross-sections.

\begin{table}[htbp]
\sisetup{round-mode=figures}
\centering
\begin{tabular}{
                 l S[table-format=5.0,round-precision=3]@{$\,\pm\,$}S[table-format=4.0]
                   S[table-format=5.0,round-precision=3]@{$\,\pm\,$}S[table-format=4.0]
}
    \toprule
    Process & \multicolumn{2}{c}{$\ell^{+}$ SR} &  \multicolumn{2}{c}{$\ell^{-}$ SR} \\ 
    \midrule
    $tq$        & 11436.2 &  469 & 17  & 1   \\
    $\bar{t}q$  &  10 & 1 & 6287.1 & 352.1     \\
    \midrule
    $t\bar{t},Wt,t\bar{b}/\bar{t}b$  & 18392 & 1126 & 18032 & 1104      \\
    \wpjets    & 18662 & 3732  & 47 & 10    \\
    \wmjets    & 25&	5	&14008	&2801     \\
    $Z,VV+\mathrm{jets}$  & 1289	&258	&1190.8	&238  \\
    Multijet   & 4523	&710	&4523&	655 \\
    \midrule
    Total expected & 54337	&3953	&44106	&3059  \\
    Data  & \multicolumn{2}{l}{\num[round-mode=off]{55800}} & \multicolumn{2}{l}{\num[round-mode=off]{44687}}\\  
    \bottomrule
\end{tabular}
\caption{\label{tab:yields} Predicted and observed event yields for the signal region~(SR).
The multijet background prediction is obtained from a binned maximum-likelihood
fit to the \MET distribution.
All the other predictions are derived using theoretical cross-sections, given
for the backgrounds in \Sect{\ref{sec:background_estimate}} and for the signal
in \Sect{\ref{sec:intro}}. The quoted uncertainties are in the predicted
cross-sections or in the number of multijet events, in case of
the multijet process.}
\end{table}

%% file: xsect_def.tex
\section{Measurement definitions}
\label{sec:xsect_def}

The paragraphs below describe the concepts and definitions on
which the cross-section measurements are based.

\subsection{Fiducial and total cross-sections}
\label{sec:fiducial:def}
Measuring a production cross-section with respect to a
fiducial volume ($\sigfid$) has the benefit of reducing systematic uncertainties
related to MC generators, since the extrapolation to the full phase space is avoided.
In the usual case of a total cross-section measurement
the measured cross-section is given by
\begin{equation}
  \label{eq:tot_xs}
  \sigtot = \frac{\hat{\nu}}{\epsilon \cdot L_\text{int}} \quad 
  \text{with} \quad \epsilon = \frac{N_\text{sel}}{N_\text{total}}\,,
\end{equation}
where $\hat{\nu}$ is the measured expectation value of the number of signal events 
and $\epsilon$ is the event selection efficiency, defined as the ratio of
$N_\text{sel}$, the number of events after applying all selection cuts
on a sample of simulated signal events, and $N_\text{total}$, the total number
of events in that sample before any cut.

When defining a fiducial phase space, which is typically chosen to be close to
the phase space of the selected data set, the fiducial acceptance is given by
\begin{equation}
  A_\text{fid} = \frac{N_\text{fid}}{N_\text{total}}\,,
\end{equation}
with $N_\text{fid}$ being the number of generated events after applying the 
definition of the fiducial volume.
The fiducial cross-section can be defined with respect to the fiducial phase space as
\begin{equation}
 \label{eq:fid_xs}
 \sigfid = \frac{N_\text{fid}}{N_\text{sel}}\cdot
 \frac{\hat{\nu}}{L_\text{int}}\,. 
\end{equation}
From \Eqn{\eqref{eq:fid_xs}} it is apparent that systematic effects which alter
$N_\text{fid}$ and $N_\text{sel}$ by the same factor do not lead to an
uncertainty in $\sigma_\text{fid}$ since the changes cancel.
Using $\sigfid$ and $A_\text{fid}$, 
\Eqn{\eqref{eq:tot_xs}} can be written as
\begin{equation}
  \label{eq:sig_tot_sig_fid}
  \sigtot = \frac{1}{A_\text{fid}}\cdot \sigfid\,,
\end{equation}
corresponding to the extrapolation of the fiducial cross-section to the full
phase space.

\subsection{Particle-level objects}
\label{sec:particle_level_objects}
The definition of a fiducial phase space requires the implementation of the
event selection at generator level. The corresponding particle-level 
objects are constructed from stable particles of the MC event record with a lifetime
larger than \SI{0.3E-10}{\second}, using the following criteria.

Particle-level leptons are defined as electrons, muons or neutrinos that
originate from a $W$-boson decay, including those emerging from a subsequent
$\tau$-lepton decay.
However, since certain MC generators do not include $W$ bosons in the MC record,
an implicit $W$-boson match is employed to achieve general applicability.
This implicit requirement excludes leptons from hadronic decays, either directly or via a $\tau$ decay.
The remaining leptons are assumed to come from a $W$-boson decay.
In $t$-channel single-top-quark events, exactly one such electron or muon and
the corresponding neutrino are present.
The selected charged-lepton four-momentum is calculated including photons within
a cone of size $\Delta R = 0.1$.

Particle-level jets are reconstructed using the \antikt algorithm with a radius parameter of $R=0.4$.
All stable particles are used to reconstruct the jets, except for the selected
electron or muon and the photons associated with them. 
Particle-level jets are identified as $b$-jets, if
the jet is within $|\eta| < 2.5$ and a $b$-hadron is associated with a
ghost-matching technique as described in \Ref{\cite{Cacciari:2008gn}}. 
Events are rejected, if a selected particle-level lepton is identified within a
cone of size $\Delta R = 0.4$ around a selected particle-level jet.

The particle-level event selection is designed to be close to the one used at reconstruction level.
Exactly one particle-level electron or muon with $\pT > \SI{25}{\GeV}$ and
$|\eta| < 2.5$ is required. There must be two particle-level jets with $\pT >
\SI{30}{\GeV}$ and $|\eta| < 4.5$; exactly one of these jets must be 
a $b$-jet.
The invariant mass of the lepton--$b$-jet system must fulfil $m(\ell b) < \SI{160}{\GeV}$.

\subsection{Pseudo top quarks}
\label{sec:pseudotop:def}

Differential cross-sections characterise the top-quark kinematics. 
To facilitate the comparison between measurements and predictions,
the top-quark objects have to closely correspond in both cases.
While parton-level definitions of the top-quark are affected by ambiguities at
NLO accuracy in calculations and incur related uncertainties, top-quark
definitions based on stable particles in MC generators form a solid foundation.
On the other hand, some calculations are only available at parton level.
Following this logic, a top-quark proxy called a pseudo top quark is
defined~\cite{TOPQ-2013-07}, based on the particle-level objects given in
\Sect{\ref{sec:particle_level_objects}}.
Variables calculated using the pseudo top quark are denoted by $\hat{t}$,
while the untagged jet is written as $\hat{j}$.

The reconstruction of the pseudo top quark starts from its decay products: the
$W$ boson and the $b$-tagged jet.
The $W$ boson is reconstructed from the charged lepton and the neutrino at
particle level.
The $z$ component of the neutrino momentum, $p_{z}(\nu)$, is calculated using the $W$-boson mass as a constraint.
If the resulting quadratic equation has two real solutions,
the one with smallest absolute value of $|p_{z}(\nu)|$ is chosen. In case of complex solutions, which
can occur due to the low \MET resolution, a kinematic fit is performed that rescales the neutrino $p_x$ and $p_y$
such that the imaginary part vanishes and at the same time the transverse
components of the neutrino momentum are kept as close as possible to the \MET.
There are two jets in the events considered and exactly one of the jets is
required to be $b$-tagged.
The pseudo top quark is then formed by adding the four-momenta of the $W$ boson
and the $b$-tagged jet.

%% file: nn.tex
\section{Separation of signal from background}
\label{sec:nn}

\begin{figure}[htbp]
  \centering
  \subfloat[][]{%
    \includegraphics[width=0.46\textwidth]{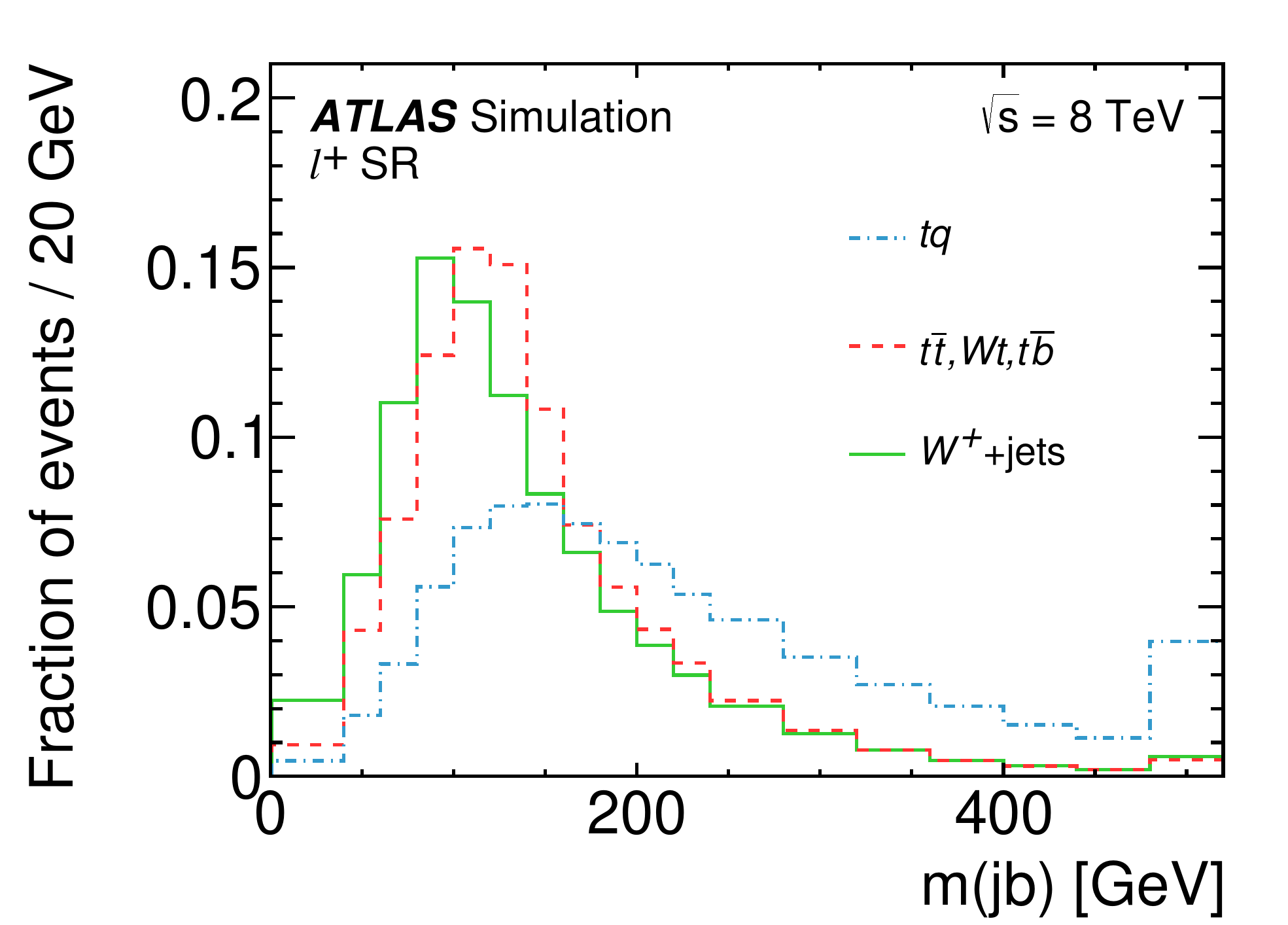}
    \label{fig:input_vars_shapes_a}
  }
  \quad
  \subfloat[][]{%
    \includegraphics[width=0.46\textwidth]{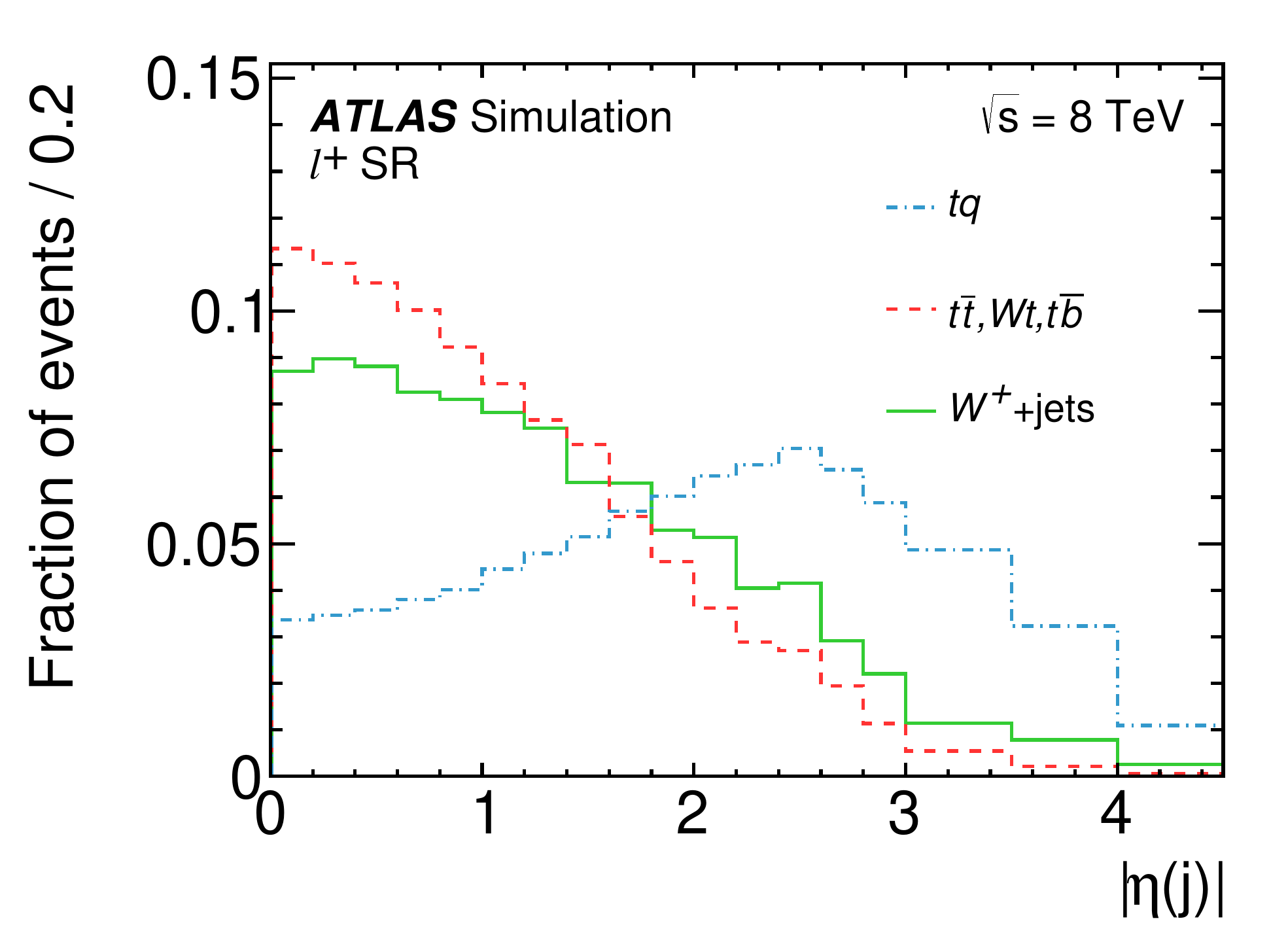}
    \label{fig:input_vars_shapes_b}
  }
  \caption{\label{fig:input_vars_shapes}
    Probability densities of the two most discriminating input variables to the
    NN:
    \protect\subref{fig:input_vars_shapes_a} the invariant mass $m(j b)$ of 
    the untagged jet and the $b$-tagged jet, and 
    \protect\subref{fig:input_vars_shapes_b} 
    the absolute value of the pseudorapidity of the untagged jet $|\eta(j)|$.
    The distributions are shown for the $tq$ signal process, the \wpjets
    background and the top-quark background in the \lp SR.
    Events beyond the $x$-axis range are included in the last bin.
  }
\end{figure}

A neural network (NN)~\cite{Feindt:2006pm} is employed to separate signal from
background events, by combining several kinematic variables into an optimised NN
discriminant (\NNout).
The reconstruction of top-quark-related kinematic variables, the ranking of
input variables according to their discriminating power, and the training
process of the NN follow closely the procedures used in previous ATLAS
publications about $t$-channel single top-quark
production~\cite{TOPQ-2011-14,TOPQ-2012-21}.
The input variables used for the NN are determined by a study in which
the expected uncertainties in the cross-section measurements are computed for
different sets of variables.
The procedure starts from an initial set of
17 variables used in previous analyses~\cite{TOPQ-2011-14,TOPQ-2012-21}. 
These variables are ranked based on the algorithm described in
Ref.~\cite{TOPQ-2011-14}.
One variable after the other is removed from the network according to the ranking, starting with the
lowest-ranked one, followed by the next-lowest-ranked one,
and so forth.
In each iteration step the full analysis is performed and the expected
uncertainty of the measurement is determined.
As a result of the study, it is found that the reduction from the set of six
highest-ranking variables to a set of five highest-ranking variables leads to a
significant increase in the uncertainty in the cross-sections. Finally, the
seven highest-ranking input variables are chosen, in order to avoid sudden changes in the
uncertainty due to statistical fluctuations.
The input variables to the NN and their definitions are given in Table~\ref{tab:NNinputVars}.
The separation between signal and the two most important backgrounds, i.e.\ the
top-quark background and the \wjets background, is illustrated in
\Fig{\ref{fig:input_vars_shapes}} for the two most discriminating variables.
\begin{table}[htbp]
  \centering
  \begin{tabular}{ll}
  \toprule
  Variable symbol & Definition\\
  \midrule
  $m(j b)$ & The invariant mass of the untagged jet ($j$) and the $b$-tagged jet ($b$). \\
  $|\eta(j)|$ & The absolute value of the pseudorapidity of the untagged jet. \\
  $m(\ell \nu b)$  & The invariant mass of the reconstructed top quark. \\
  $m_{\mathrm{T}}(\ell\MET)$ & The transverse mass of the lepton--\MET system, as defined in \Eqn{\eqref{eq:mTW}}. \\
  $|\Delta \eta(\ell \nu,b)|$ & The absolute value of $\Delta \eta$ between the
  reconstructed $W$ boson and the $b$-tagged jet. \\
  $m(\ell b)$ &  The invariant mass of the charged lepton ($\ell$) and the $b$-tagged jet. \\
  $\cos \theta^*(\ell,j)$ & The cosine of the angle, $\theta^*$, between the charged lepton and the untagged \\    
    & jet in the rest frame of the reconstructed top quark. \\
  \bottomrule
  \end{tabular}
  \caption{The seven input variables to the NN ordered by their discriminating
  power. The jet that is not $b$-tagged is referred to as \textit{untagged}
  jet.}
  \label{tab:NNinputVars}
\end{table}

The training of the NN is done with a sample of simulated events that comprises 
events with leptons of positive and negative charge.
This approach gives the same sensitivity as a scenario in
which separate NNs are trained in the \lp SR and in the \lm SR.
The modelling of the input variables is checked in the \wcr and in the \tcr;
see \Sect{\ref{sec:selection}} for the definition. 
In the \tcr both jets are $b$-tagged, which poses the question how to define
variables which are using the untagged jet in the SR. The two $b$-jets are
sorted in $|\eta|$ and the jet with the highest $|\eta|$ is assigned to mimic the untagged
jet of the SR. The distributions of all input variables are found to be well
modelled in the VRs.

\begin{figure}[htbp]
  \centering
  \subfloat[][]{
    \includegraphics[width=0.46\textwidth]{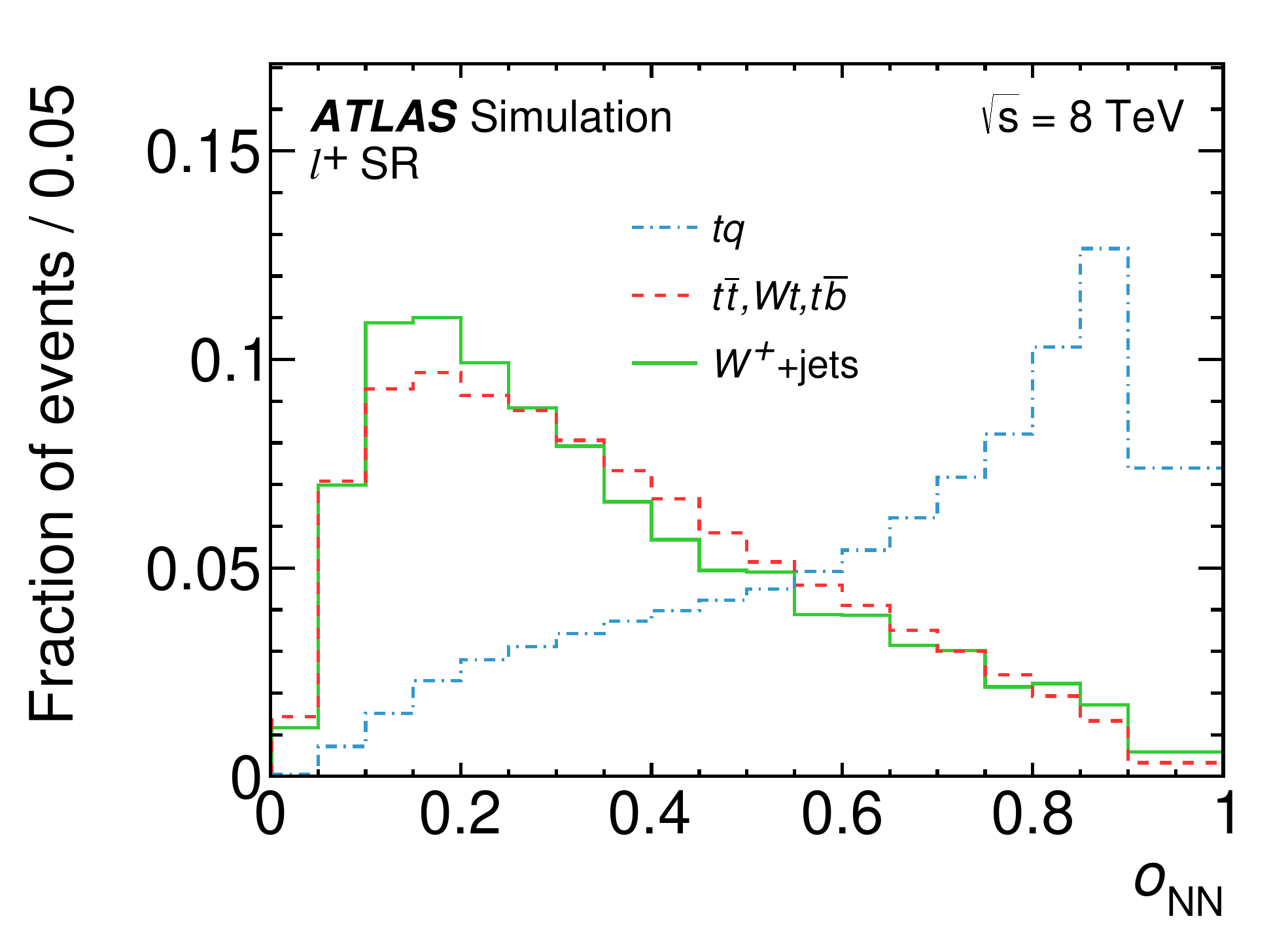}
    \label{subfig:2j_plus_NNshape}
  }
  \quad
  \subfloat[]{
    \includegraphics[width=0.46\textwidth]{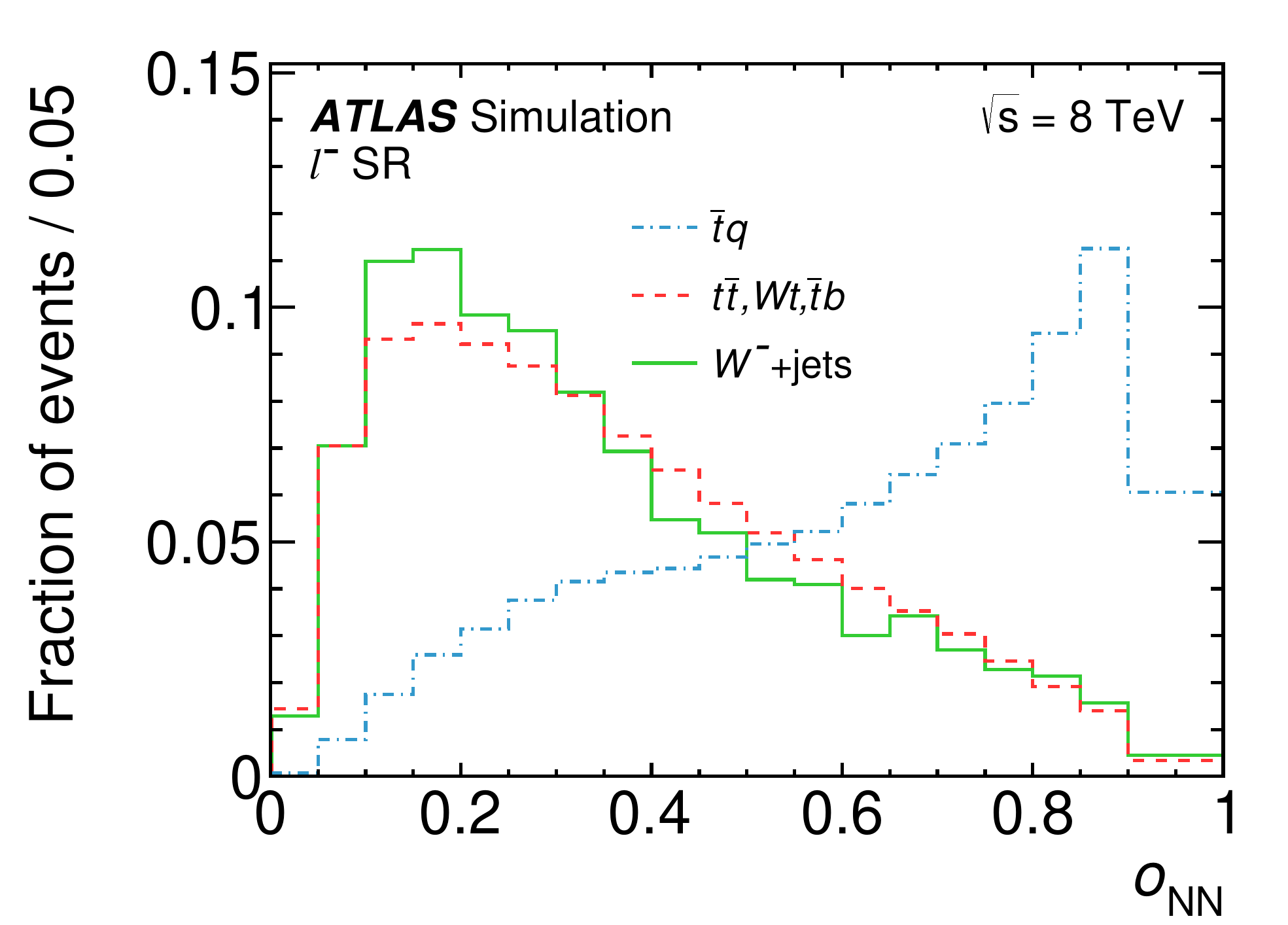}
    \label{subfig:2j_minus_NNshape}
  }
  \caption{\label{fig:nn_templates}
    Probability densities of the NN discriminants in the signal region (SR) for 
    the $tq$ and $\bar{t}q$ signal processes, the \wjets background and the top-quark background: 
    \protect\subref{subfig:2j_plus_NNshape} in the \lp SR and
    \protect\subref{subfig:2j_minus_NNshape} in the \lm SR.}
\end{figure}

\begin{figure}[t]
  \centering
  \subfloat[][]{
    \includegraphics[width=0.48\textwidth]{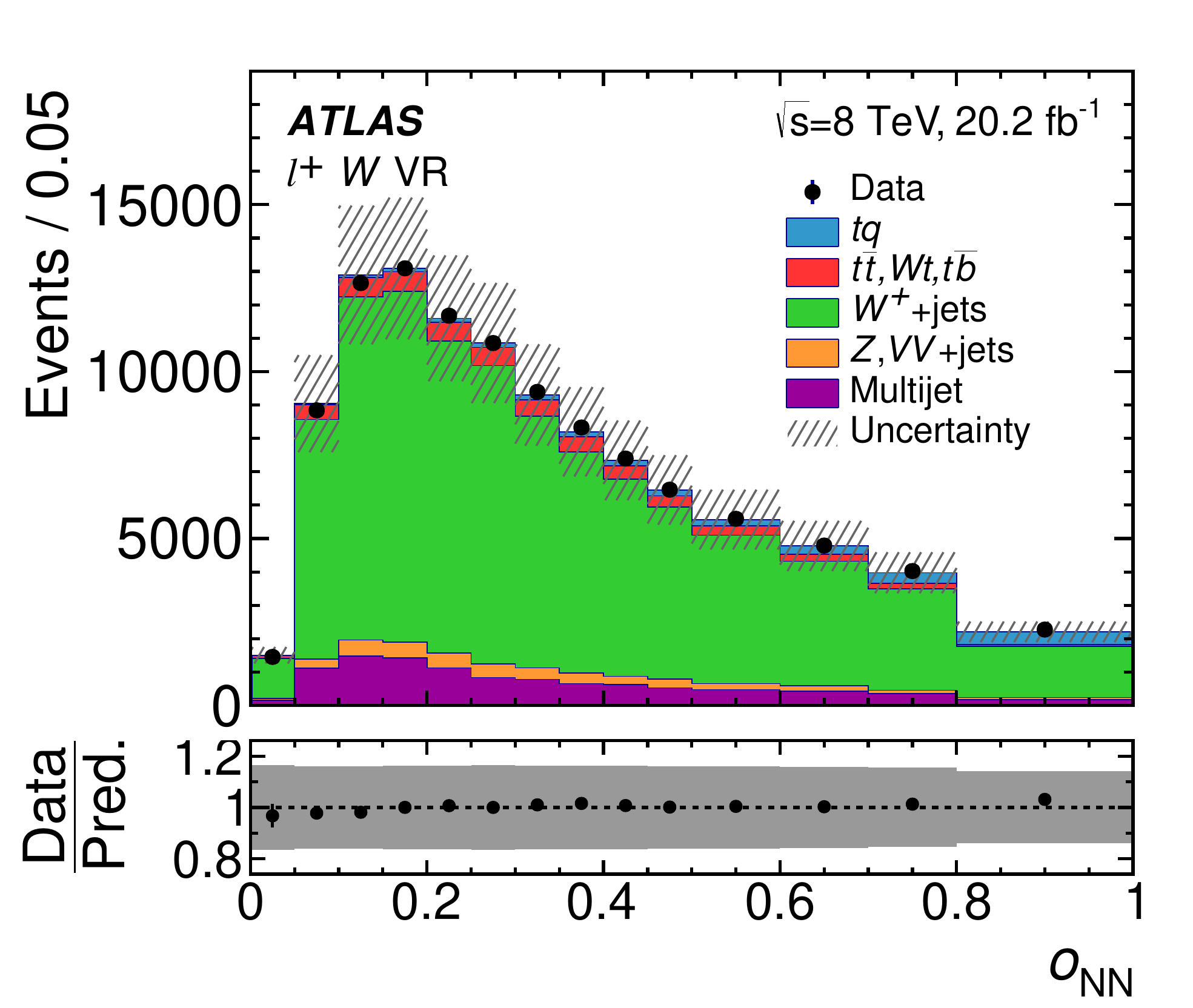}
    \label{subfig:NN_Wjet_plus}
    }
  \subfloat[][]{
    \includegraphics[width=0.48\textwidth]{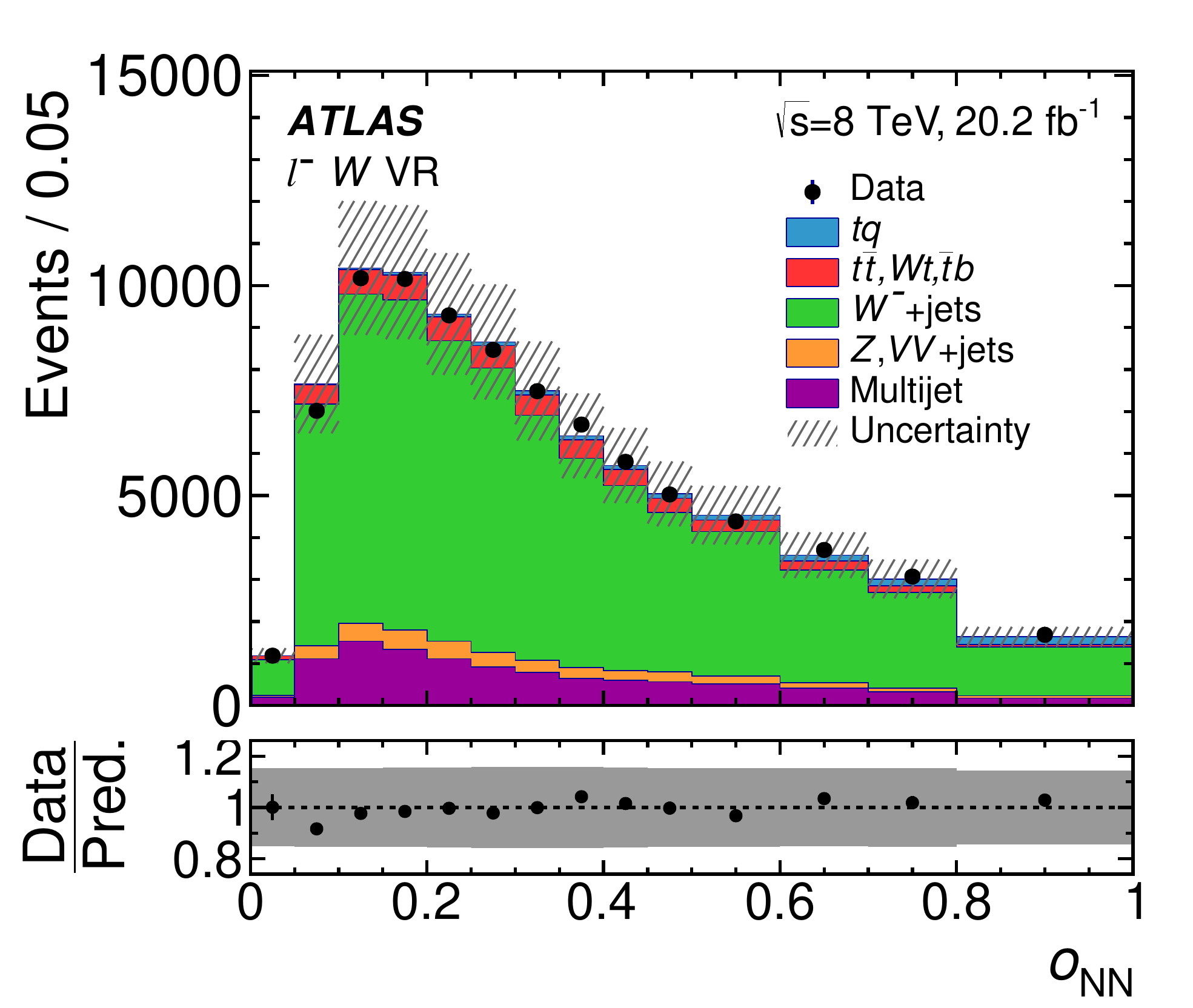}
    \label{subfig:NN_Wjet_minus}
    } \vspace*{3mm} \\
  \subfloat[]{
    \includegraphics[width=0.48\textwidth]{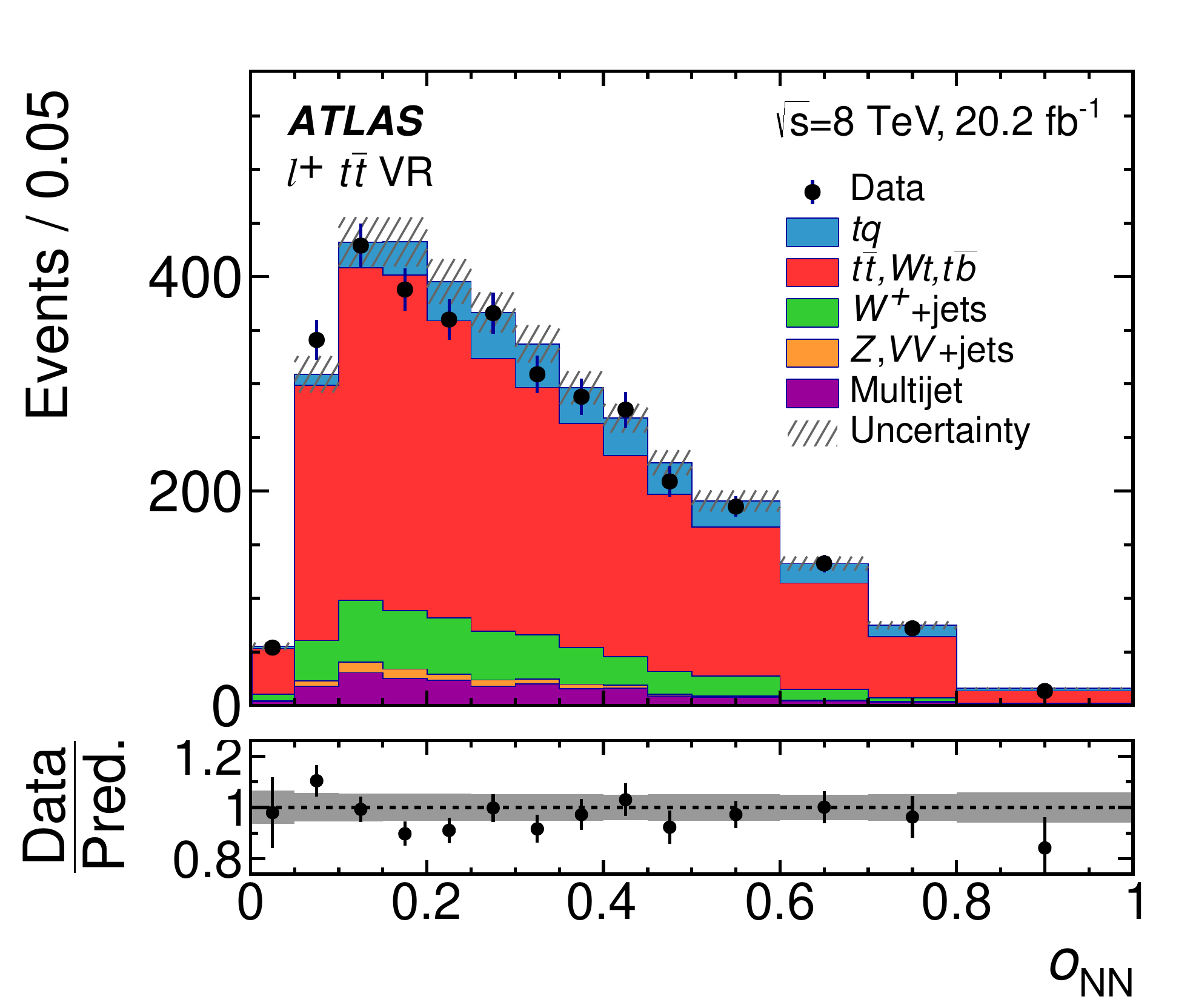}
    \label{subfig:NN_ttbar_plus}
    }
  \subfloat[]{
    \includegraphics[width=0.48\textwidth]{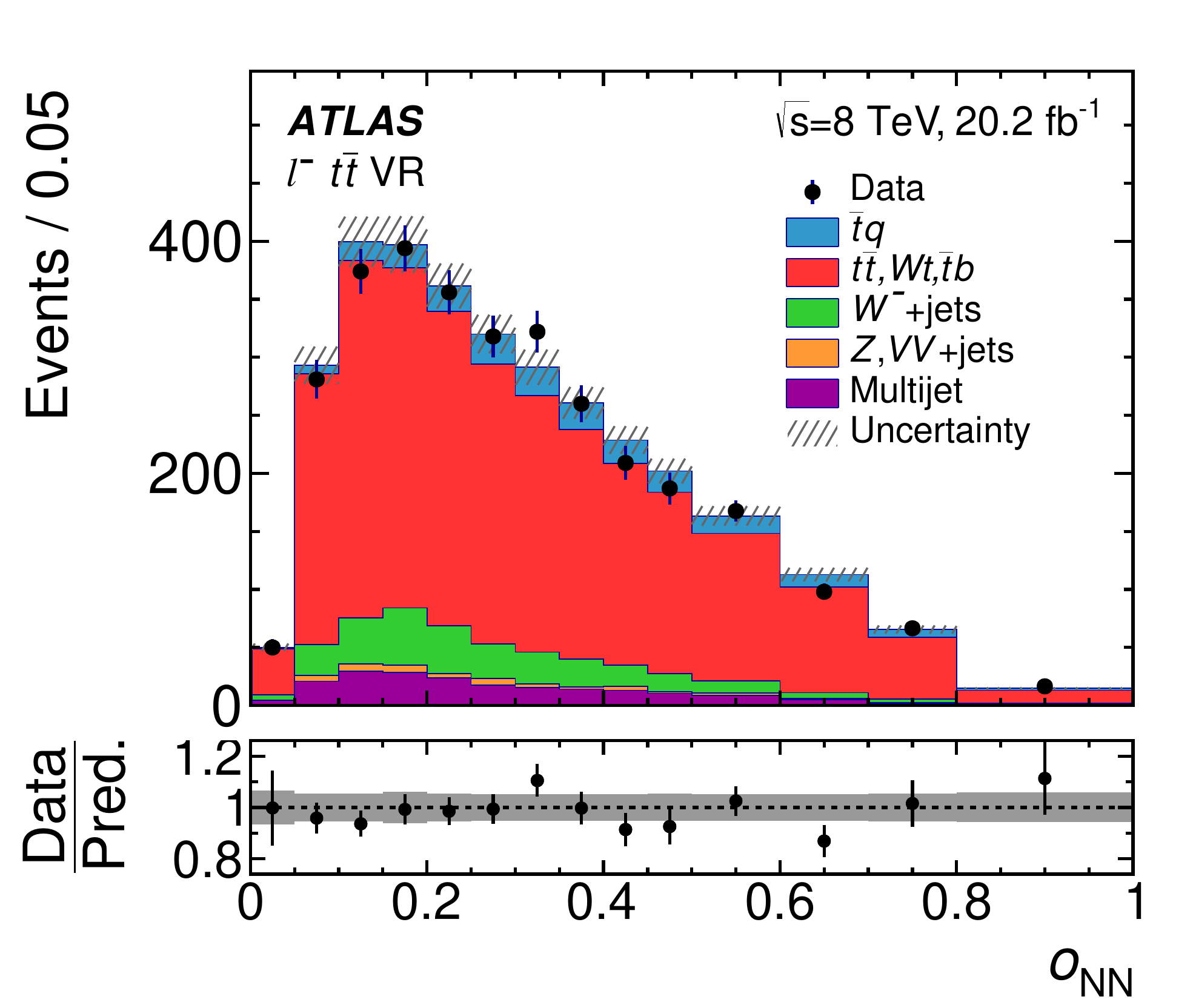}
    \label{subfig:NN_ttbar_minus}
    }
  \caption{\label{fig:nnCR}
    Observed \NNout distributions
    (a, b) in the \wcr
    and (c, d) in the \tcr
    compared to the model obtained from simulated events.
    The simulated distributions are normalised to the event rates obtained by
    the fits of the \MET distributions as described in
    \Sect{\protect\ref{sec:background_estimate}}.
    The hatched uncertainty band represents the uncertainty in the
    pre-fit process cross-sections and the bin-by-bin MC statistical
    uncertainty, added in quadrature.
    The lower panels show the ratio of the observed to the expected number of
    events in each bin.}
\end{figure}

In \Fig{\ref{fig:nn_templates}}, the probability densities of the resulting
\NNout distributions are shown for the signal, the top-quark background, and the
\wjets~background.
The modelling of collision data with simulated events is further tested by
applying the NNs in the validation regions.
The corresponding  distributions are shown in \Fig{\ref{fig:nnCR}}.
Good agreement between the model and the measured distributions is found.

%% file: sys.tex
\section{Systematic uncertainties}
\label{sec:sys}

Many sources of systematic uncertainty affect the individual top-quark and top-antiquark 
cross-section measurements and their ratio.
The uncertainties are split into the following categories:
 
\paragraph{Object modelling:}

Systematic uncertainties due to the residual differences between 
data and MC simulation, for reconstructed jets, electrons and muons 
after calibration, and uncertainties in corrective scale 
factors are propagated through the entire analysis.
The main source of object modelling uncertainty is the jet energy
scale (JES).

Uncertainties in the lepton trigger, reconstruction, and selection efficiencies
in simulations are estimated from measurements of the efficiency using 
$Z\rightarrow\ell^{+}\ell^{-}$ decays. 
To evaluate uncertainties in the lepton momentum scale and resolution,
the same processes are used~\cite{PERF-2013-05}. The uncertainty in the 
charge misidentification rates was studied and found to be negligible
for this analysis. 

The jet energy scale was derived using
information from test-beam data, LHC collision data and simulation. Its uncertainty increases
with $\eta$ and decreases with the $\pT$ of the reconstructed
jet~\cite{PERF-2012-01}.
The JES uncertainty has various components originating from the calibration method, the calorimeter response,
the detector simulation, and the specific choice of parameters in the parton shower and fragmentation 
models employed in the MC event generator. Additional contributions 
come from the modelling of pile-up effects, differences between
$b$-quark-induced jets and light-quark or gluon-induced jets.
Included in the JES components are 
also uncertainties in the flavour composition of the
jets and the calorimeter response to jets of different flavours.
Both JES flavour uncertainties are reduced by using
actual gluon-fractions of the untagged jet obtained from simulated signal samples. 
A parameterisation with 22 uncorrelated components is used, as described in
\Ref{\cite{PERF-2012-01}}.

Small uncertainties arise from the modelling of the jet energy resolution and the
missing transverse momentum, which accounts for contributions of calorimeter cells 
not matched to any jets, low-\pT jets, and pile-up.
The effect of uncertainties associated with the jet-vertex fraction is also considered for each jet.

Since the analysis makes use of $b$-tagging, the uncertainties in the $b$- and
$c$-tagging efficiencies and the mistag rates~\cite{ATLAS-CONF-2014-046,ATLAS-CONF-2014-004} 
are taken into account and called flavour tagging uncertainty.
Since the interaction of matter and antimatter with the detector material is
different, the difference in the $b$-tagging efficiency between jets initiated
by a $b$-quark and a $b$-antiquark is estimated and results to be
$\sim$\SI{1}{\%} based on simulated \tq and \tbarq events .

\paragraph{Monte Carlo generators and parton densities:}
Systematic uncertainties from MC modelling are estimated by comparing different
generators and varying parameters for the event generation.
These uncertainties are estimated for all processes involving top quarks, and
taken to be correlated among the \tq and \tbarq processes and uncorrelated
between these two and the top-quark background (\ttbar, $Wt$, $t\bar{b}$, and
$\bar{t}b$).

The uncertainty due to the choice of 
factorisation scale and renormalisation scale in the ME computation of the
MC generators is estimated by varying these scales independently by factors of
one half and two using the \POWHEGBOX generator. In addition, a different set of
tuned parameters of the \PYTHIA parton shower with modified \alphas is
used to match the scale variation in the ME. The detailed list of modified
parameters is given in~\Ref{\cite{Skands:2010ak}}. The uncertainty is defined by
the envelope of all independent variations.

Systematic uncertainties in the matching of the NLO matrix calculation and the 
parton shower are estimated by
comparing samples produced with \MCatNLO and with \POWHEGBOX,
in both cases interfaced to the \HERWIG parton shower.
For the $tq$ and $\tbar q$ processes, \MGMCatNLO is used instead of \MCatNLO.

The uncertainty from the parton shower and hadronisation modelling is estimated
by comparing samples produced with \POWHEGBOX + \HERWIG  and \POWHEGBOX + \PYTHIA. 

Systematic uncertainties related to the PDFs are taken into account for all
processes, except for the \zjets, due to the small yield, and multijet contributions.
The uncertainty is estimated following the PDF4LHC
recommendation~\cite{Butterworth:2015oua}, using the \textsc{PDF4LHC15\_NLO} PDF
set.
In addition, the acceptance difference between \textsc{PDF4LHC15\_NLO}
and \ct is considered,
since the latter PDF set is not covered by the uncertainty obtained with \textsc{PDF4LHC15\_NLO}.
The total PDF uncertainties are dominated by the
acceptance differences between \ct and \textsc{PDF4LHC15\_NLO}. For the two
signal processes the correlation coefficient of the total PDF uncertainties is
found to be close to one.

Modelling uncertainties in the \wjets sample are investigated 
using particle-level distributions obtained with the \SHERPA event generator 
by varying simultaneously the factorisation and renormalisation scales.
The corresponding fractional changes with respect to the nominal
particle-level $\pT(W)$ distribution are applied to the reconstructed $\pT(W)$
distribution and modified \NNout distributions are obtained.
The effect on the measured $t$-channel cross section is found to be negligible.

Finally, the MC statistical uncertainty is included.

\paragraph{Background normalisation:}

The uncertainties in the normalisation of the various background processes are
estimated by using the uncertainties in the theoretical cross-section
predictions as detailed in \Sect{\ref{sec:background_estimate}}.

For the \wjets and \zjets backgrounds, an uncertainty of \SI{21}{\%} is assigned. 
This uncertainty is estimated based on parameter variations in the generation of
the \SHERPA samples.
It was found that a correlated variation of the factorisation and renormalisation scales has 
the biggest impact on the kinematic distributions and produces variations covering the 
unfolded $Z/W$+jets data and their uncertainties~\cite{STDM-2012-24}.

The multijet background estimate has an uncertainty of \SI{15}{\%},
based on comparisons of the default method with the yield obtained 
with the matrix method~\cite{ATLAS-CONF-2014-058}. Additionally an uncertainty
in the shape of distributions is defined in the same way.

\paragraph{Luminosity:}
The absolute luminosity scale is derived from beam-separation scans performed in November 2012. 
The uncertainty in the integrated luminosity is \SI{1.9}{\%}~\cite{newlumi}.

%% file: results_inclusive_xs.tex
\section{Fiducial and total cross-section measurements}
\label{sec:xsect_tot}

The signal yields $\hat{\nu}(\tq)$ and $\hat{\nu}(\tbarq)$ (see
\Eqn{\eqref{eq:tot_xs}}) are extracted by performing a binned 
maximum-likelihood fit to the \NNout distributions in the \lp SR and 
in the \lm SR.
The production of \tq and \tbarq are treated independently.
The signal rates, the rate of the combined top-quark background (\ttbar, $Wt$, $t\bar{b}$, and $\bar{t}b$), and the rate of the combined $W$\,+\,light-jets, $W$+$c\bar{c}$, and $W$+$b\bar{b}$ background,
are fitted simultaneously.
The rates of $W^++$ jets and $W^-+$ jets are independent parameters in the fit.
The event yields of the multijet background and the $Z,VV+\mathrm{jets}$
background are fixed to the estimates given in Table~\ref{tab:yields}.
The multijet background is determined in a data-driven way, see
\Sect{\ref{sec:background_estimate}}, and is therefore not subject to the fit
of the signal yields. 
The $Z,VV+\mathrm{jets}$ background is relatively small and cannot be
further constrained by the fit. 

The maximum-likelihood function is given by the
product of Poisson probability terms for the individual histogram bins (see \Ref{\cite{TOPQ-2011-14}}).
Gaussian prior probability distributions are included multiplicatively in the
maximum-likelihood function to constrain the background rates, which are subject
to the fit, to their predictions given the associated uncertainties.
The event yields estimated in the fit are given in Table~\ref{tab:nu_hat_fit}. 

\begin{table}[htbp]
 \sisetup{round-mode=figures, retain-explicit-plus}
 \centering
\begin{tabular}{l 
                S[table-format=5.0,round-precision=3]@{$\,\pm\,$}S[table-format=4.0]
                S[table-format=5.0,round-precision=3]@{$\,\pm\,$}S[table-format=4.0]}
 \toprule
 Process              & \multicolumn{2}{c}{$\hat{\nu}(\ell^+)$} 
                      & \multicolumn{2}{c}{$\hat{\nu}(\ell^-)$}  \\
 \midrule
 $tq$                 & 11848 & 195          & 17  & 1 \\
 $\bar{t}q$           & 11 & 1 & 6921 & 174    \\
\midrule
 $\ttbar,Wt,t\bar{b}/\bar{t}b$ & 19265 & 741 & 18888 & 727  \\
 $W^++$ jets          & 18843 & 776 &    48 & 2 \\
 $W^-+$ jets          &    23 &   1 & 13119 & 737 \\
 $Z,VV+\mathrm{jets}$ & 
 \multicolumn{1}{S[table-format=5.0, round-precision=3]@{\phantom{$\,\pm\,$}}}{1291} 
 & & 
 \multicolumn{1}{S[table-format=5.0, round-precision=3]@{\phantom{$\,\pm\,$}}}{1193} & \\
 Multijet & 
 \multicolumn{1}{S[table-format=5.0,round-precision=3]@{\phantom{$\,\pm\,$}}}{4523} 
 & & 
 \multicolumn{1}{S[table-format=5.0,
 round-precision=3]@{\phantom{$\,\pm\,$}}}{4523} &
 \\
\midrule
 Total estimated      & 55802 & 1092 & 44709 & 1050  \\
 Data                 & \multicolumn{1}{S[table-format=5.0, round-mode=off]@{\phantom{$\,\pm\,$}}}{55800} &
                      & \multicolumn{1}{S[table-format=5.0, round-mode=off]@{\phantom{$\,\pm\,$}}}{44687} & \\
\bottomrule
\end{tabular}
\caption{\label{tab:nu_hat_fit}Event yields for the different processes
  estimated with the fit to the \NNout distribution 
  compared to the numbers of observed events.
  Only the statistical uncertainties are quoted.
  The $Z,VV+\mathrm{jets}$ contributions and the multijet
  background are fixed in the fit; therefore no uncertainty is quoted for these processes.}
\end{table}

\begin{figure}[htbp]
  \centering
  \subfloat[][]{
  \includegraphics[width=0.48\textwidth]{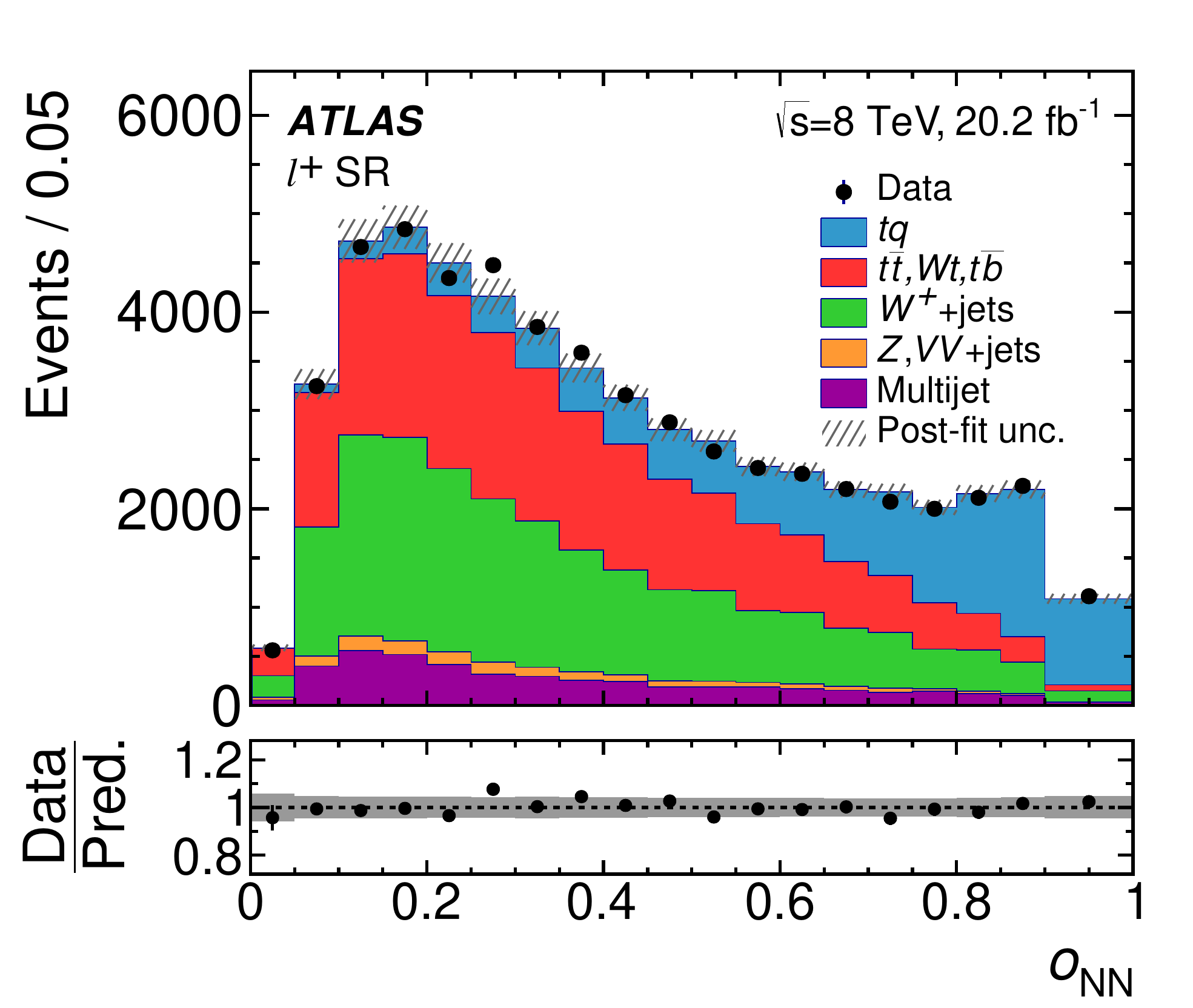}
  }
  \subfloat[][]{
  \includegraphics[width=0.48\textwidth]{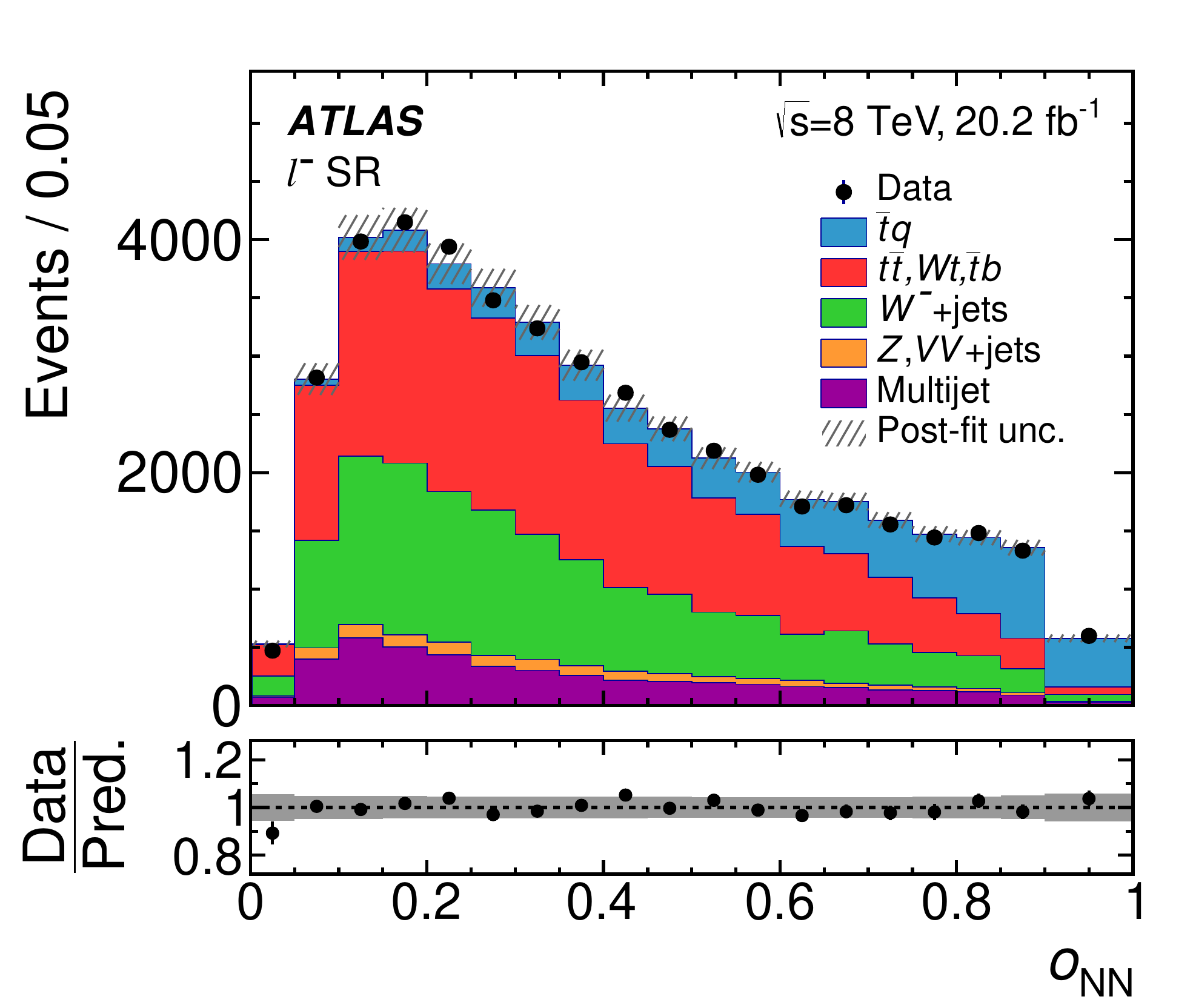}}
  \caption{\label{fig:NN_fitresult}%
    Observed \NNout distributions in  
    (a) the \lp SR and in (b) the \lm SR
    compared to the model obtained from simulated events.
    The simulated distributions are normalised to
    the event rates obtained by the fit to the discriminants.
    \postfitCaptionUncSentence  \ratioPanelSentence}
\end{figure}

\begin{figure}[htbp]
  \centering
  \vspace*{-8mm}

  \subfloat[][]{
    \includegraphics[width=0.44\textwidth]{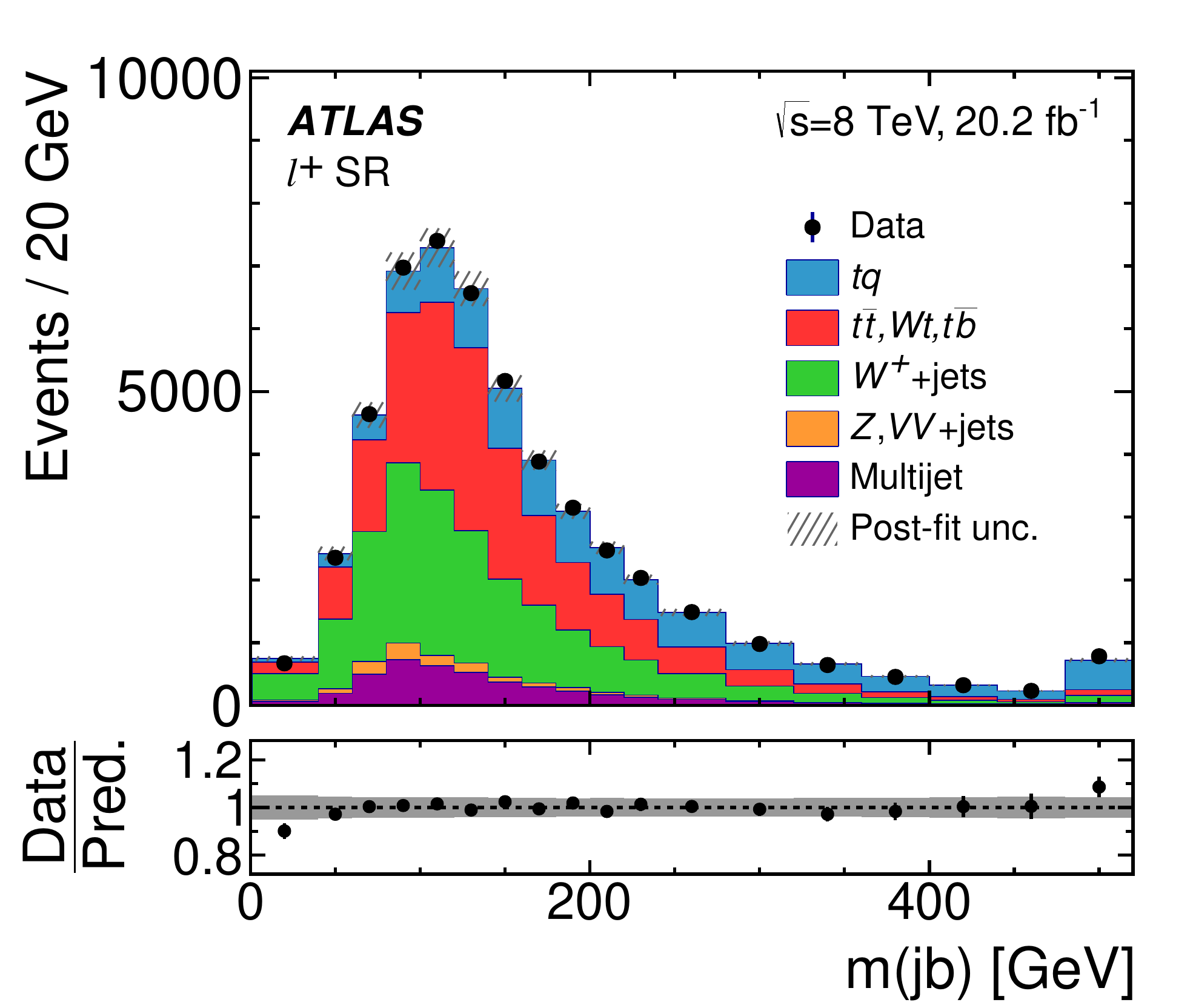}
    \label{subfig:mjb_plus}
  }
  \subfloat[][]{
    \includegraphics[width=0.44\textwidth]{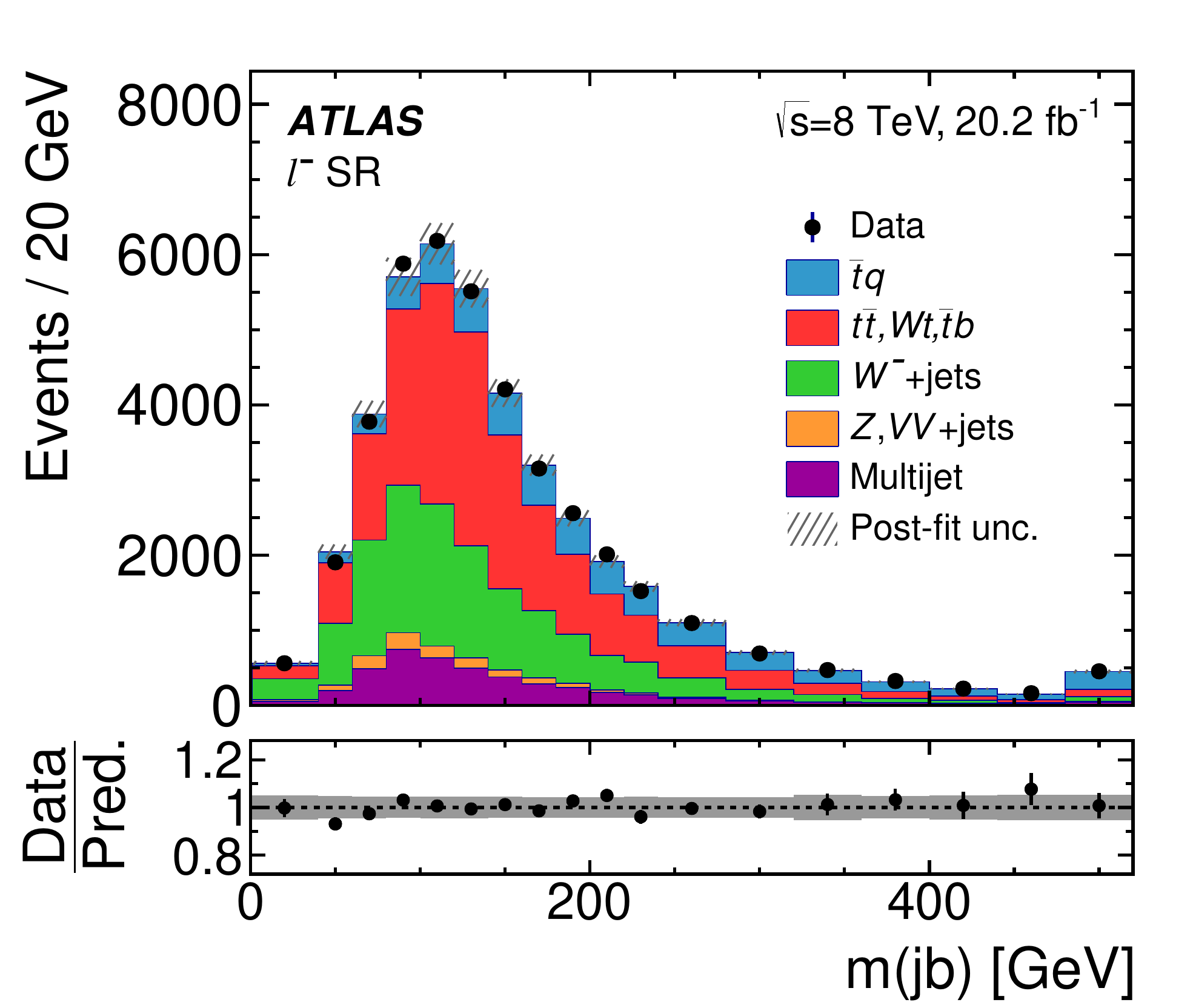}
    \label{subfig:mjb_minus}
    }\\
  \vspace*{-3mm}
      
  \subfloat[][]{
    \includegraphics[width=0.44\textwidth]{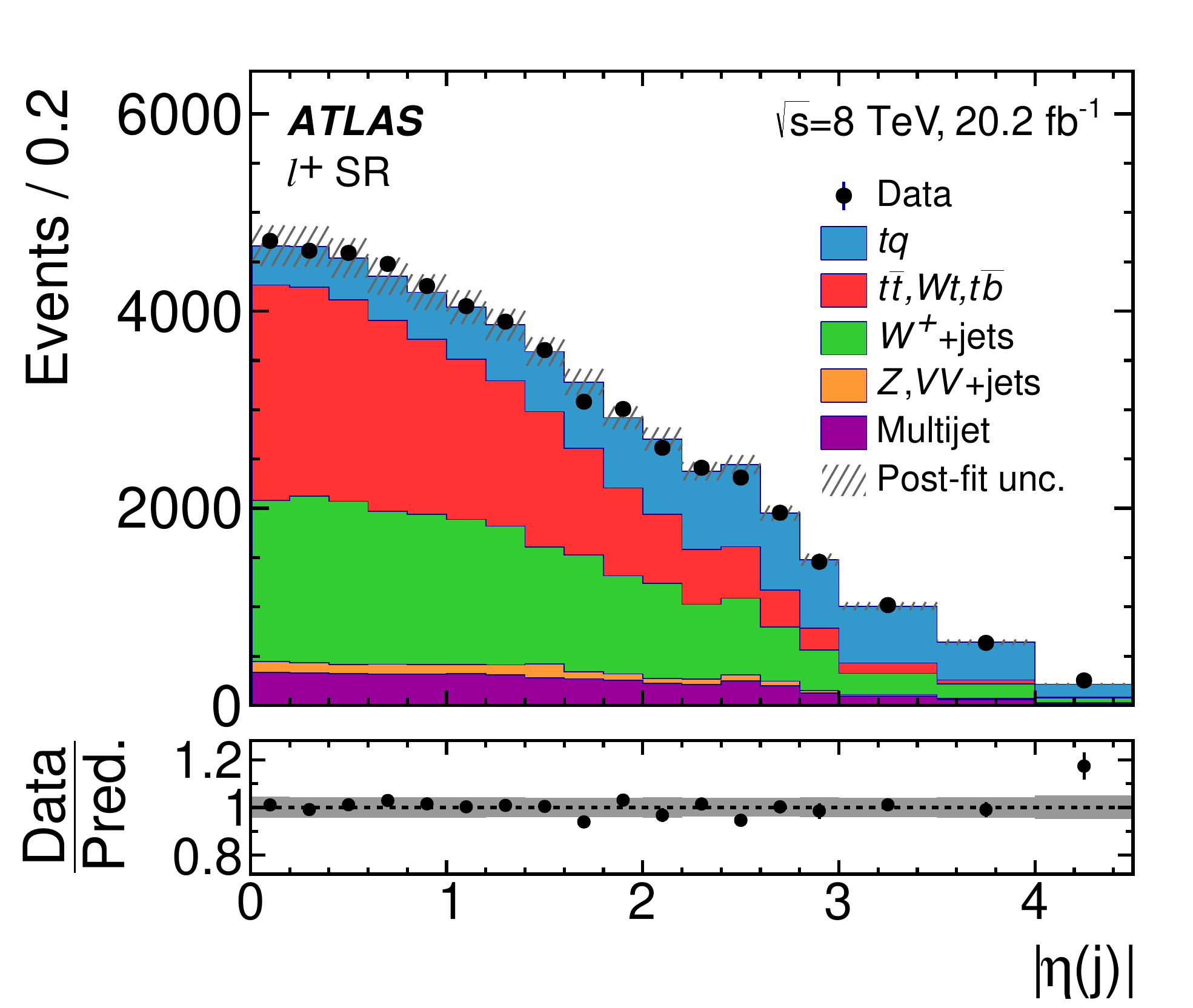}
    }
  \subfloat[][]{
    \includegraphics[width=0.44\textwidth]{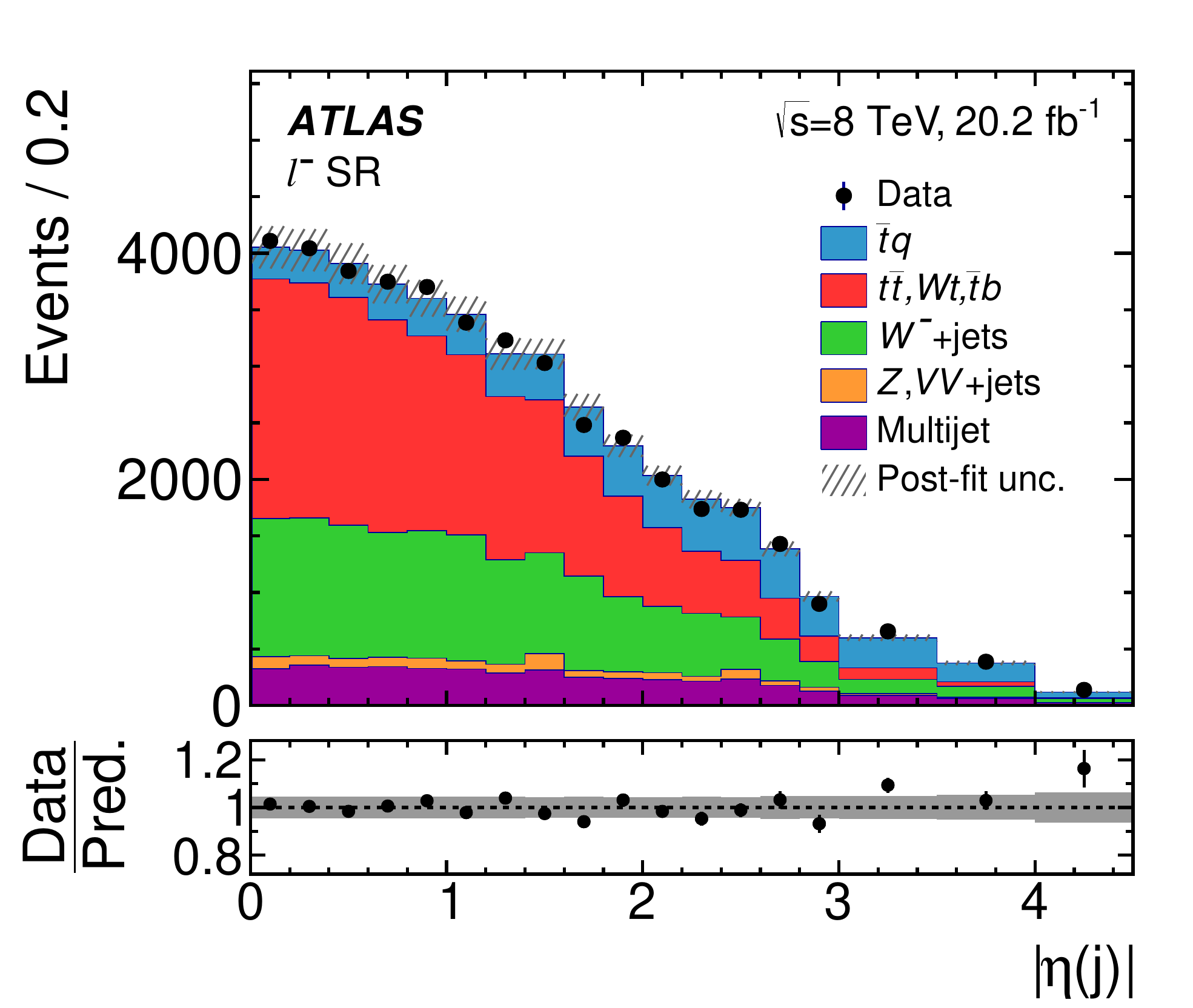}
    \label{subfig:untagged_eta_minus}
  }\\
  \vspace*{-3mm}

  \subfloat[][]{
    \includegraphics[width=0.44\textwidth]{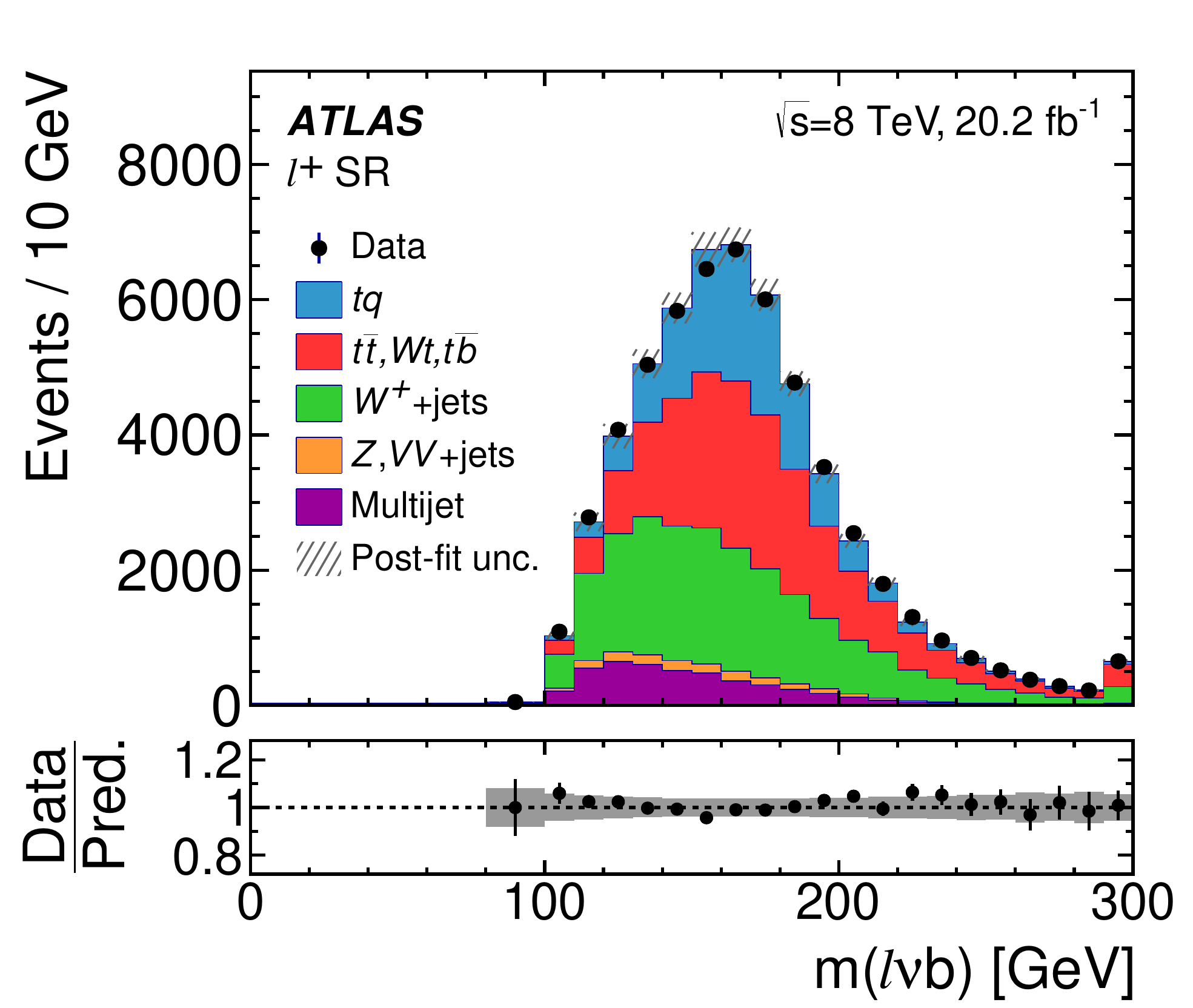}
    \label{subfig:mlnub_plus}
  }  
  \subfloat[][]{
    \includegraphics[width=0.44\textwidth]{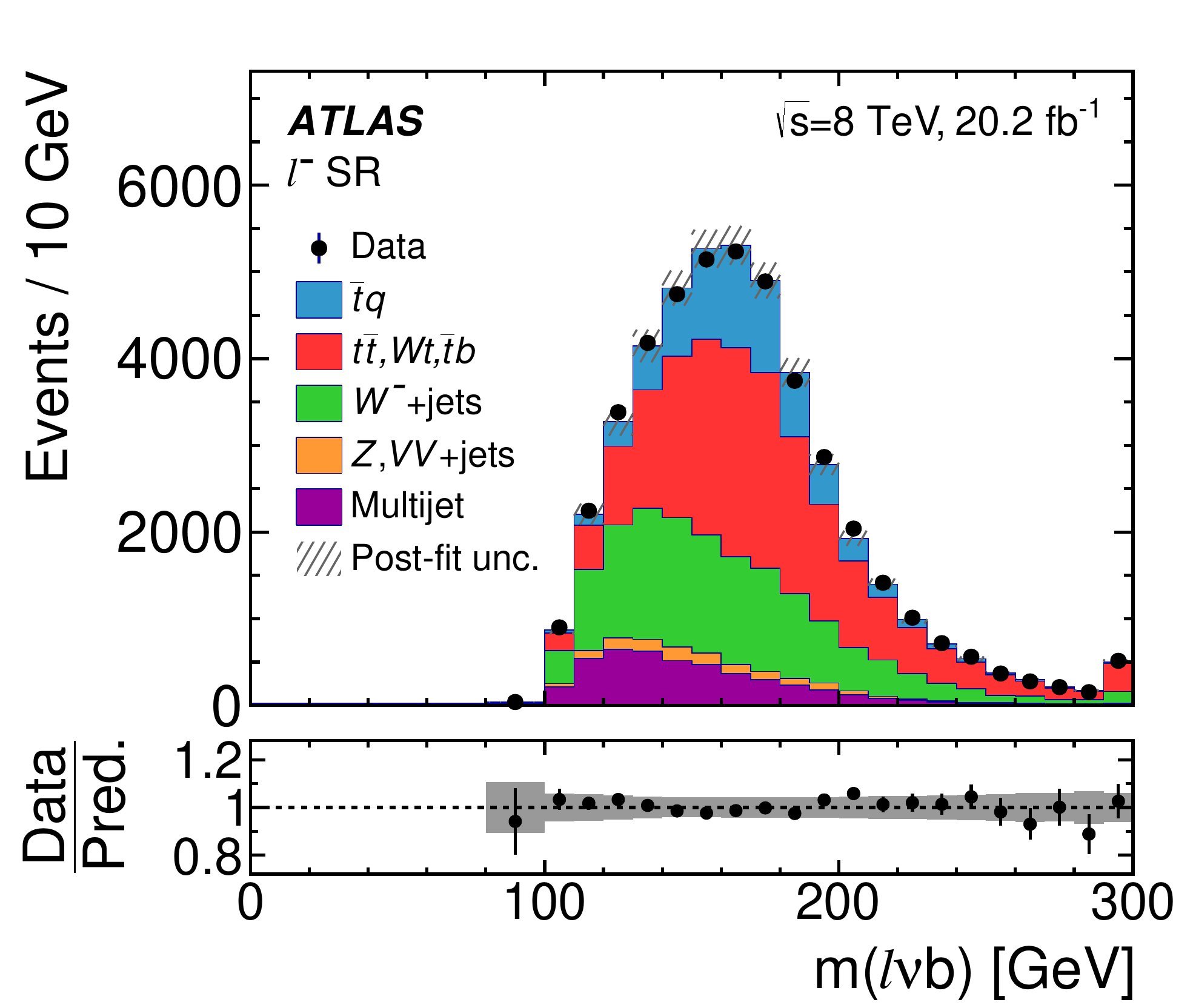}
    \label{subfig:mlnub_minus}
  }

  \vspace*{-2mm}
  \caption{\label{fig:input_vars_postfit}
    Observed distributions of the three most important input variables to the NN
    in the SR compared to the model obtained with simulated events. The
    definitions of the variables can be found in \Tab{\ref{tab:NNinputVars}}.
    The simulated distributions are normalised to the event rates obtained by 
    the maximum-likelihood fit to the NN discriminants.
    \postfitCaptionUncSentence  \ratioPanelSentence
    Events beyond the $x$-axis range in \protect\subref{subfig:mjb_plus},
    \protect\subref{subfig:mjb_minus}, \protect\subref{subfig:mlnub_plus} and \protect\subref{subfig:mlnub_minus}
    are included in the last bin.}
\end{figure}

In \Fig{\ref{fig:NN_fitresult}}, the observed \NNout 
distributions are shown and are compared to the compound model of signal and
background normalised to the fit result. 
Figure~\ref{fig:input_vars_postfit} displays the observed distributions of
the three most discriminating variables compared to the distributions obtained
with simulated events normalised to the fit result. 
Differences between data and prediction are covered by the
normalisation uncertainty of the different fitted processes.

\begin{figure}[htbp]
  \centering
  \subfloat[][]{
    \includegraphics[width=0.48\textwidth]{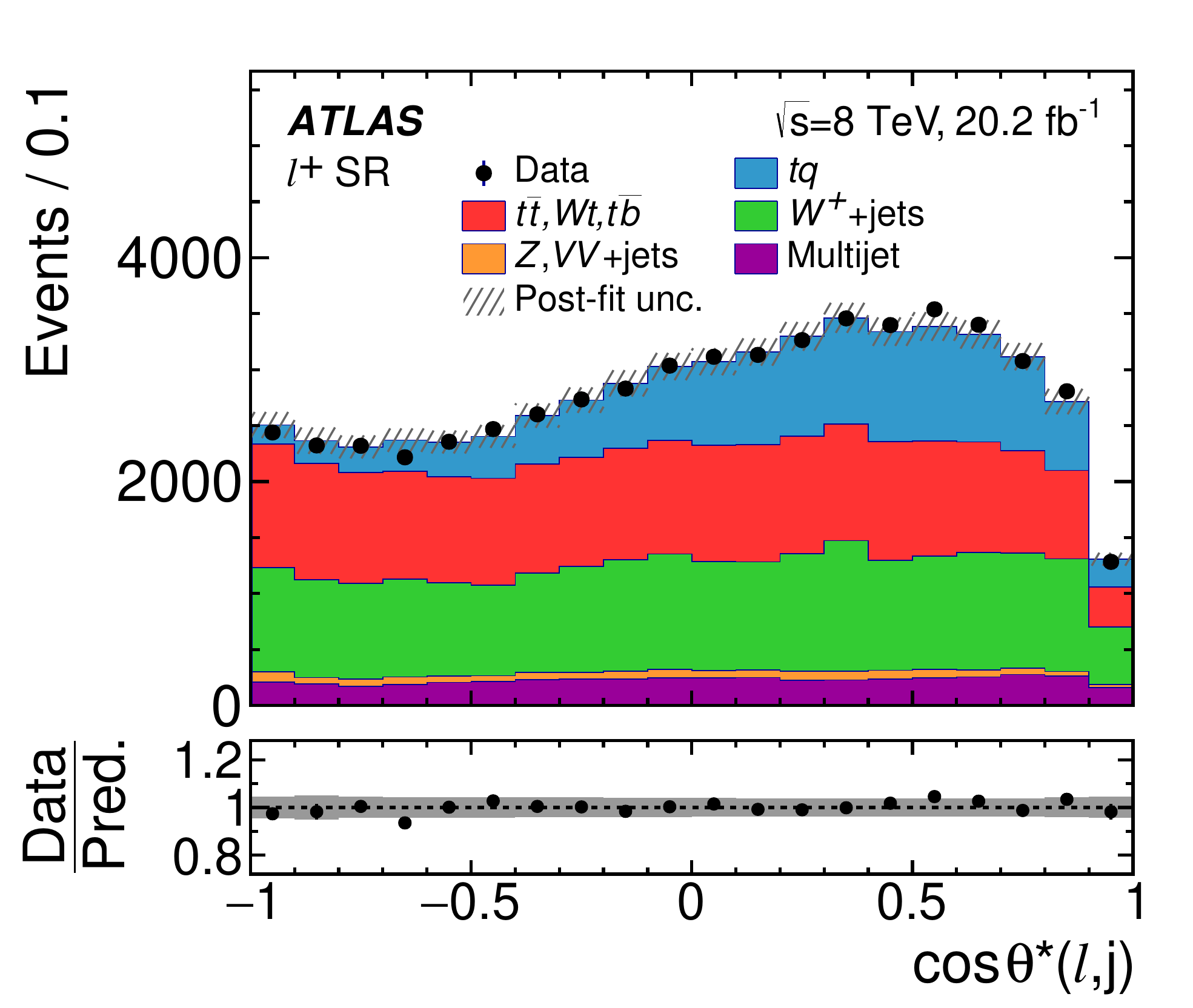}
    \label{subfig:cos_theta_plus}
    }
  \subfloat[][]{
    \includegraphics[width=0.48\textwidth]{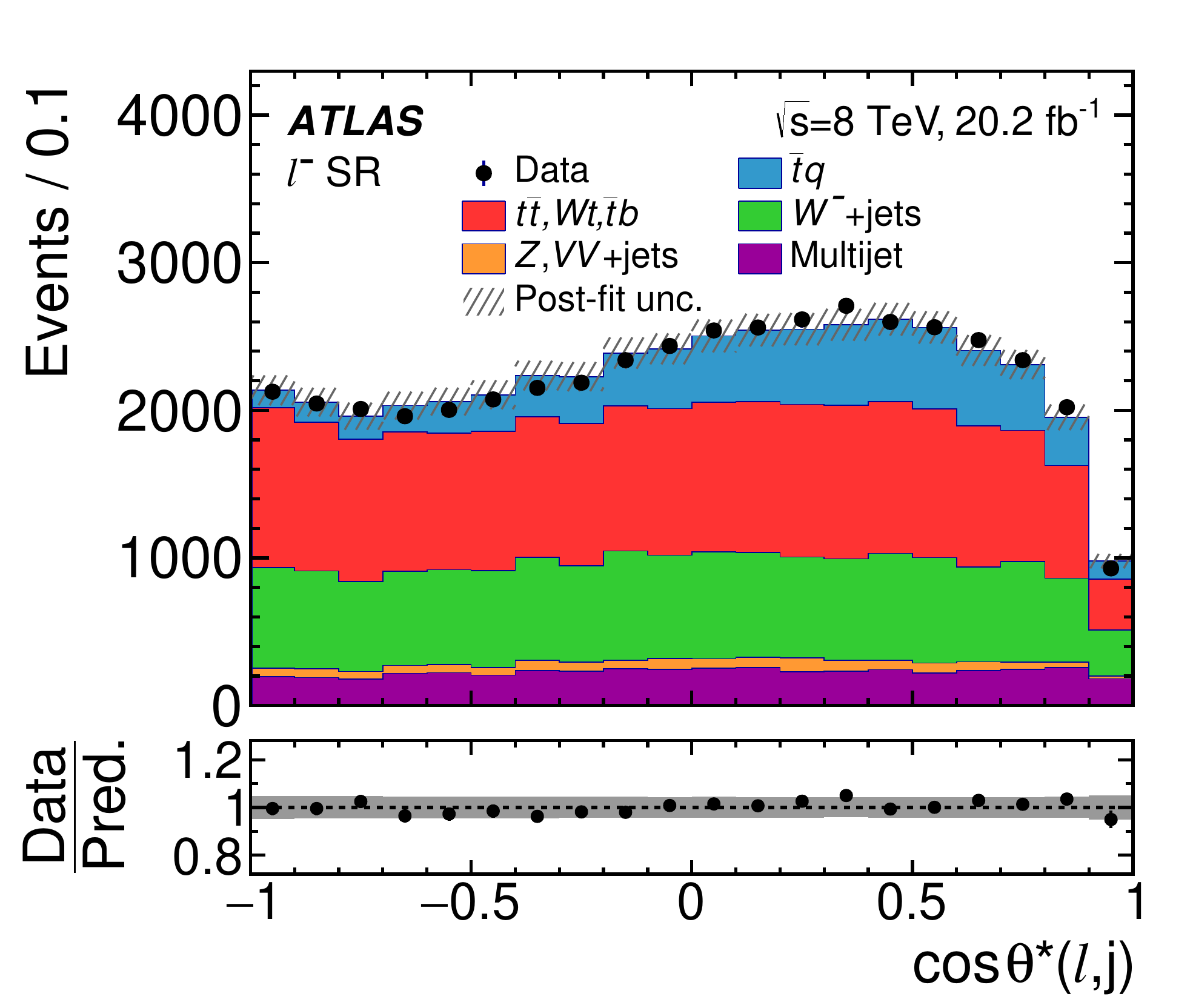}
    \label{subfig:cos_theta_minus}
    }
  \caption{\label{fig:cos_theta_postfit}
    Observed distributions of $\cos \theta^*(\ell, j)$
    in \protect\subref{subfig:cos_theta_plus} the \lp SR and in 
    \protect\subref{subfig:cos_theta_minus} the \lm SR
    compared to the model obtained from simulated events.
    The simulated distributions are normalised to the event rates obtained by
    the fit to the \NNout distributions.
    \postfitCaptionUncSentence  \ratioPanelSentence}
\end{figure}

Since single top-quarks are produced via the charged-current weak interaction
($W$-boson exchange), they are polarised. The polarisation is most prominently
visible in the distribution of $\cos \theta^*(\ell, j)$ shown in
\Fig{\ref{fig:cos_theta_postfit}}. The good modelling of the observed
distribution of this characteristic variable by
simulated distributions scaled to the fitted event rates serves as
further confirmation of the fit result.

\subsection{Fiducial cross-section measurements}
\label{sec:fiducial_cross_section_result}

The fiducial cross-sections are calculated using \Eqn{\eqref{eq:fid_xs}}, yielding
\begin{linenomath}
\begin{align}
\sigfid(\tq) &=  9.78 \pm 0.16 \, (\text{stat.}) \pm 0.52 \,
(\text{syst.}) \pm 0.19 \, (\text{lumi.})~\text{pb} \\
 &= 9.78 \pm 0.57~\text{pb}\nonumber
\intertext{and}
\sigfid(\tbarq) &= 5.77 \pm 0.14 \, (\text{stat.}) \pm 0.41 \,
(\text{syst.}) \pm 0.11 \, (\text{lumi.})~\text{pb} \\
 &= 5.77 \pm 0.45~\text{pb}. \nonumber
\end{align}
\end{linenomath}
The uncertainties in the measured expectation values of the number of signal
events, $\hat{\nu}(\tq)$ and $\hat{\nu}(\tbarq)$ in \Eqn{\eqref{eq:fid_xs}}, are
obtained from pseudo-experiments, employing the same technique as in Ref.~\cite{TOPQ-2012-21}, and are
propagated to the measured cross-sections.
The systematic uncertainties discussed in \Sect{\ref{sec:sys}} cause variations
of the signal acceptance, the background rates and the shape of the NN discriminant. 
Only significant shape uncertainties are taken into account in the statistical
analysis. Shape uncertainties are considered significant if their magnitude
exceeds the statistical uncertainty in at least one bin of the
\NNout distribution.
In order to dampen statistical fluctuations a median filter is applied to
the distribution of the bin-wise relative uncertainty.
The filter uses a five-bin-wide sliding window and is by construction not
applied to the first and the last two bins of a histogram. After applying this
procedure, shape uncertainties are considered for the following sources: two JES
uncertainty components, jet energy resolution, \MET modelling, the
modelling of the multijet background, and all MC-generator-related
uncertainties.

Since the $tq$ and $\tbar q$ production cross-sections are measured in a
fiducial region, systematic uncertainties in the event rates affect only
\nselfidNoFrac~in \Eqn{\eqref{eq:fid_xs}}, thereby reducing the uncertainties
related to the choice of PDF, signal MC generator and parton-shower 
by about 1 percentage point each. The uncertainties in the
scale choice of the signal generator and the NLO matching are reduced by about
2 percentage points each.
Contributions of the various sources of systematic uncertainty to the
measured values of $\sigfid(tq)$ and $\sigfid(\tbar q)$ 
are shown in Table~\ref{tab:xs-uncertainty}. 
\begin{table}[htbp]
\centering
\begin{tabular}{lcc}
    \toprule
     Source & $\Delta \sigfid(tq)\ / \ \sigfid(tq)$ & 
              $\Delta \sigfid(\tbar q)\  / \  \sigfid(\tbar q)$ \\
            & [\%] & [\%] \\
    \midrule
    Data statistics            & $\pm$ 1.7 & $\pm$ 2.5  \\
    Monte Carlo statistics     & $\pm$ 1.0 & $\pm$ 1.4  \\
    \midrule
    Background normalisation   & $<0.5$    & $<0.5$     \\ 
    Background modelling       & $\pm$ 1.0 & $\pm$ 1.6  \\  
    Lepton reconstruction      & $\pm$ 2.1 & $\pm$ 2.5  \\
    Jet reconstruction         & $\pm$ 1.2 & $\pm$ 1.5  \\
    Jet energy scale           & $\pm$ 3.1 & $\pm$ 3.6  \\
    Flavour tagging            & $\pm$ 1.5 & $\pm$ 1.8  \\
    \MET modelling             & $\pm$ 1.1 & $\pm$ 1.6  \\
    $b/\bbar$ tagging efficiency  & $\pm$ 0.9 & $\pm$ 0.9  \\
    PDF                        & $\pm$ 1.3 & $\pm$ 2.2  \\
    $tq$ ($\tbar q$) NLO matching     & $\pm$ 0.5 & $<0.5$     \\
    $tq$ ($\tbar q$) parton shower    & $\pm$ 1.1 & $\pm$ 0.8  \\
    $tq$ ($\tbar q$) scale variations & $\pm$ 2.0 & $\pm$ 1.7  \\
    \ttbar NLO matching        & $\pm$ 2.1 & $\pm$ 4.3  \\
    \ttbar parton shower       & $\pm$ 0.8 & $\pm$ 2.5  \\
    \ttbar scale variations    & $<0.5$    & $<0.5$     \\
    Luminosity                 & $\pm$ 1.9 & $\pm$ 1.9  \\
    \midrule
    Total systematic           & $\pm$ 5.6 & $\pm$ 7.3  \\
    Total (stat.\ + syst.)     & $\pm$ 5.8 & $\pm$ 7.8  \\
    \bottomrule
   \end{tabular}
   \caption{Detailed list of the contribution from each source of
   uncertainty to the total uncertainty in the measured values of $\sigfid(\tq)$
   and $\sigfid(\tbarq)$.
   The estimation of the systematic uncertainties has a statistical uncertainty
   of \SI{0.3}{\%}.
   Uncertainties contributing less than \SI{0.5}{\%} are marked with \enquote{$<0.5$}.}
   \label{tab:xs-uncertainty}
\end{table}
The relative combined uncertainties, including the statistical and 
systematic uncertainties, are $\pm$\SI{5.8}{\%} for
$\sigfid(\tq)$ and $\pm$\SI{7.8}{\%} for $\sigfid(\tbarq)$. 
The three largest sources of uncertainty are the uncertainty in the JES
calibration, the choice of matching method used for the NLO generator of the top-quark
background and the uncertainty in the lepton reconstruction.

\begin{figure}[!htpb]
\centering
 \subfloat[][]{
  \includegraphics[width=0.47\textwidth]{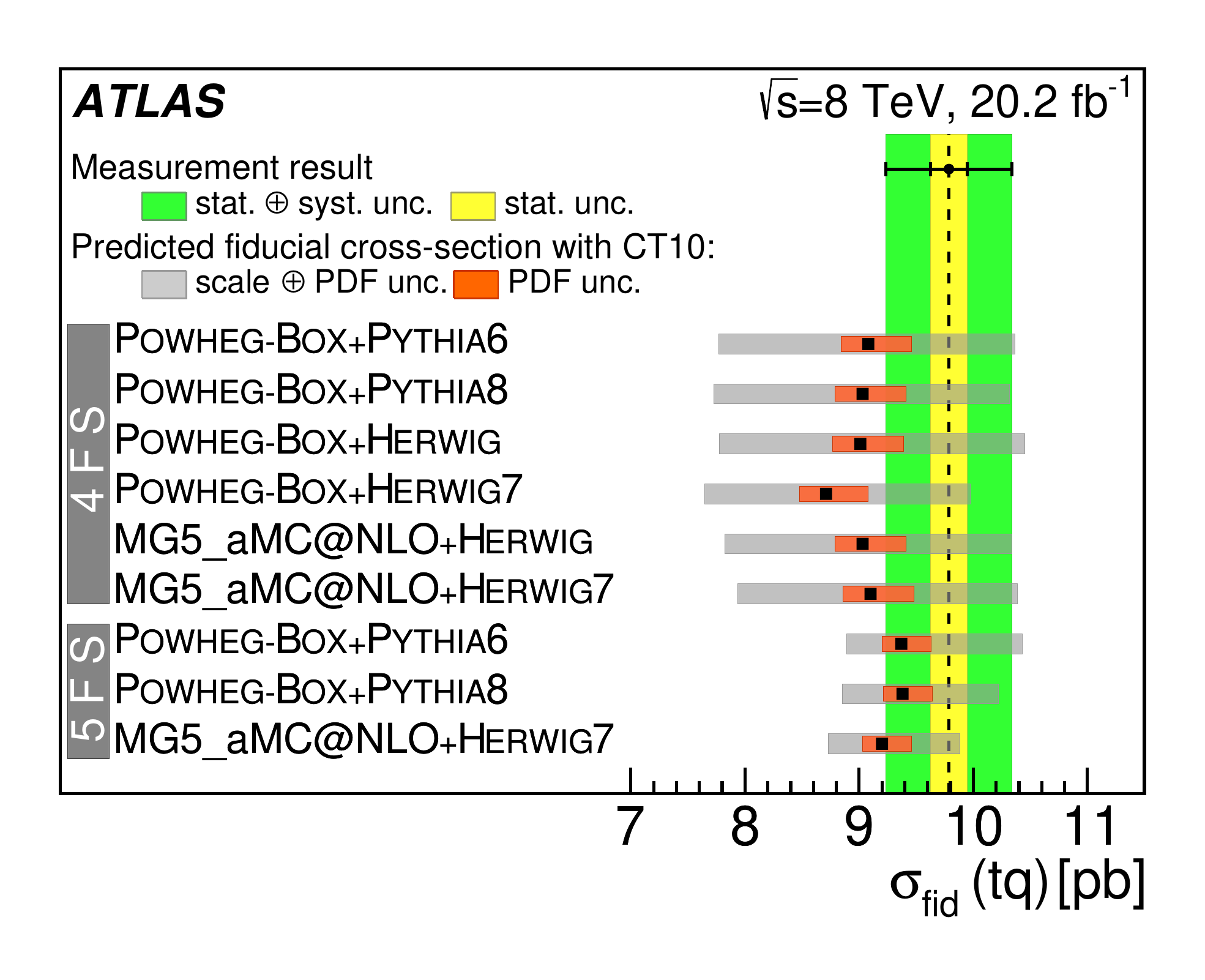}
  \label{subfig:top_fid_xs}
  }
\quad
 \subfloat[][]{
  \includegraphics[width=0.47\textwidth]{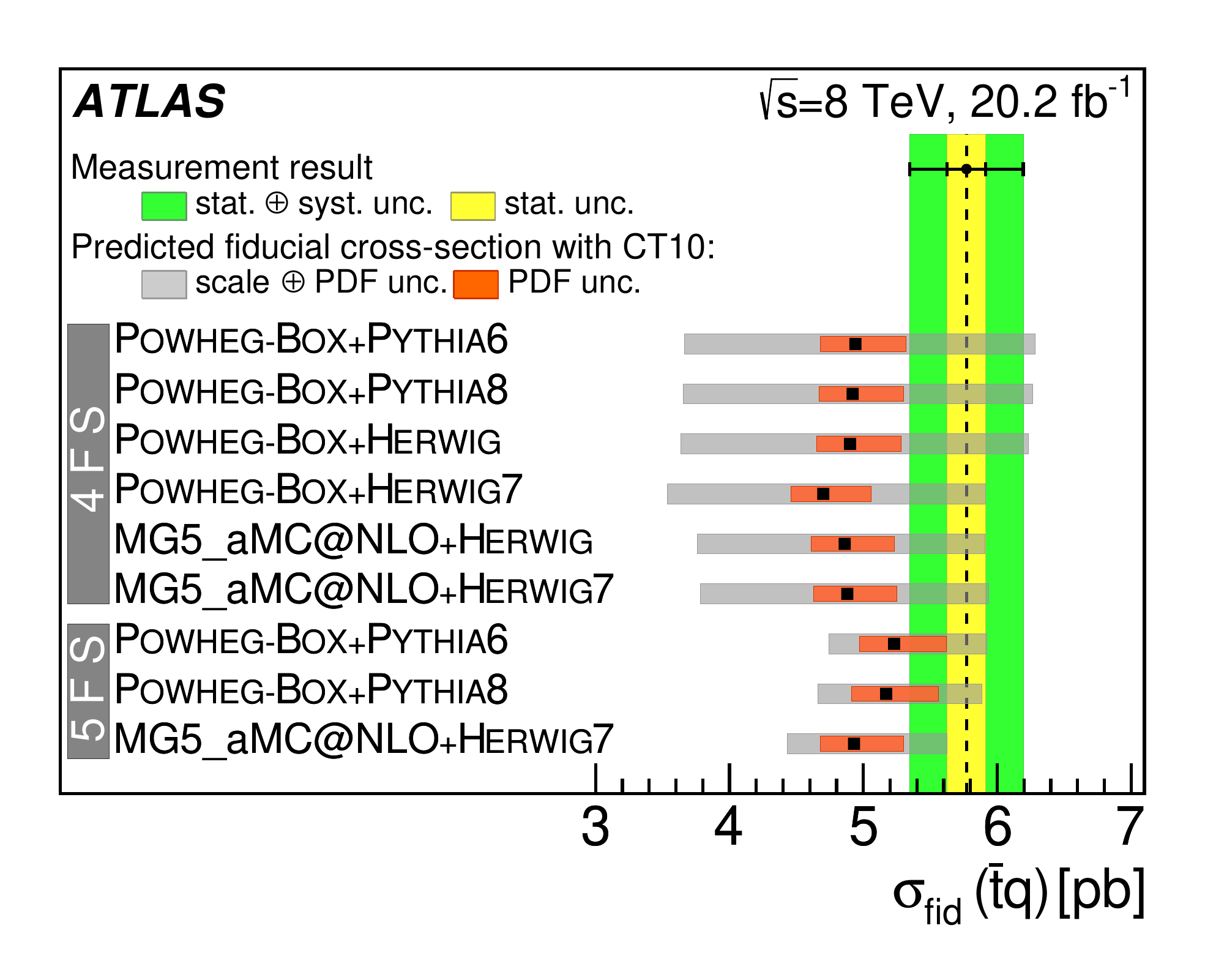}
  \label{subfig:antitop_fid_xs}
  }
\caption{Measured $t$-channel \protect\subref{subfig:top_fid_xs}
single-top-quark and \protect\subref{subfig:antitop_fid_xs} single-top-antiquark
fiducial cross-sections compared to predictions by the NLO MC generators \POWHEGBOX and \MGMCatNLO in the
four-flavour scheme (4FS) and five-flavour scheme (5FS) combined with different
parton-shower models.
The uncertainties in the predictions include the uncertainty due to the scale
choice using the method of independent restricted scale variations and the intra-PDF uncertainty 
in the \ct PDF set.}
\label{fig:fid_measured_xs}
\end{figure}

Figure~\ref{fig:fid_measured_xs} shows the measured fiducial cross-sections in 
comparison to the predictions by the NLO MC generators \POWHEGBOX and \MGMCatNLO
combined with the parton-shower programs \PYTHIAV{6}~(v6.428), 
\PYTHIAV{8} (v8.2)~\cite{Sjostrand:2007gs}, \HERWIG (v6.5.20) and
\HERWIGV{7} (v7.0.1)~\cite{Bellm:2015jjp}.
The 4FS and the 5FS are explored.
The predictions are computed with the \ct PDF set and
include the uncertainty in the scale choice using the 
method of independent restricted scale variations as described  in \Sect{\ref{sec:intro}} 
and the uncertainty in the PDFs, using the intra-PDF uncertainties of \ct.
The predictions based on the 5FS feature strongly reduced scale uncertainties
compared to those based on the 4FS. When computing the predictions of
$\sigfid$ based on~\Eqn{\eqref{eq:sig_tot_sig_fid}}, the uncertainties in the
predictions of $\sigtot$ are treated as correlated with the scale and PDF
uncertainties in $A_\text{fid}$.
For the \PYTHIAV{6} parton shower the value of \alphas in the set of tuned parameters is
also modified consistently with the change of the scale in the ME. 
PDF uncertainties are obtained by reweighting to eigenvectors of their respective error sets.
The predictions of all setups agree with each other and also with the measured values.

\subsection{Total cross-section measurements}

\begin{figure}[htpb]
 \centering
 \subfloat[][]{
 \includegraphics[width=0.75\textwidth]{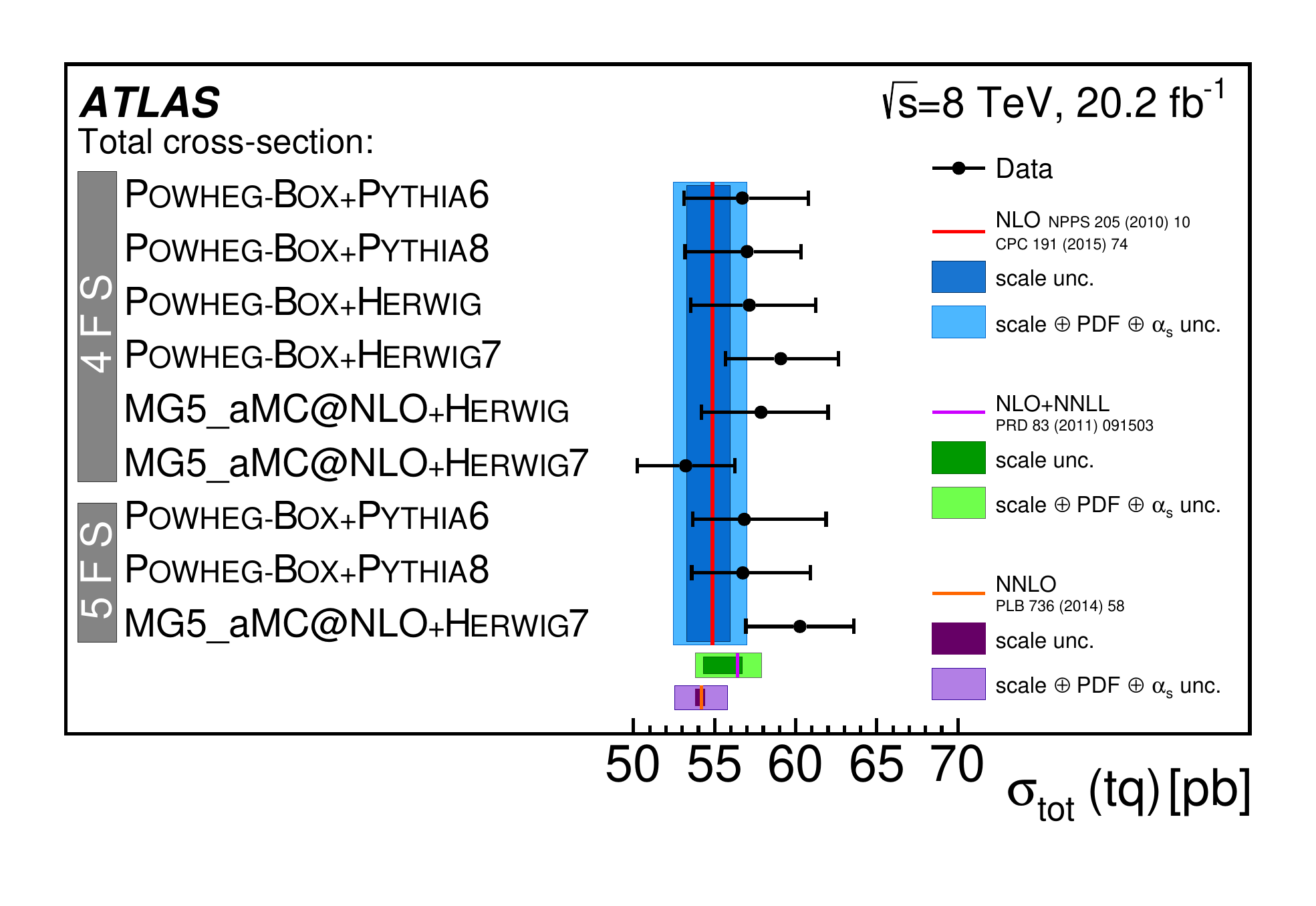}
 \label{subfig:top_total_xs}
 }
\\
 \subfloat[][]{
 \includegraphics[width=0.75\textwidth]{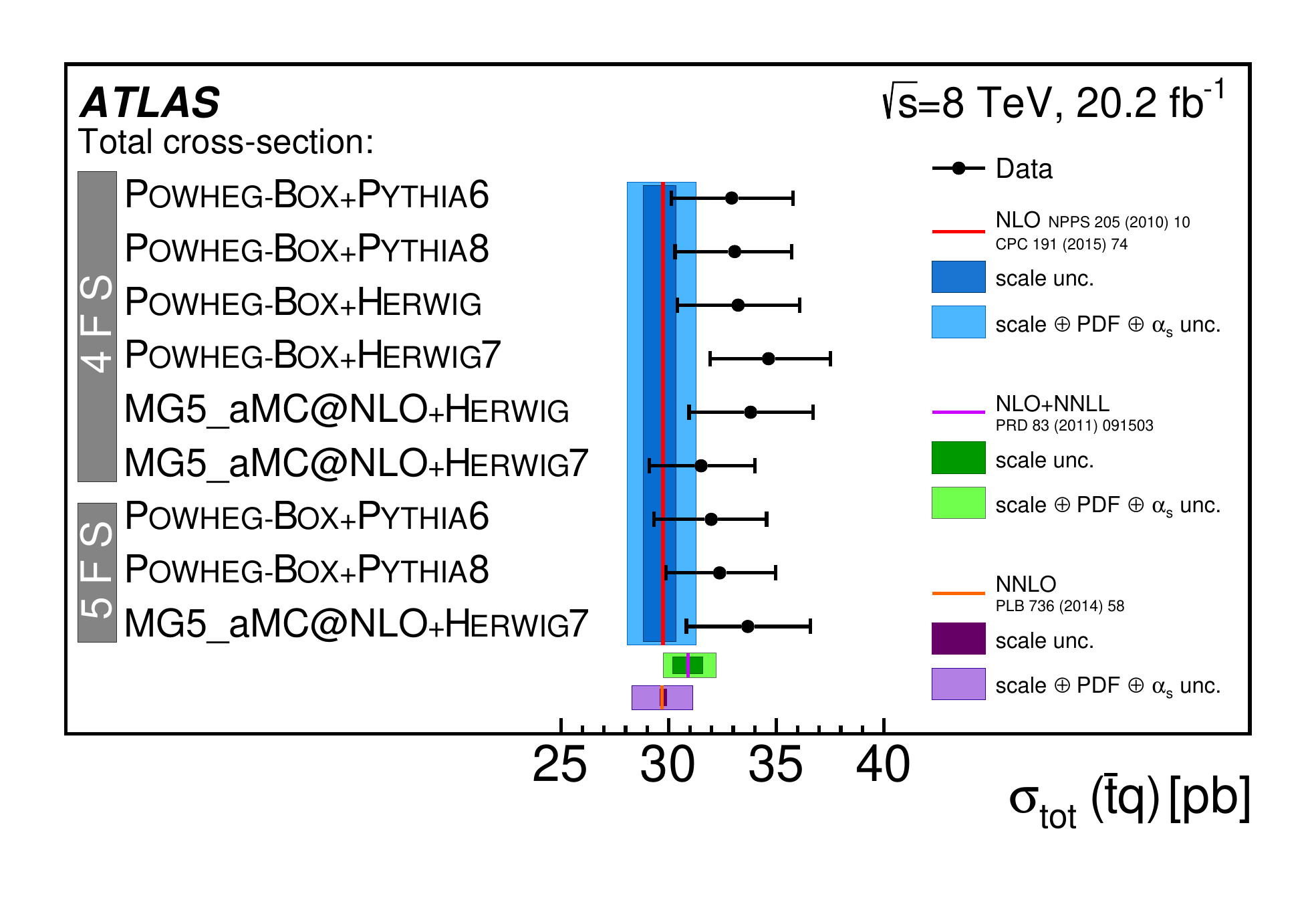}
 \label{subfig:antitop_total_xs}
 }
 \caption{Extrapolated $t$-channel 
 \protect\subref{subfig:top_total_xs} single-top-quark
 and \protect\subref{subfig:antitop_total_xs} single-top-antiquark
 production cross-sections for
 different MC-generator setups compared to fixed-order NLO
 calculations. For the three calculations, the uncertainty from the 
 renormalisation and factorisation scales are indicated in darker shading, and
 the total uncertainties, including the renormalisation and factorisation scale 
 as well as the PDF+$\alphas$ uncertainties, are indicated in lighter shading. 
 For the NNLO prediction, only the renormalisation and factorisation 
 scale uncertainty is provided in \Ref{\cite{Brucherseifer:2014ama}}. 
 For comparison, the PDF+$\alphas$ uncertainties from the NLO
 prediction~\cite{Kant:2014oha} are added to the NNLO renormalisation and factorisation scale uncertainty reflected in the lighter shaded uncertainty band.
 For this comparison, the uncertainty in the extrapolation does not include the
 contribution from the NLO-matching method and from the choice of parton-shower
 model.}
 \label{fig:inclusive_result}
\end{figure}

Using the predictions of $A_\text{fid}$ by different MC generators, 
the fiducial cross-sections are extrapolated to the full phase space and
compared to fixed-order calculations. The PDF and scale uncertainties in
$A_\text{fid}$ are included and correlated with the PDF and scale uncertainty in 
$\sigfid$.
Figure~\ref{fig:inclusive_result} shows the total cross-sections obtained 
by the extrapolation, based on $A_\text{fid}$ from \POWHEGBOX and \MGMCatNLO for
the 4FS and 5FS and for different parton-shower MC programs.
Since the extrapolation from the fiducial to the total cross-sections is
performed for different MC generators, the uncertainty in the NLO-matching
method and the uncertainty due to the choice of the parton-shower program are
not considered for the extrapolation part, but these uncertainties are kept for the fiducial
cross-sections entering the extrapolation.
The measured values
are compared with fixed-order perturbative QCD
calculations~\cite{Kant:2014oha,Campbell:2009ss,Kidonakis:2011wy,Brucherseifer:2014ama}.
For the default generator \POWHEGBOX + \PYTHIAV{6} the fiducial acceptances are
determined to be $A_\text{fid}(\tq) = (17.26^{+0.46}_{-0.21})\,\%$ and
$A_\text{fid}(\tbarq) = (17.52^{+0.45}_{-0.20})\,\%$, thereby yielding
\begin{linenomath}
\begin{align}
 \sigtot(tq) &= 56.7 \pm 0.9 \, (\text{stat.}) \pm 2.7 \, (\text{exp.})
 \,^{+2.7}_{-1.7} \, (\text{scale}) \pm 0.4 \, (\text{PDF}) \label{eq:extrapolated_top_xs}\\
 &\quad\pm 1.0 \, (\text{NLO-matching method}) \, \pm 1.1 \, (\text{parton shower})
 \pm 1.1\,(\text{lumi.})\;\si{pb} \nonumber\\
 &= 56.7^{+4.3}_{-3.8}\;\si{\pb}\nonumber
 \intertext{and}
 \sigtot(\tbar q) &= 32.9  \pm 0.8 \, (\text{stat.}) \pm 2.3 \, 
 (\text{exp.}) \,^{+1.4}_{-0.8} \, (\text{scale}) \pm 0.3 \, (\text{PDF}) \label{eq:extrapolated_antitop_xs}\\
 &\quad\,^{+0.7}_{-0.6} \, (\text{NLO-matching method})\,\pm 0.6 \, (\text{parton shower})
 \pm 0.6\,(\text{lumi.})\;\si{\pb} \nonumber\\
 &= 32.9^{+3.0}_{-2.7}\;\si{\pb}\,.\nonumber
\end{align}
\end{linenomath}

The experimental systematic uncertainty~(exp.) contains the uncertainty in the
fiducial cross-sections, without the scale, PDF, NLO-matching method and
parton-shower components, which are quoted separately and include both the
uncertainties in $\sigfid$ and $A_\text{fid}$.
The relative total uncertainty is $^{+7.6}_{-6.7}\,\%$ for $\sigtot(\tq)$ and $^{+9.1}_{-8.4}\,\%$ for
$\sigtot(\tbarq)$.

The total inclusive cross-section is obtained by adding $\sigtot(tq)$ and
$\sigtot(\tbar q)$ in
\Eqns{\eqref{eq:extrapolated_top_xs}}{\eqref{eq:extrapolated_antitop_xs}}:
\begin{linenomath}
\begin{align}
 \label{eq:xs_comb}
 \begin{split}
 \sigtot(tq+\tbar q) &= 89.6  \pm 1.2 \, (\text{stat.}) \pm 5.1 \, (\text{exp.})\,^{+4.1}_{-2.5} \, (\text{scale}) \pm 0.7 \, (\text{PDF})\\
 &\quad\,^{+1.7}_{-1.6} \, (\text{NLO-matching method})\,\pm 1.6 \, (\text{parton shower}) \pm 1.7 \, (\text{lumi.})~\text{pb}
 \end{split}\\
 &= 89.6^{+7.1}_{-6.3}~\text{pb}\;. \nonumber
\end{align}
\end{linenomath}
The systematic uncertainties are assumed to be \SI{100}{\%} correlated between
$tq$ and $\tbar q$, except for the MC statistical uncertainty.
Therefore, the uncertainties are added linearly component by component. The
data statistical uncertainties of $\sigtot(tq)$ and $\sigtot(\tbar q)$ are
added in quadrature to obtain the data statistical uncertainty of
$\sigtot(tq+\tbar q)$. The same is done for the MC statistical uncertainty.
The experimental systematic uncertainty~(exp.) contains the uncertainty in the
fiducial cross-sections, without the scale, PDF, NLO-matching method and
parton-shower components. 

\subsection{\rt measurement}
The ratio of the measured total cross-sections for top-quark
and top-antiquark production in the $t$-channel is determined to be
\begin{equation}
  \rt = \frac{\sigtot(\tq)}{\sigtot(\tbarq)}=1.72 \pm 0.05\, (\text{stat.})\,
  \pm 0.07\, (\text{exp.}) = 1.72 \pm 0.09.
\end{equation}
The correlation of uncertainties in $\sigtot(\tq)$ and $\sigtot(\tbarq)$ is
taken into account in the pseudo-experiments used to determine the
uncertainties in $\hat{\nu}(\tq)$ and $\hat{\nu}(\tbarq)$, see
\Sect{\ref{sec:fiducial_cross_section_result}}.
Significant sources of systematic uncertainty in the
measured values of \rt are shown in \Tab{\ref{tab:rt-uncertainty}}.
\begin{table}[htbp]
\centering
\begin{tabular}{lcc}
    \toprule
    Source & $\Delta \rt / \rt$ [\%]\\
    \midrule
    Data statistics             & $\pm$ 3.0 \\
    Monte Carlo statistics      & $\pm$ 1.8 \\
    \midrule
    Background modelling        & $\pm$ 0.7  \\ 
    Jet reconstruction          & $\pm$ 0.5  \\
    $\MET$ modelling            & $\pm$ 0.6  \\
    \tq (\tbarq) NLO matching & $\pm$ 0.5  \\
    \tq (\tbarq) scale variations & $\pm$ 0.7  \\
    \ttbar NLO matching         &  $\pm$ 2.3 \\
    \ttbar parton shower        & $\pm$ 1.7 \\
    PDF                         & $\pm$ 0.7 \\
    \midrule
    Total systematic            & $\pm$ 3.9 \\
    Total (stat.\ + syst.)      & $\pm$ 5.0 \\
    \bottomrule
   \end{tabular}
   \caption{\label{tab:rt-uncertainty} Significant contributions to
   the total relative uncertainty in the measured value of \rt.
   The estimation of the systematic uncertainties has a statistical uncertainty
   of \SI{0.3}{\%}.
   Uncertainties contributing less than \SI{0.5}{\%} 
   are not shown.
   }
\end{table}

\begin{figure}[htbp]
  \centering
  \includegraphics[width=0.6\textwidth]{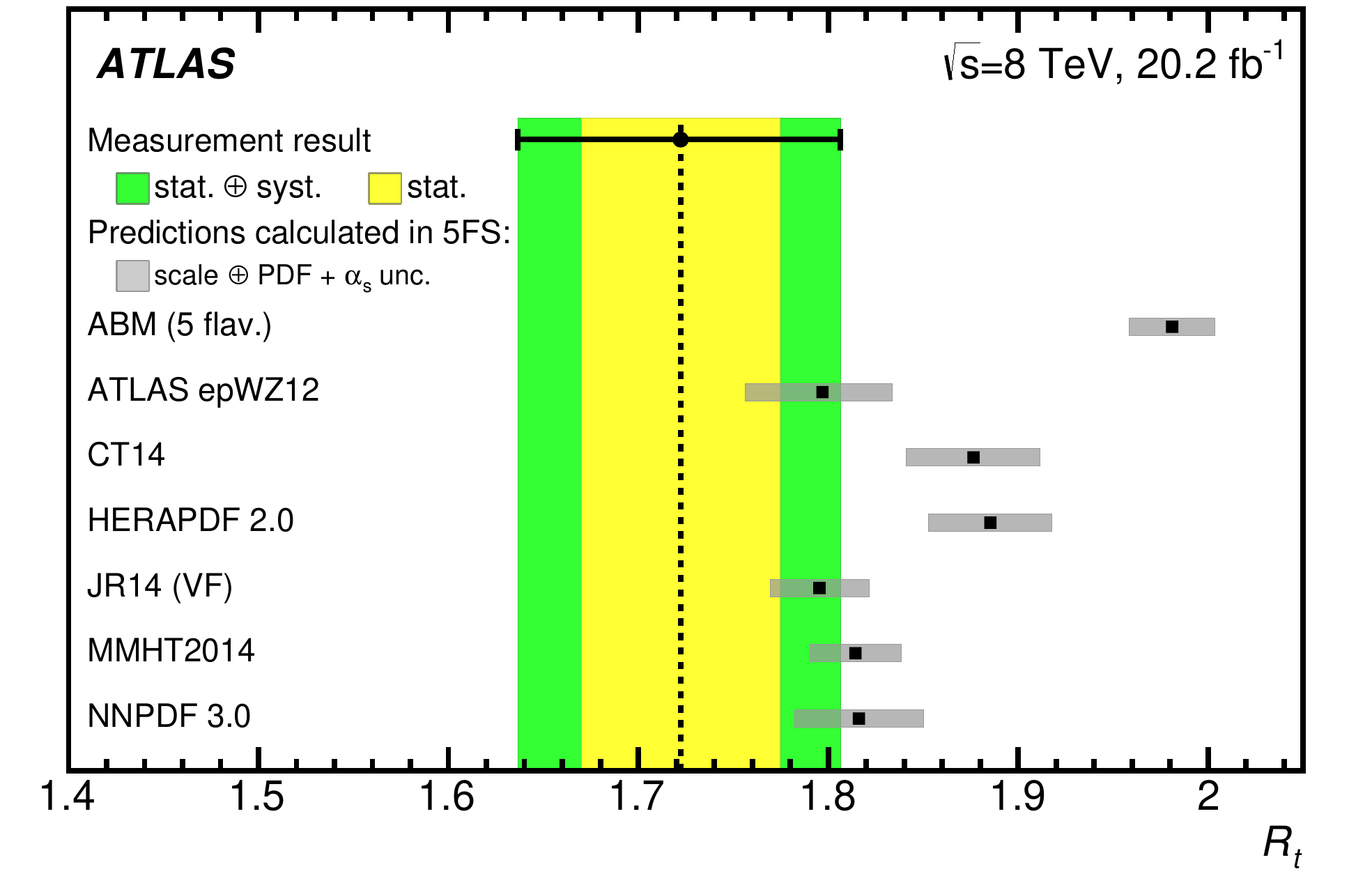}
  \caption{\label{fig:rtop}
    Predicted values of $\rt=\sigtot(\tq)/\sigtot(\tbarq)$ calculated with
    \HATHOR~\cite{Kant:2014oha} at NLO accuracy in
    QCD~\cite{Campbell:2009ss} in the 5FS using different NLO PDF 
    sets~\cite{Alekhin:2012ig,STDM-2011-43,Dulat:2015mca,Abramowicz:2015mha,Jimenez-Delgado:2014twa,Harland-Lang:2014zoa,Ball:2014uwa} 
    compared to the measured value.
    The error bars on the predictions 
    include the uncertainty in the renormalisation and factorisation scales and  
    the combined internal PDF and \alphas uncertainty.
    The dashed black line indicates the central value of the measured \rt value.
    The combined statistical and systematic uncertainty of the measurement is shown in green,
    while the statistical uncertainty is represented by the yellow error band. 
    The uncertainty in the measured \rt value does not include the PDF
    components for this comparison.
  }
\end{figure}

Figure~\ref{fig:rtop} compares the observed value of \rt to predictions
based on several different PDFs.
For this comparison the uncertainty in the measured \rt value does not include
the PDF components. The uncertainties in the predictions include the uncertainty
in the renormalisation and factorisation scales and the combined internal PDF
and \alphas uncertainty.
Most predictions agree at the $1\,\sigma$ level
with the measured value; only the prediction based on ABM (5 flav.)~\cite{Alekhin:2012ig}
is about $2.5\,\sigma$ above the measurement. The main differences of the ABM
PDF set compared to the other sets are the treatment of the $b$-quark PDF and
the value of $\alphas$.

\subsection{Estimation of top-quark mass dependence}
\label{app:mtop}

The $t$-channel cross-section results given above are obtained for a
top-quark mass of $\mtop = \SI{172.5}{\GeV}$. The dependence of the measured
cross-sections on $\mtop$ is estimated by repeating the measurement 
with different mass assumptions. The MC samples for all processes containing
top quarks are reproduced for six different values of 
$\mtop$, namely $165, 167.5, 170, 175, 177.5$ and \SI{180}{\GeV}.
The samples comprise the \tq and \tbarq signal as well as the background
samples for $\ttbar, Wt, t\bar{b}$ production.
The dependences of the resulting cross-sections on $\mtop$ are fitted
with a first-order polynomial, for which the constant term is given by the
central value at $\mtop = \SI{172.5}{\GeV}$
\begin{equation}
 \sigma(\mtop) = \sigma(\SI{172.5}{\GeV}) + a \cdot \Delta\mtop [\si{\GeV}]\,,
\end{equation}
where $\Delta\mtop = \mtop - \SI{172.5}{\GeV}$.
The fitted parameters $a$, the slopes, are given in \Tab{\ref{tab:paratopmass_app}} for all measured cross-sections.

\begin{table}[htbp]
\centering
  \begin{tabular}{l|c}
    \toprule
     Measurement& $a\left[\frac{\si{\pb}}{\si{\GeV}}\right]$\\
    \midrule 
    $\sigma_{\text{fid}}(tq)$ & $-0.06\pm 0.01$  \\
    $\sigma_{\text{fid}}(\tbar q)$ & $-0.04\pm 0.01$  \\
    $\sigma_{\text{tot}}(tq)$ & $-0.59\pm 0.08$  \\
    $\sigma_{\text{tot}}(\tbar q)$ & $-0.37\pm 0.06$  \\
    $\sigma_{\text{tot}}(tq+\tbar q)$ & $-0.96\pm 0.13$  \\
    \bottomrule
  \end{tabular}
  \caption{Slopes $a$ of the mass dependence of the measured cross-sections.}
  \label{tab:paratopmass_app}
\end{table}

\subsection{Determination of $|V_{tb}|$}
\label{sec:Vtb}

Single top-quark production in the $t$-channel proceeds via a $Wtb$ vertex and
the measured cross-section is proportional to \flvtbsqr.
In the SM, $|V_{tb}|$ is very close to one and $\fl$ is exactly one, but 
new-physics contributions could alter the value of $\fl$ significantly.
The determination of \flvtb based on single-top-quark
cross-section measurements is independent of assumptions about the number of
quark generations and the unitarity of the CKM matrix.
The only assumptions required are that $|V_{tb}|\gg|V_{td}|,|V_{ts}|$
and that the $Wtb$ interaction involves a left-handed weak coupling as in the
SM. 

The value of \flvtbsqr is extracted by dividing the measured total inclusive
cross-section $\sigtot(tq+\bar{t}q)$ by the SM expectation given in
\Eqn{\eqref{eq:predicted_combined_xs}}.
When calculating \flvtbsqr, the experimental and theoretical
uncertainties are added in quadrature.
The uncertainty in \mtop is also considered, assuming 
$\Delta\mtop = \SI{\pm 1}{\GeV}$.
The result obtained is
\begin{linenomath}
\begin{align}
  \label{eq:Vtb}
  \begin{split}
  \flvtb & = 1.029 \pm 0.007 \, (\text{stat.}) \pm 0.029 \,
    (\text{exp.}) \,^{+0.023}_{-0.014} \, (\text{scale})\pm 0.004 \, (\text{PDF})\\
  &\quad\pm 0.010 \, (\text{NLO-matching method})\pm 0.009 \, (\text{parton shower}) \pm 0.010 \, (\text{lumi.})\\
  &\quad\pm 0.005 \, (m_t) \pm 0.024 \, (\text{theor.}) 
  \end{split}\\
  & = 1.029\pm 0.048\,.  \nonumber
\end{align}
\end{linenomath}
The uncertainty in \flvtb is broken down in the first terms,
reflecting the uncertainties in the combined total cross-section, as well as the
uncertainty in the top-quark mass and the uncertainty in the theoretical
cross-section calculation.
The result is in full agreement with the SM prediction.
Restricting the range of $|V_{tb}|$ to the
interval $[0, 1]$ and assuming $\fl=1$, as required by the SM, a lower limit on
$|V_{tb}|$ is extracted:
$|V_{tb}|>0.92$ at \SI{95}{\%} confidence level.

%% file: xsect_diff.tex
\section{Differential cross-section measurements}
\label{sec:xsect_diff}

The measured differential distributions are unfolded, so that they can be directly compared to theoretical predictions.
Two sets of unfolded cross-sections are derived: particle level and parton level.
Particle-level cross-sections are measured in the fiducial volume defined in \Sect{\ref{sec:xsect_def}}.
Parton-level cross-sections are measured in the whole kinematic range using the MC simulation to extrapolate from the acceptance phase space.
Particle-level cross-sections are measured as a function of the transverse
momentum, \pTthat, and absolute value of the rapidity, \absythat,  of the pseudo
top quark and pseudo top antiquark.
In addition, they are measured as a function of the transverse momentum, \pTjhat, and the absolute value of the rapidity, \absyjhat,
of the accompanying jet in the $t$-channel exchange, by assuming this jet is the untagged jet in the event.
Parton-level cross-sections are measured as a function of the transverse momentum, \pTt, and absolute value of the rapidity, \absyt,  of the top quark and top antiquark.

Differential cross-sections are extracted from an event sample enriched in
signal events, which is obtained by cutting on \NNout.
The cut value is set to $\NNout > 0.8$ (see \Fig{\ref{fig:NN_fitresult}}),
which achieves a good signal-to-background ratio and thereby reduces the impact
of the systematic uncertainties on the backgrounds, while maintaining enough
data events to keep the data statistical uncertainties at an acceptable level.

\begin{figure}[htbp]
  \centering
  \subfloat[][]{
    \includegraphics[trim={0 0 0 30pt}, clip,
    width=0.42\textwidth]{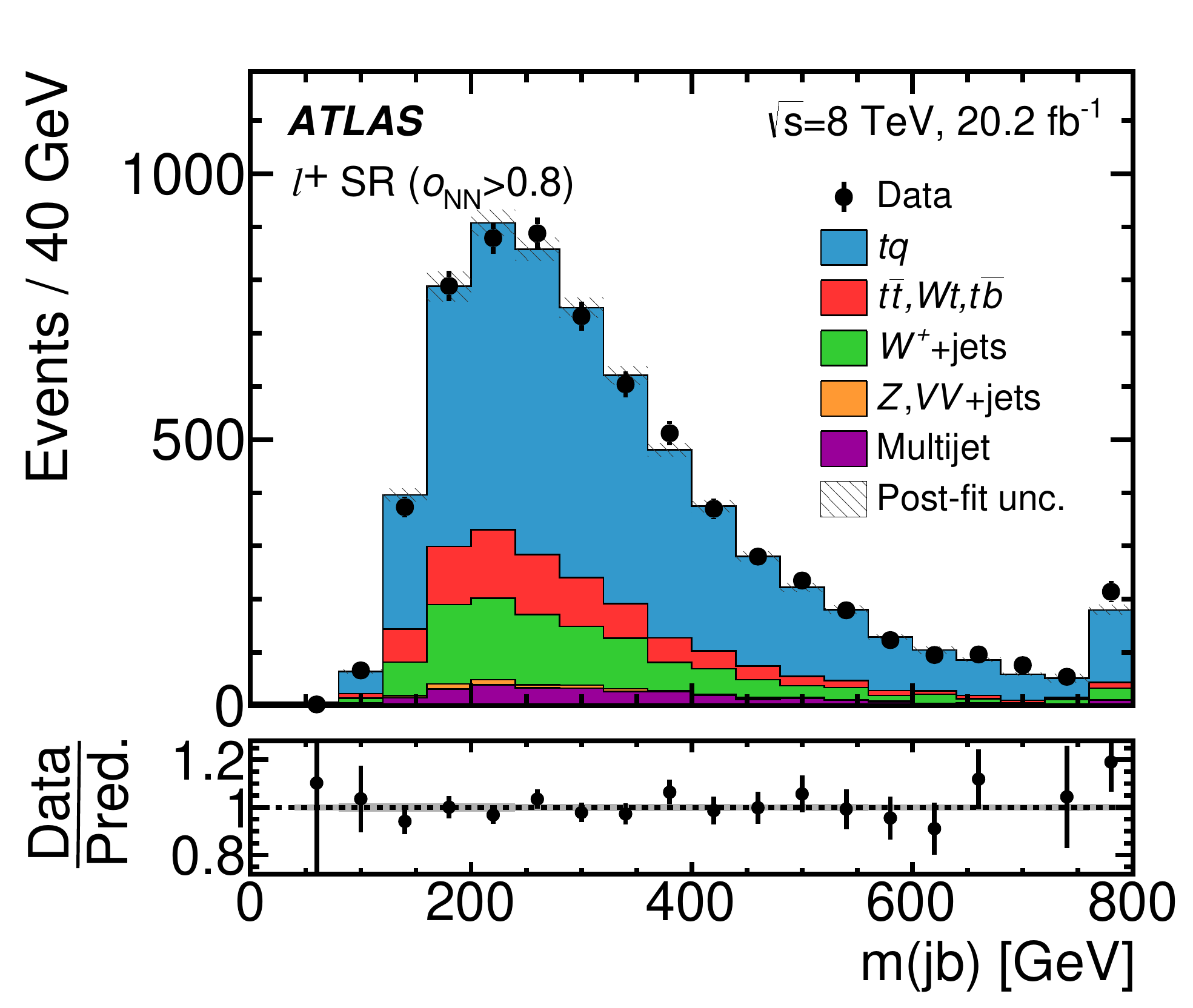} }
  \subfloat[][]{
    \includegraphics[trim={0 0 0 30pt}, clip,
    width=0.42\textwidth]{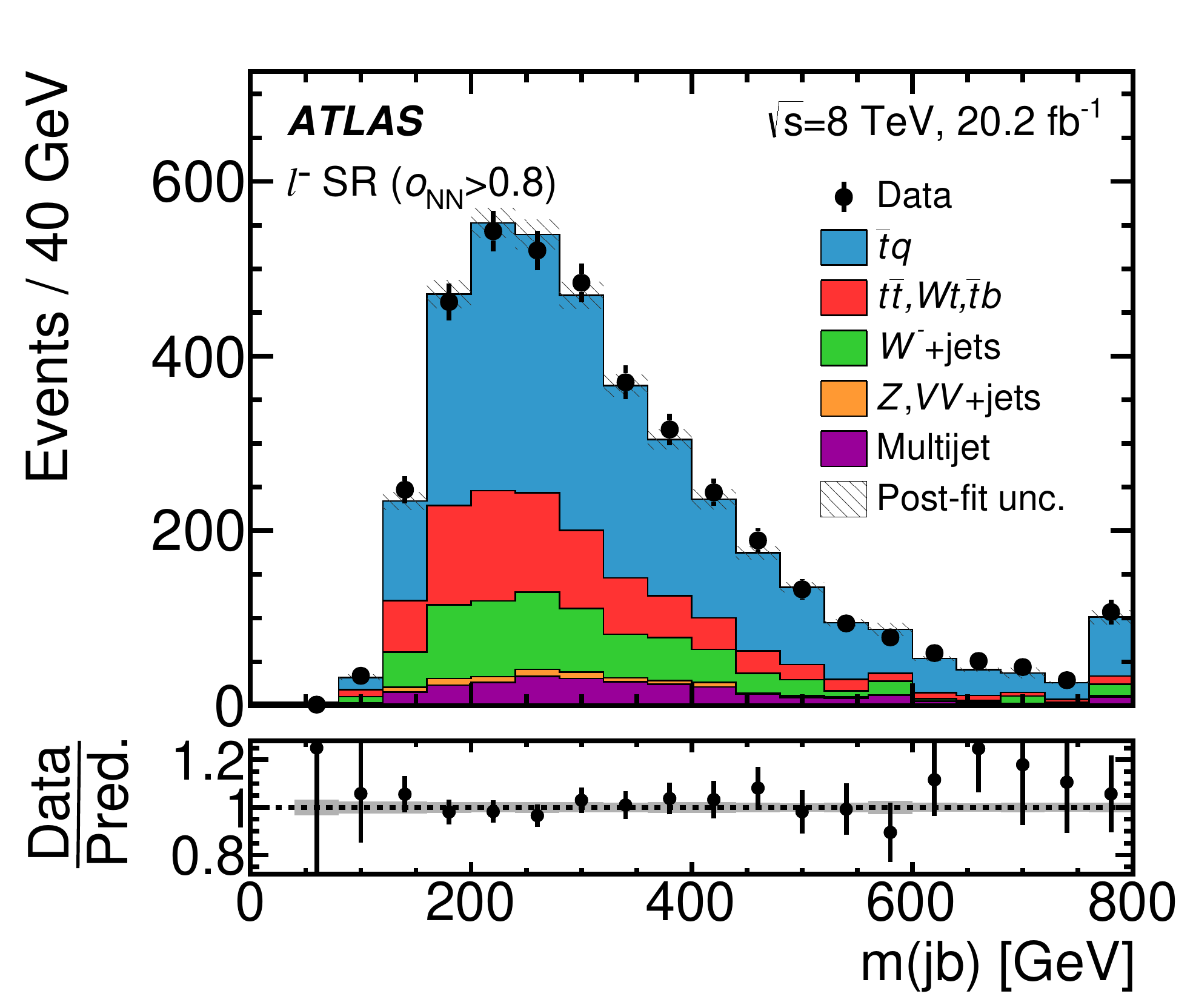} }\\
  \subfloat[][]{
    \includegraphics[trim={0 0 0 30pt}, clip,
    width=0.42\textwidth]{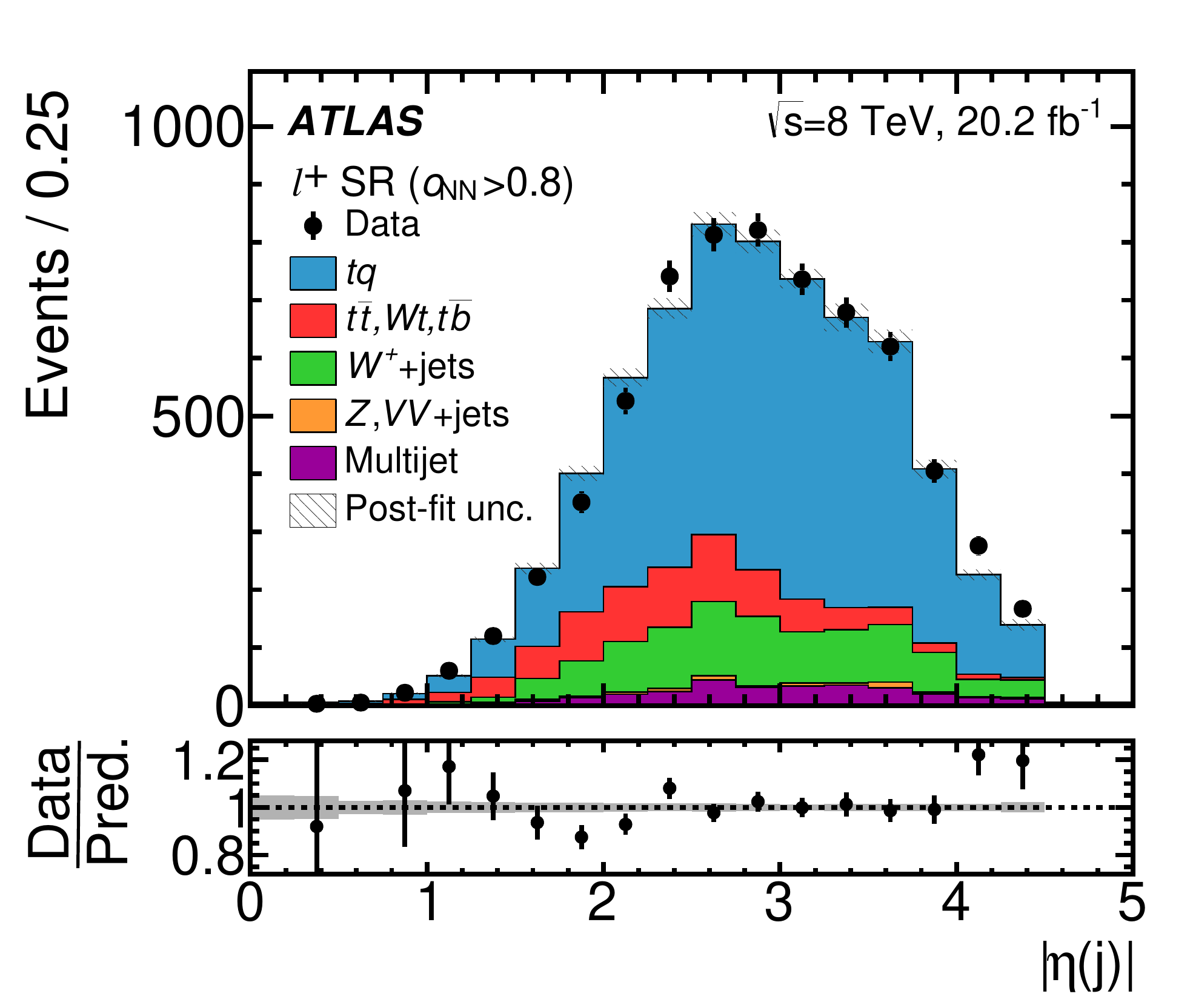} }
  \subfloat[][]{
    \includegraphics[trim={0 0 0 30pt}, clip,
    width=0.42\textwidth]{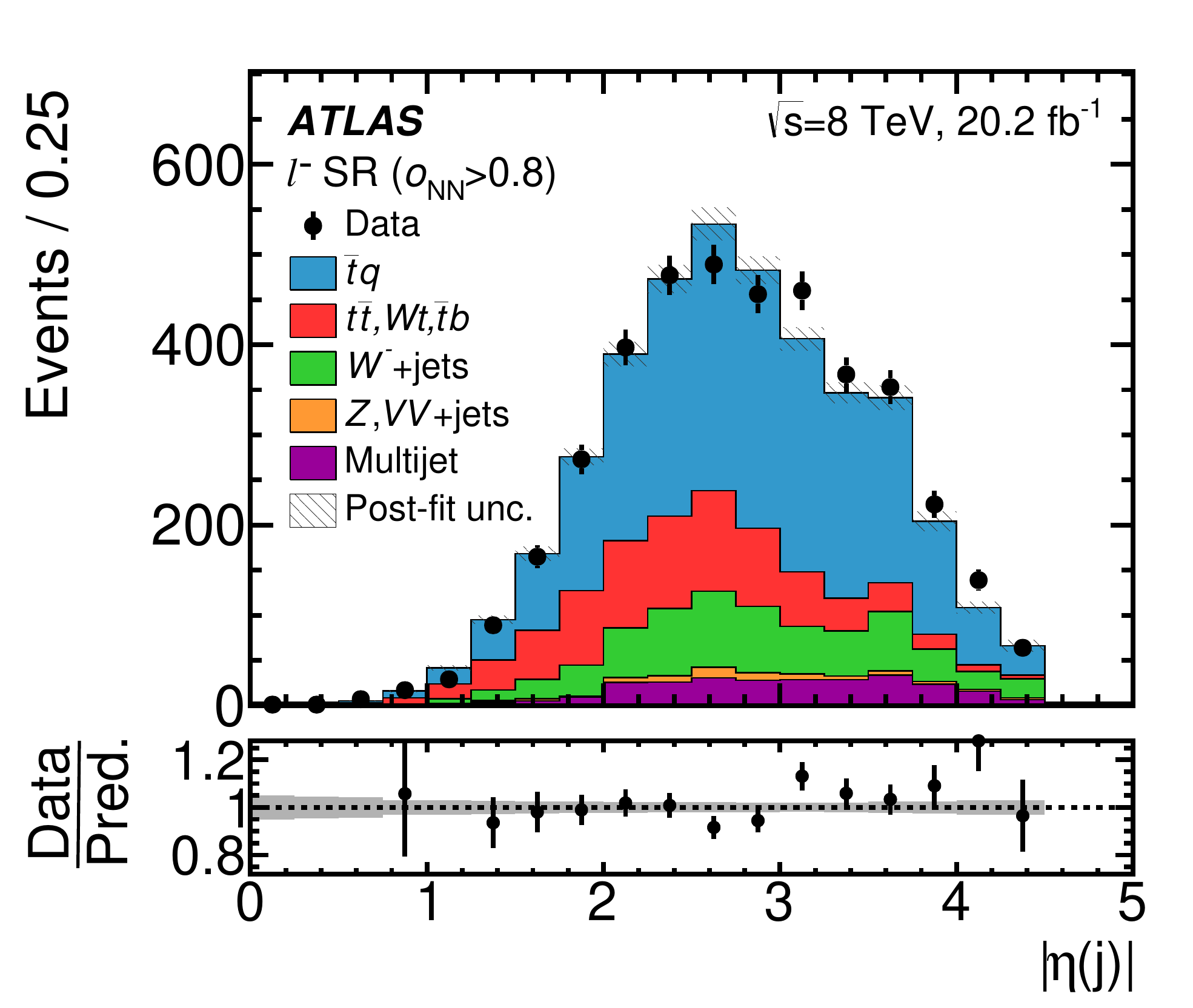} }\\
  \subfloat[][]{
    \includegraphics[trim={0 0 0 30pt}, clip,
    width=0.42\textwidth]{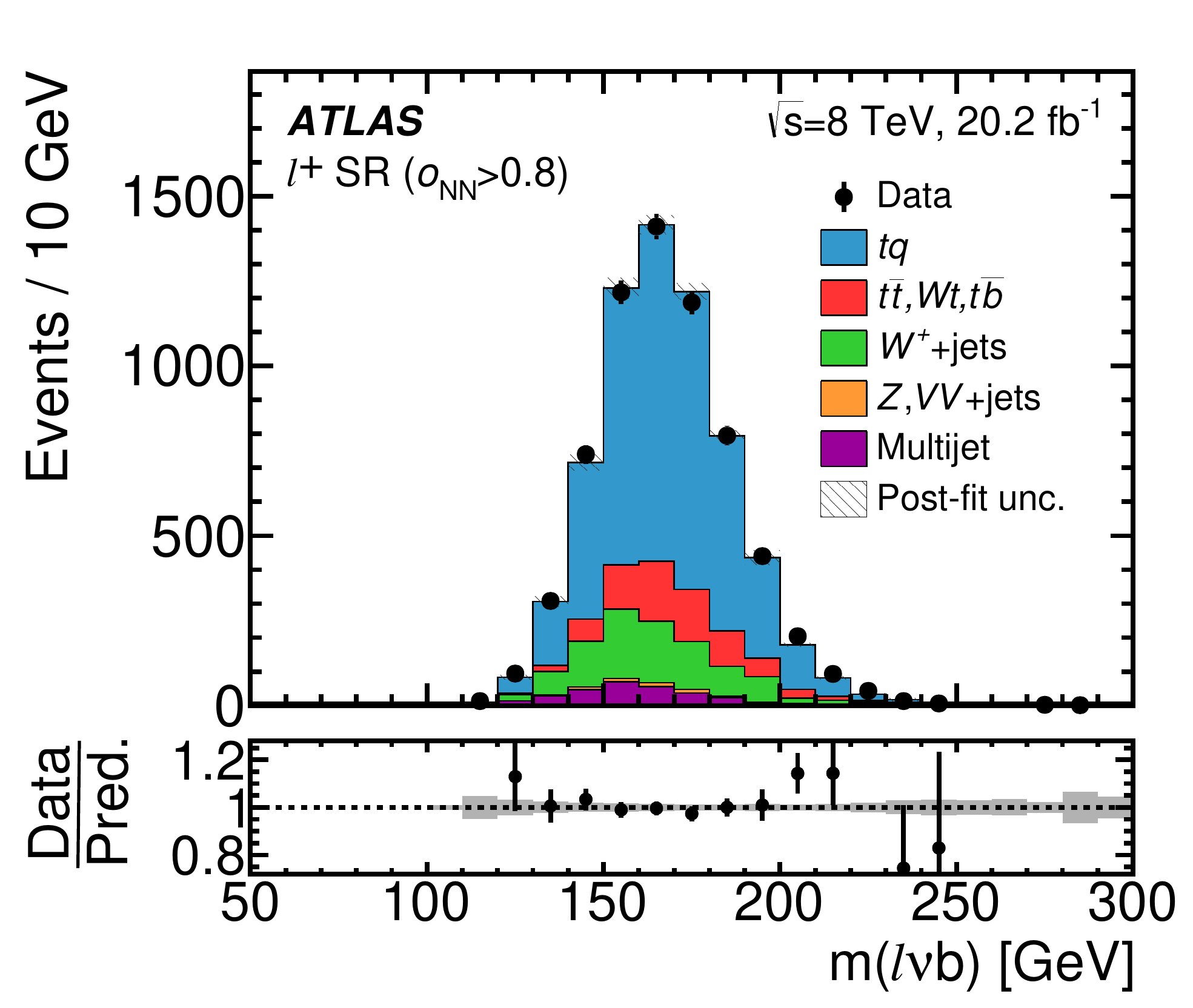} }
  \subfloat[][]{
    \includegraphics[trim={0 0 0 30pt}, clip,
    width=0.42\textwidth]{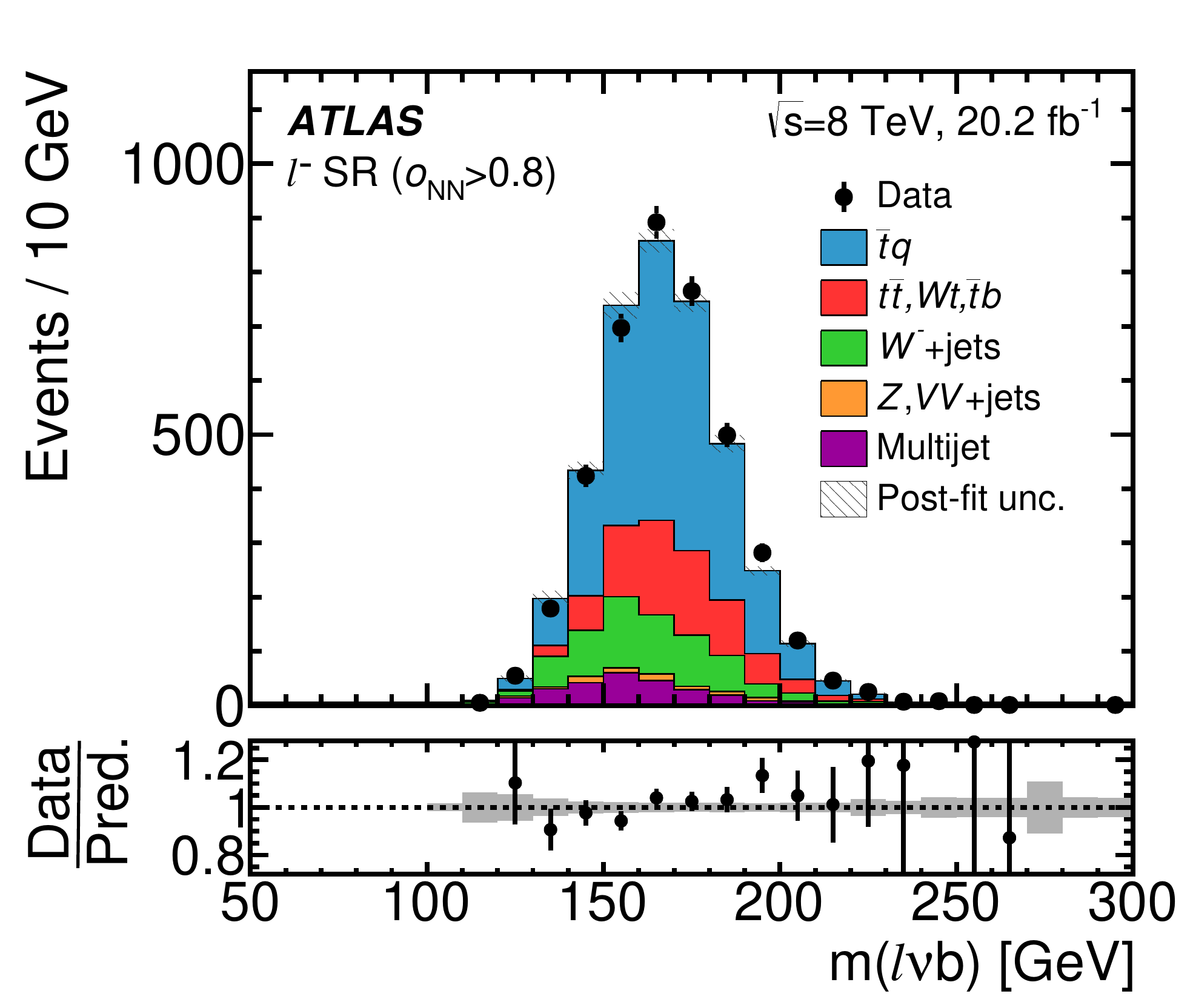} }
  \caption{Observed distributions of the first three input variables to the default neural network
    in the signal region (SR), after a cut of $\NNout > 0.8$ on the network output.
    The distributions are compared to the model obtained from simulated events.
    The simulated distributions are normalised to
    the event rates obtained by the fit to the discriminants.
    The definitions of the variables can be found in \Tab{\ref{tab:NNinputVars}}.
    \postfitCaptionUncSentence 
    Events beyond the $x$-axis range in (a) and (b) are included in the last bin.
    The lower panels show the ratio of the observed to the expected number of
    events in each bin to illustrate the goodness-of-fit.
  }
  \label{fig:CP_NN1a}
\end{figure}

Table~\ref{tab:evtyield} lists the numbers of events after the selection,
including the cut on $\NNout$,
separated into the \lp SR and the \lm SR.
Both signal and backgrounds, except for the multijet background, are normalised
to their fit value resulting from the binned maximum-likelihood fit to the whole
\NNout distribution, which was used to extract the total $t$-channel
cross-sections described in \Sect{\ref{sec:xsect_tot}}.
The multijet background normalisation is derived from the fit to the \MET distribution described in \Sect{\ref{sec:background_estimate}}.
Distributions of the three most discriminating input variables to the default NN (introduced in \Sect{\ref{sec:nn}}) after the cut on \NNout are shown in \Fig{\ref{fig:CP_NN1a}}.

\begin{table}[htbp]
  \centering
  \sisetup{round-mode=figures}
  \begin{tabular}{l
    S[table-format=6.0, round-precision=3]@{$\,\pm\,$}S[table-format=3.0, round-precision=2, table-number-alignment=left]
    S[table-format=6.0, round-precision=3]@{$\,\pm\,$}S[table-format=3.0, round-precision=2, table-number-alignment=left]}
  \toprule
  Process & \multicolumn{2}{c}{$\ell^{+}$ SR $(\NNout > 0.8)$} & \multicolumn{2}{c}{$\ell^{-}$ SR $(\NNout > 0.8)$} \\
  \midrule
  \tq       &  4472 & 183  & 5 & 0   \\
  \tbarq   &   3   &  0   & 2271 & 127 \\
  \midrule
  $\ttbar, Wt, t\bar{b}/\bar{t}b$  &  754 & 45  & 753 & 45 \\
  \wpjets     &  \multicolumn{1}{S[table-format=6.0, round-precision=2]@{$\,\pm\,$}}{963} & 193  &  1 &  0 \\
  \wmjets     &   1  &  0  &   \multicolumn{1}{S[table-format=6.0, round-precision=2]@{$\,\pm\,$}}{607} &  121 \\
  $Z, VV$\,+\,jets  &  52   &  10   &   60 &  12 \\
  Multijet         &  291  & 46  &  267 &  39 \\
  \midrule 
  Total estimated   & 6536 & 274 & 3964 & 186  \\
  Data  & \multicolumn{1}{S[table-format=6.0, round-mode=off]@{\phantom{$\,\pm\,$}}}{6567} & & \multicolumn{1}{S[table-format=6.0, round-mode=off]@{\phantom{$\,\pm\,$}}}{4007} \\
  \bottomrule
  \end{tabular}
  \caption{Predicted (post-fit) and observed event yields for the signal region (SR),
    after the requirement on the neural network discriminant, $\NNout > 0.8$. 
    The multijet background prediction is obtained from the fit to the \MET
    distribution described in \Sect{\ref{sec:background_estimate}},
    while all the other predictions and uncertainties are derived from the total
    cross-section measurement.
    An uncertainty of 0 means that the value is $<0.5$.}
\label{tab:evtyield}
\end{table}

\begin{table}[htbp]
  \centering
  \sisetup{round-mode=figures}
  \begin{tabular}{l
    S[table-format=6.0, round-precision=3]@{$\,\pm\,$}S[table-format=3.0, round-precision=2, table-number-alignment=left]
    S[table-format=6.0, round-precision=3]@{$\,\pm\,$}S[table-format=3.0, round-precision=2, table-number-alignment=left]}
  \toprule
  Process & \multicolumn{2}{c}{$\ell^{+}$ SR $(\NNoutII > 0.8)$} & \multicolumn{2}{c}{$\ell^{-}$ SR $(\NNoutII > 0.8)$} \\
  \midrule
  \tq       &  3443 & 141  & 3 &  0 \\
  \tbarq   &   2   &  0    & 1862 &  104 \\
  \midrule
  $\ttbar, Wt, t\bar{b}/\bar{t}b$   &  \multicolumn{1}{S[table-format=6.0, round-precision=4]@{$\,\pm\,$}}{1072} & 64   & \multicolumn{1}{S[table-format=6.0, round-precision=4]@{$\,\pm\,$}}{1057} & 63 \\
  \wpjets     &  770 & 154  &  0 & 0  \\
  \wmjets     &  0  & 0   &   494 &  99 \\
  $Z, VV$\,+\,jets      &  43   &  9   &   48 & 10 \\
  Multijet          &  192  & 30  &  186 & 27 \\
  \midrule 
  Total estimated   & 5522 & 221 & 3650 & 159 \\
  Data  & \multicolumn{1}{S[table-format=6.0, round-mode=off]@{\phantom{$\,\pm\,$}}}{5546}  & & \multicolumn{1}{S[table-format=6.0, round-mode=off]@{\phantom{$\,\pm\,$}}}{3647} \\
  \bottomrule
  \end{tabular}
  \caption{Predicted (post-fit) and observed event yields for the signal region (SR),
    after the requirement on the second neural network discriminant, $\NNoutII > 0.8$. 
    The multijet background prediction is obtained from the fit to the \MET
    distribution described in \Sect{\ref{sec:background_estimate}},
    while all the other predictions and uncertainties are taken from the total
    cross-section measurement.
    An uncertainty of 0 means that the value is $<0.5$.}
  \label{tab:evtyield_2ndNN}
\end{table}

\begin{figure}[htbp]
  \centering
  \subfloat[][]{
    \includegraphics[width=0.48\textwidth]{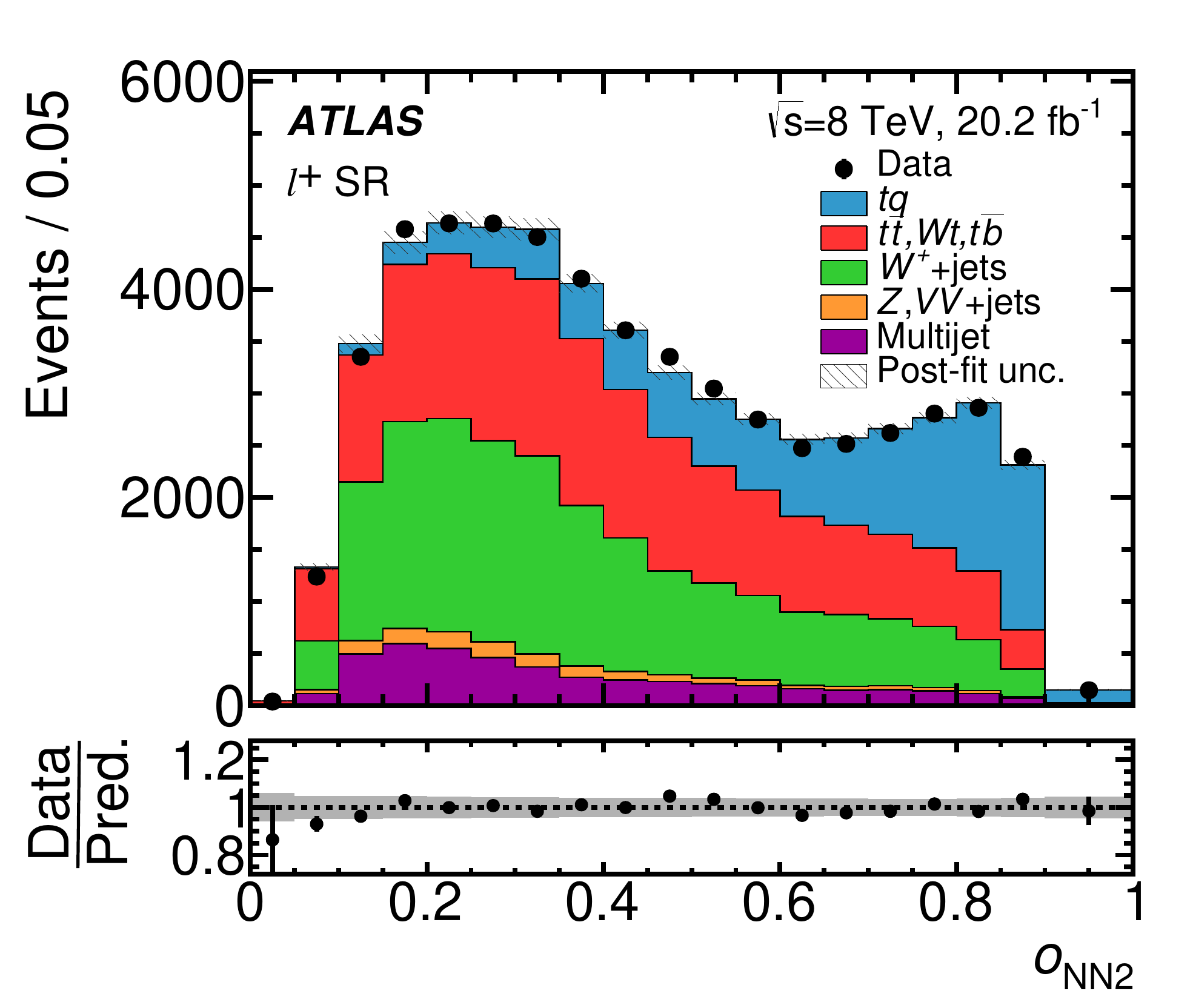}
    }
  \subfloat[][]{
    \includegraphics[width=0.48\textwidth]{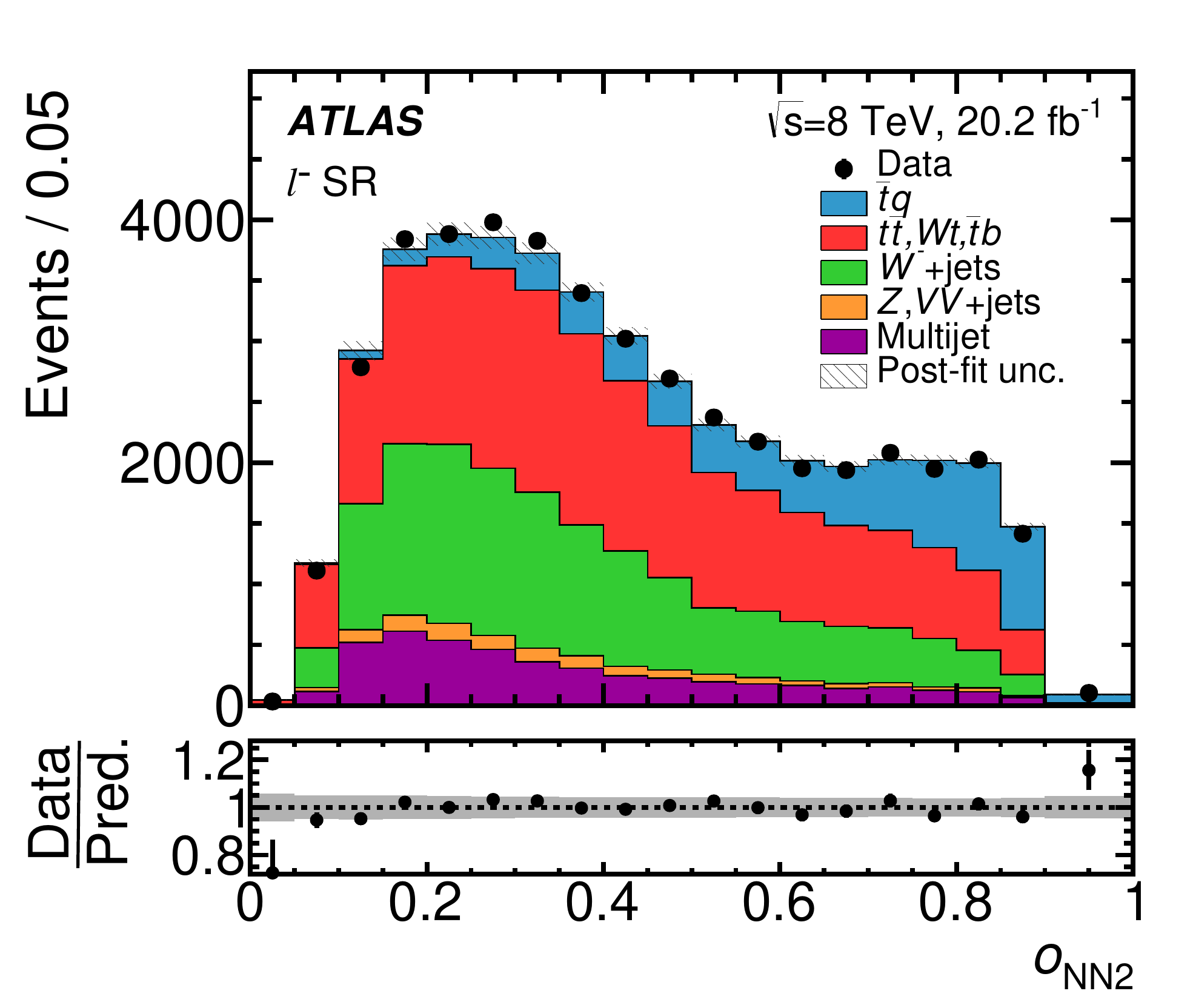}
    }
  \caption{Neural network output distribution (\NNoutII) of the neural network 
    without $|\eta(j)|$ normalised to the fit results of the default network
    for (a) the \lp and (b) the \lm signal region (SR).
    The distributions are compared to the model obtained from simulated events.
    The simulated distributions are normalised to
    the event rates obtained by the fit to the discriminants.
    \postfitCaptionUncSentence 
  }
  \label{fig:NNouts2network}
\end{figure}

For the measurement of the \absyjhat distribution, a second neural network (NN2) is trained omitting the variable $|\eta(j)|$,
in order to reduce the distortion of the \absyjhat distribution as a result of cutting
on the NN output.
The distribution of the neutral network output variable \NNoutII is shown in \Fig{\ref{fig:NNouts2network}}
for both the \lp and \lm signal regions.
\newcommand*{\DiffXsectFigNNCaption}{%
  The distributions are compared to the model obtained from simulated events.
  The simulated distributions are normalised to
  the event rates obtained by the fit to the discriminants.
  \postfitCaptionUncSentence 
  The lower panels show the ratio of the observed to the expected number of
  events in each bin to illustrate the goodness-of-fit.
}
A cut $\NNoutII > 0.8$ is placed on the NN output to select the events used in the unfolding.
The event yields after the event selection with this network are shown in
Table~\ref{tab:evtyield_2ndNN}.
Very good agreement between the data and the predictions can be seen for both networks,
indicating that the variables are also well described in the region where signal dominates.

\begin{figure}[htbp]
  \centering
  \subfloat[][]{
    \includegraphics[width=0.46\textwidth]{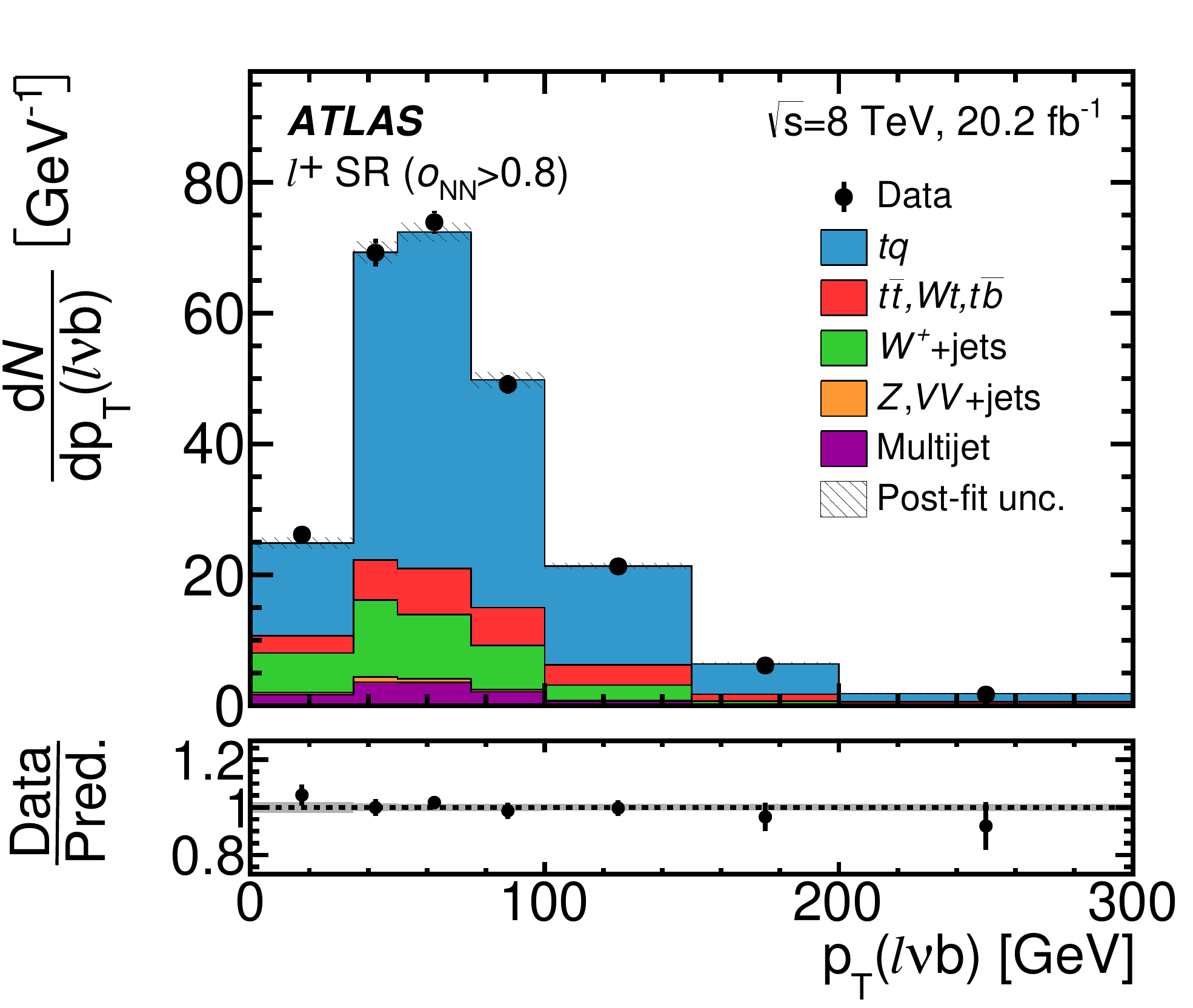}
    }
  \subfloat[][]{
    \includegraphics[width=0.46\textwidth]{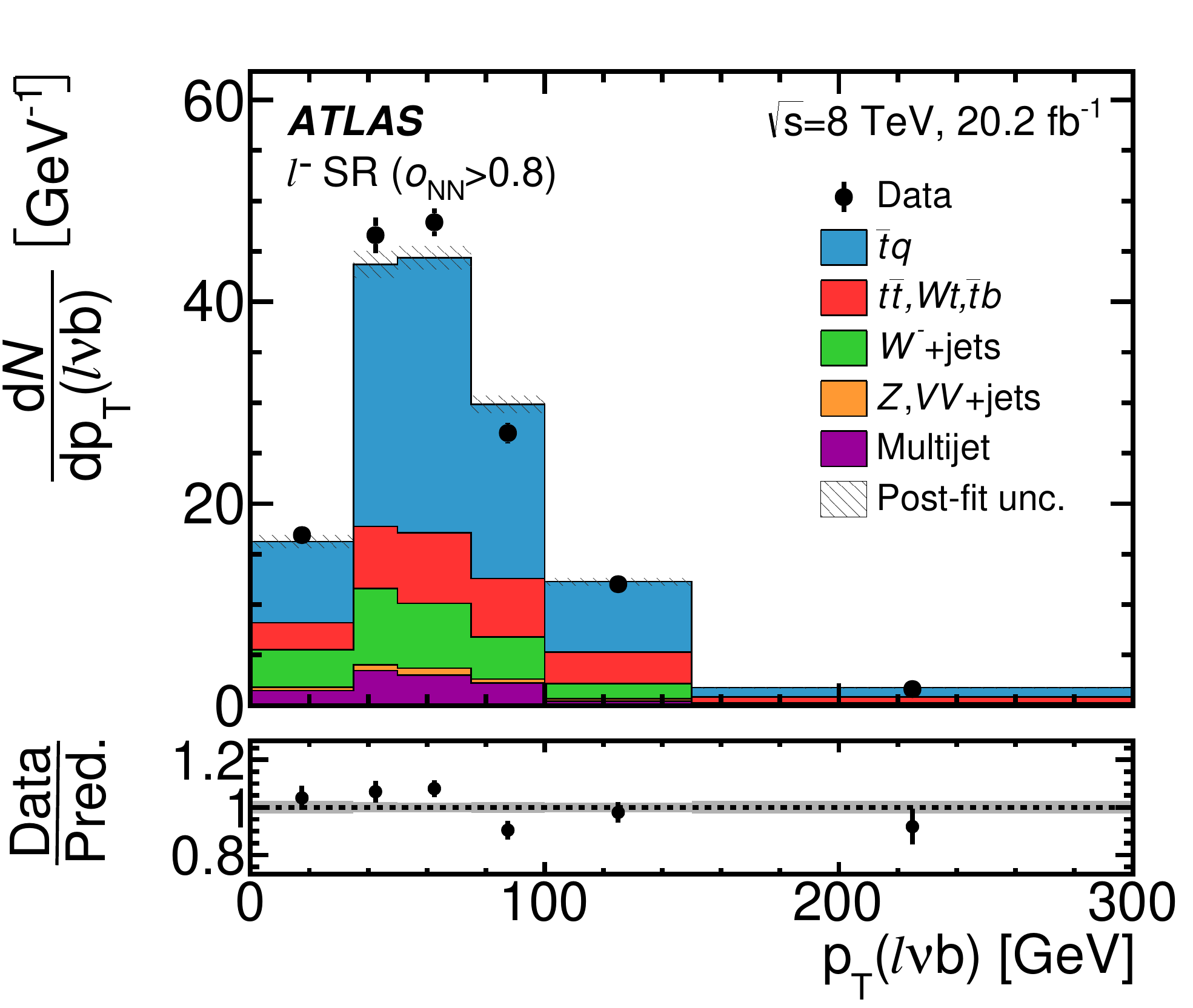}
    }\\
  \subfloat[][]{
    \includegraphics[width=0.46\textwidth]{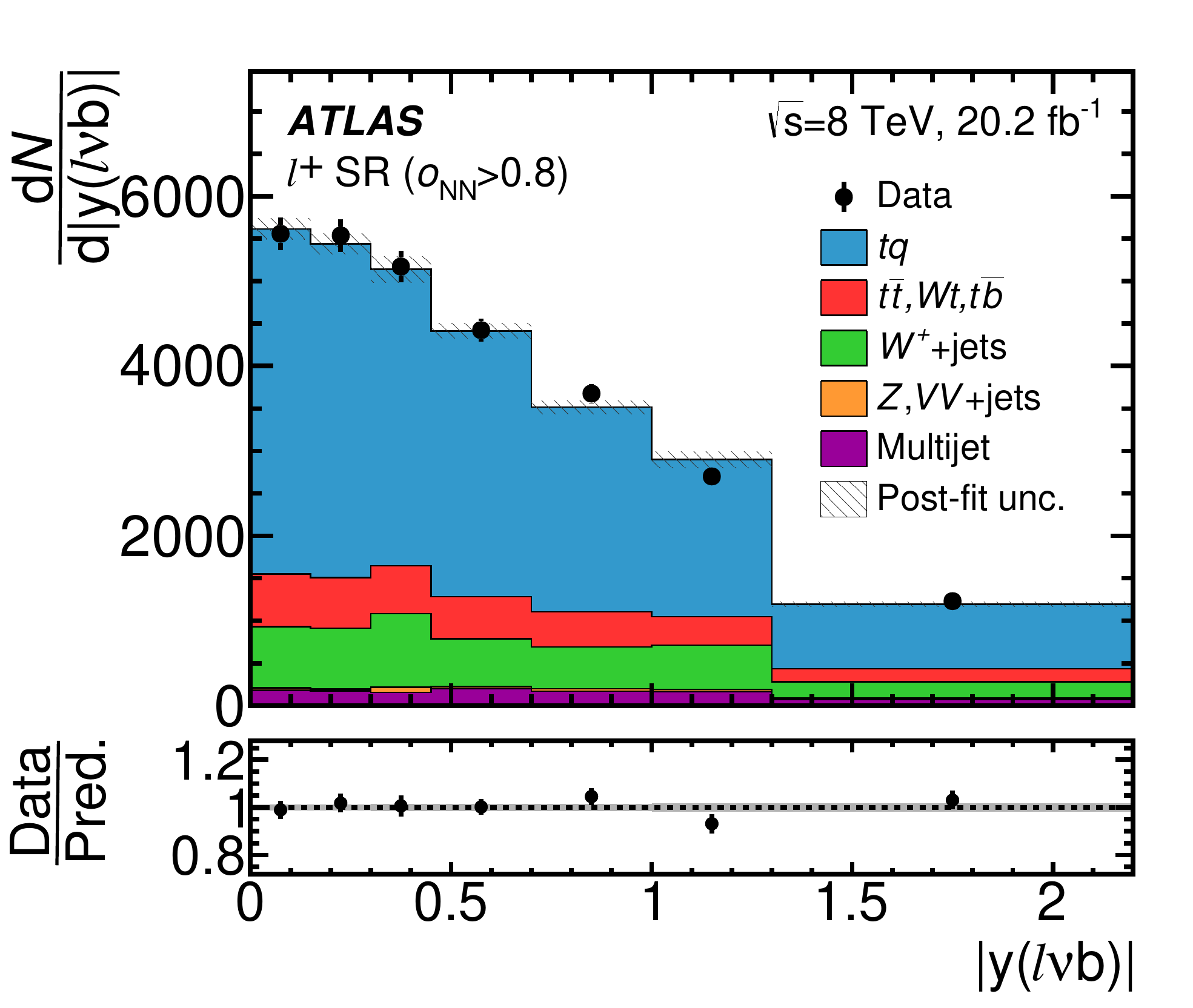}
    }
  \subfloat[][]{
    \includegraphics[width=0.46\textwidth]{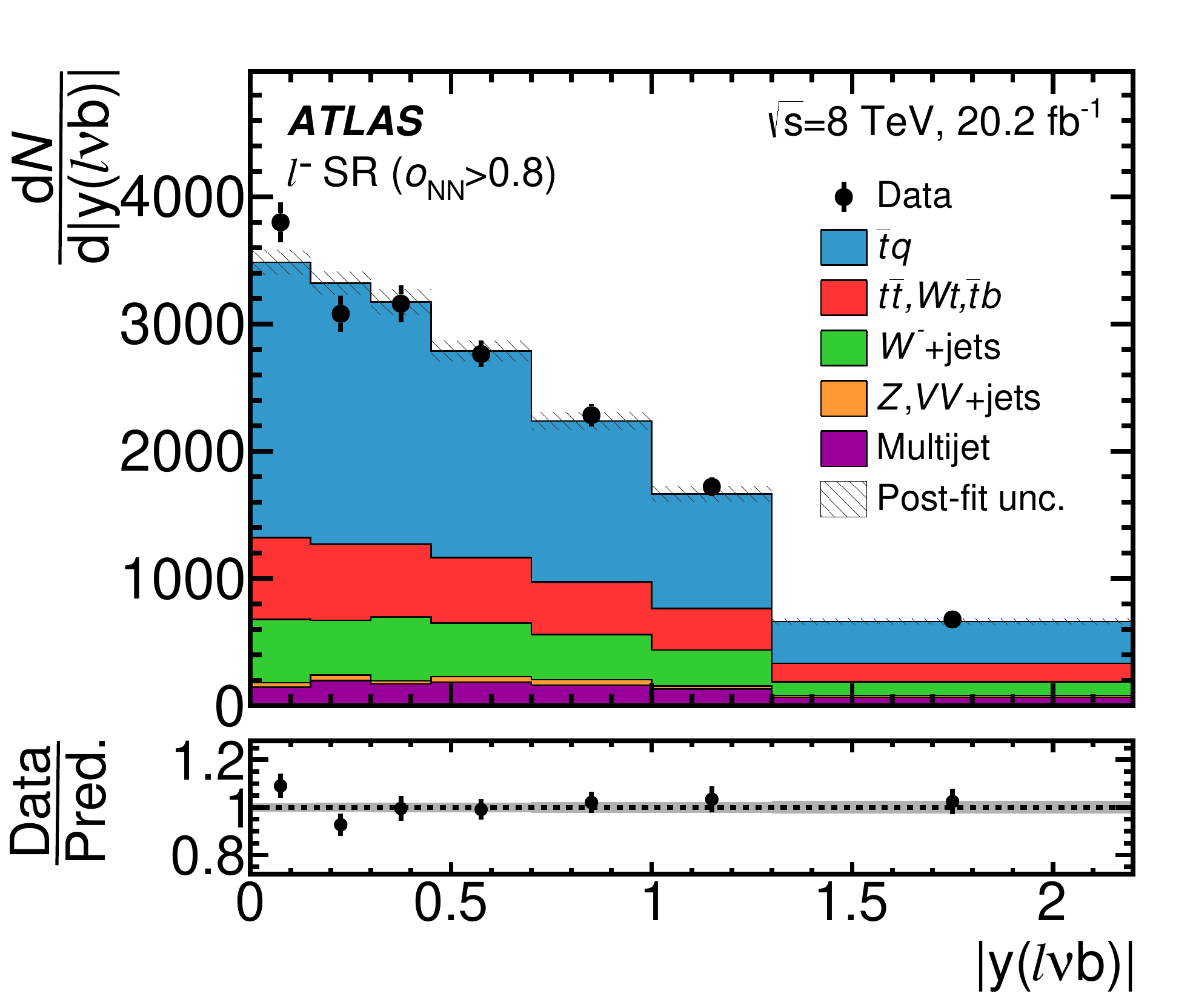}
    }
  \caption{Measured distributions of (a, b) \pTtreco and (c, d) \absytreco for 
    (a, c) \lp and (b, d) \lm events in the signal region (SR) after a cut of $\NNout > 0.8$.
    \DiffXsectFigNNCaption
  }
  \label{fig:binned_pseutop}
\end{figure}

\begin{figure}[htbp]
  \centering
  \subfloat[][]{
    \includegraphics[width=0.46\textwidth]{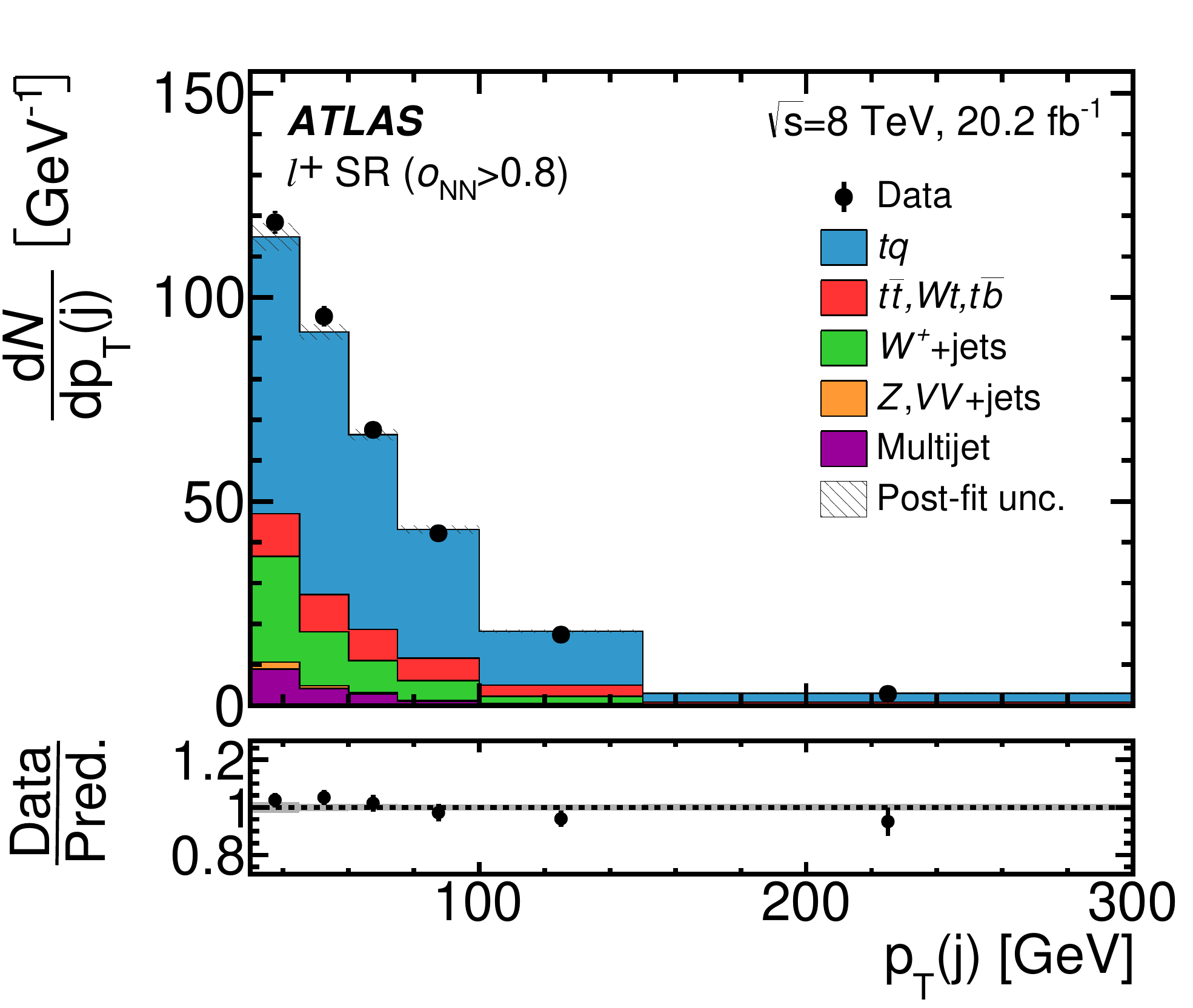}
    }
  \subfloat[][]{
    \includegraphics[width=0.46\textwidth]{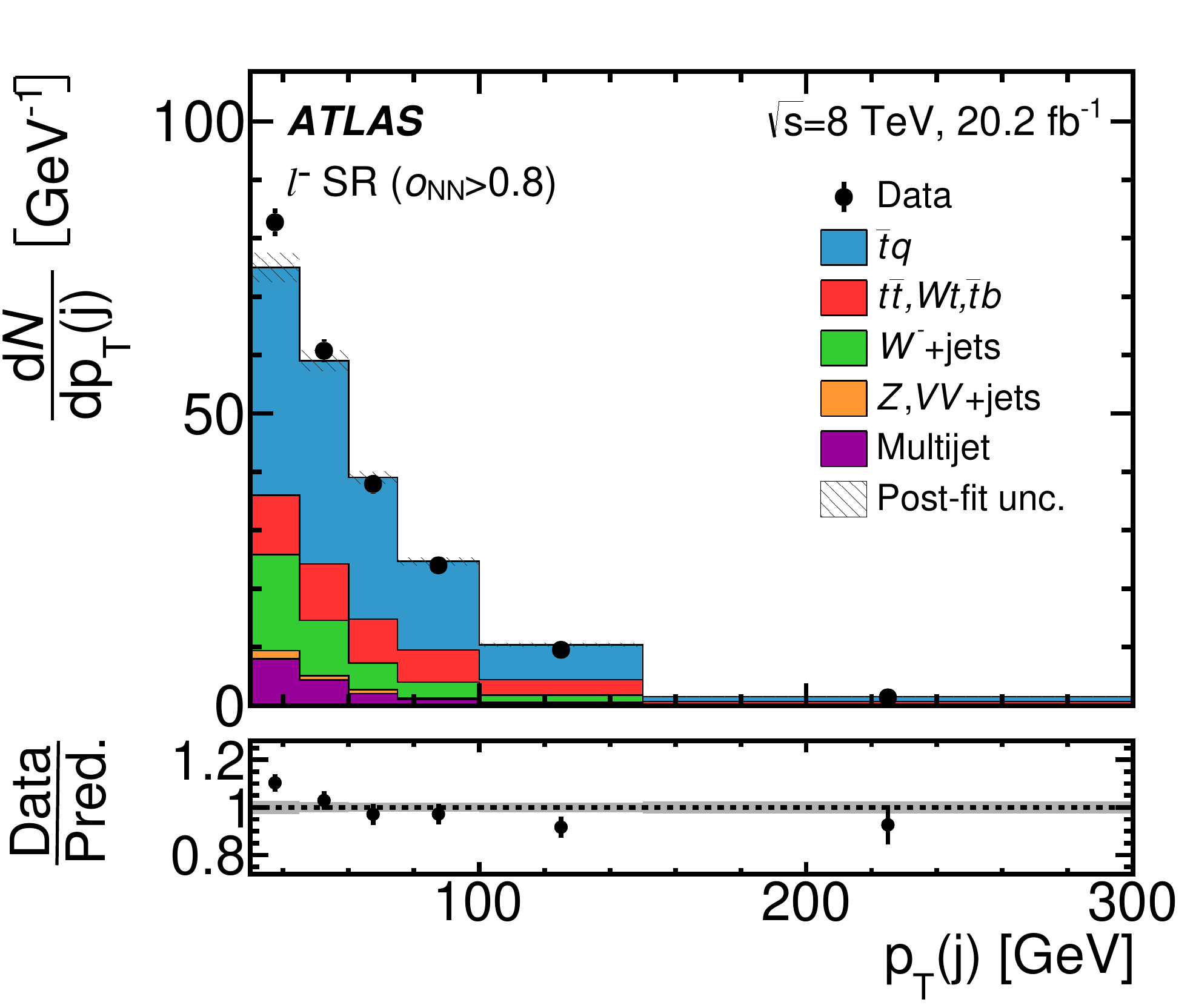}
    }\\
  \subfloat[][]{
    \includegraphics[width=0.46\textwidth]{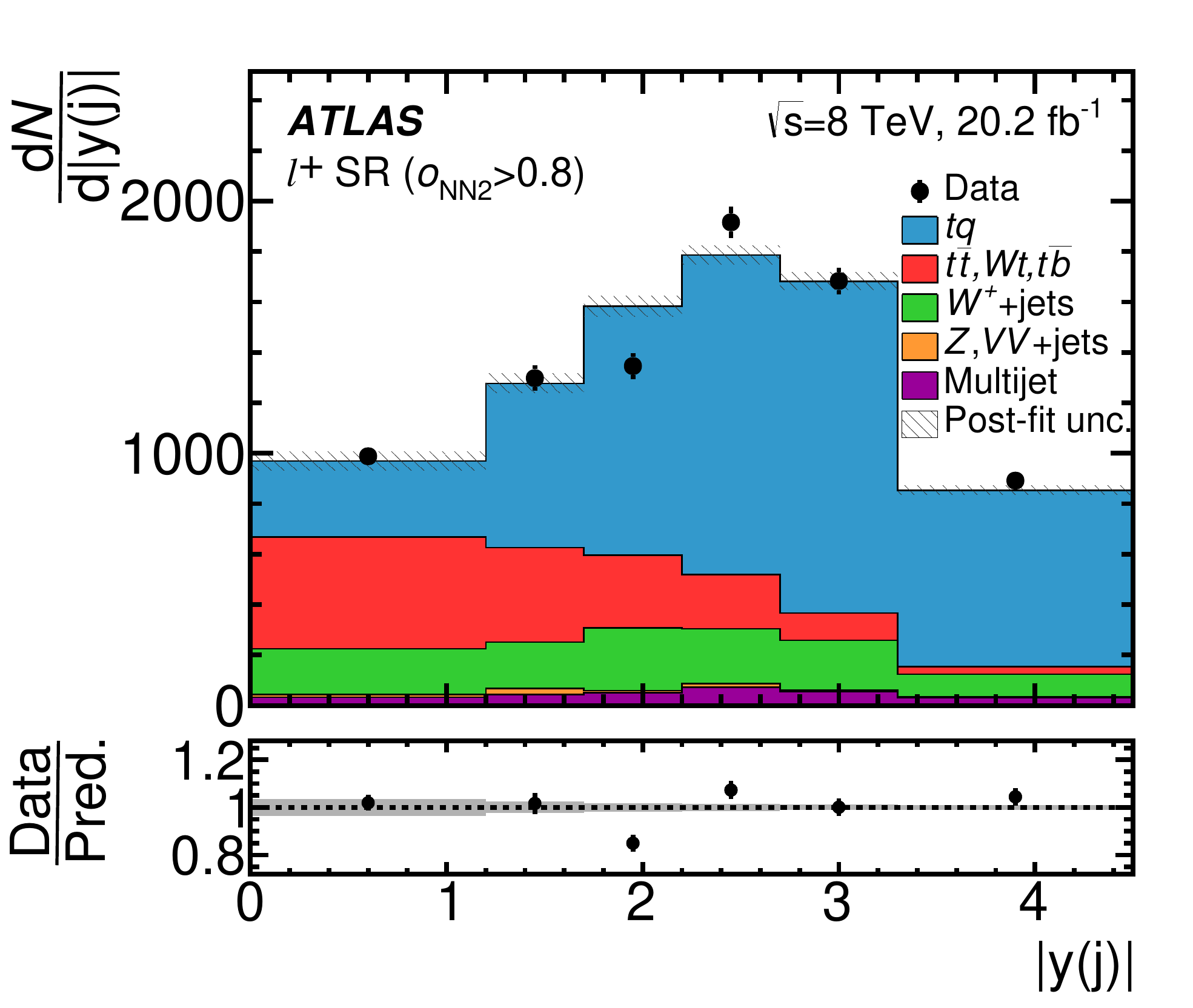}
    }
  \subfloat[][]{
    \includegraphics[width=0.46\textwidth]{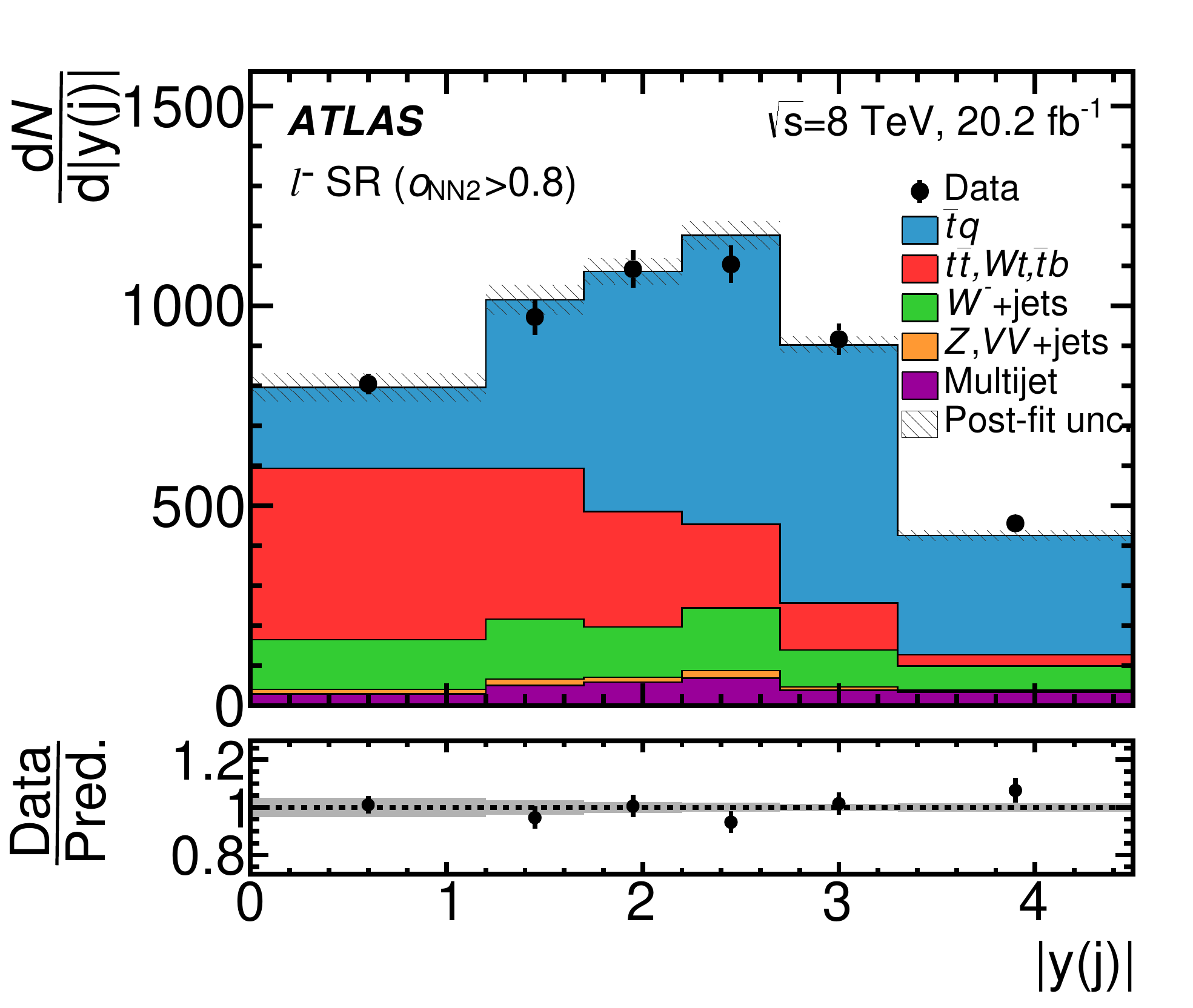}
    }
  \caption{Measured distributions of (a, b) \pTj and (c, d) \absyj at reconstruction level
    for (a, c) \lp and (b, d) \lm events in the signal region (SR) after a cut of $\NNout (\NNoutII) > 0.8$.
    \DiffXsectFigNNCaption
  }
  \label{fig:binned_lightjet}
\end{figure}

The measured differential distributions used in the unfolding are shown in \Figs{\ref{fig:binned_pseutop}}{\ref{fig:binned_lightjet}}.

Normalised differential cross-sections are evaluated by dividing the cross-section in each bin
by the sum of the cross-sections in all bins for a given variable.
The uncertainty in the normalised cross-section in each bin is determined
from the coherent variation of the cross-section in that bin and the total cross-section
when a variation reflecting a systematic uncertainty is applied.

\subsection{Unfolding technique}
\label{sec:xsect_diff:unfold}

D'Agostini's iterative approach~\cite{DAgostini1995487},
implemented in RooUnfold~\cite{Adye:2011gm}, is used to unfold the distributions.
The method is based on picturing the problem with an \enquote{effect} and a \enquote{cause}.
The number of reconstructed measured $t$-channel single-top-(anti)quark events
in bin $j$
is the effect,
while the number of produced $t$-channel events in a $pp$ collision in bin $k$, $N_{k}$, corresponds to the cause.
As indicated, the bins of the measured distribution are labelled with $j$,
while the bins of the generator-level distribution are labelled with $k$.

The unfolding starts from the reconstructed measured distributions.
The aim is to correct these distributions for resolution and efficiency effects.
The observed number of events in each bin $j$ of the measured distribution can be described by:
\begin{equation}
N^{\text{data}}_j = \sum_k M_{jk} \epsilon_k L_{\text{int}}\cdot \dif\hat{\sigma}_k + \hat{B}_j\,,
\end{equation}
where $\dif\hat{\sigma}_k$ is the estimated cross-section in each bin $k$,
$M_{jk}$ is the migration matrix, 
$\epsilon_k$ is the efficiency for an event to be selected in bin $k$ and
$\hat{B}_j$ is the sum of all background contributions.

\begin{figure}[htbp]
  \centering
  \subfloat[][]{\includegraphics[width=0.46\textwidth]{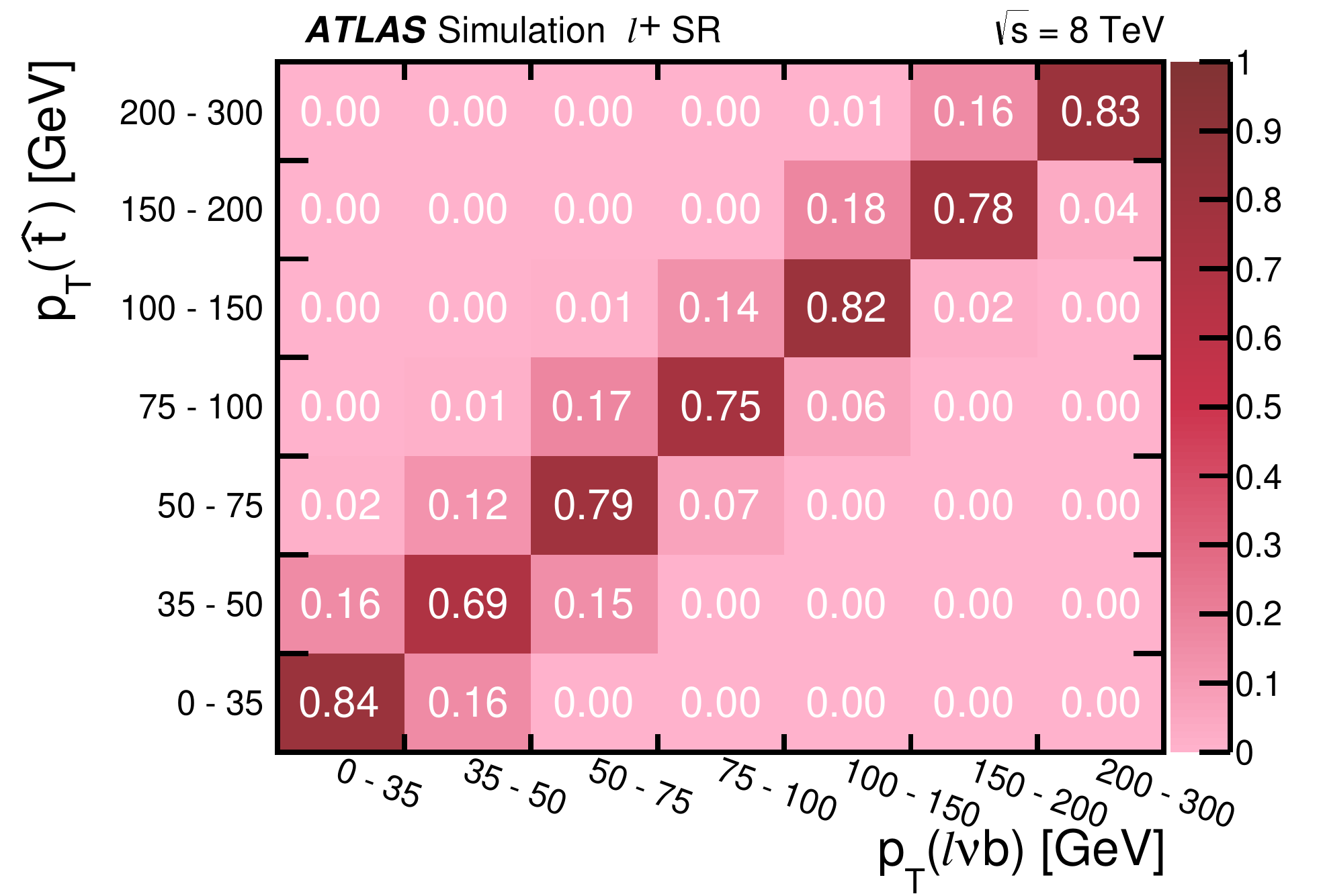}}\quad
  \subfloat[][]{\includegraphics[width=0.46\textwidth]{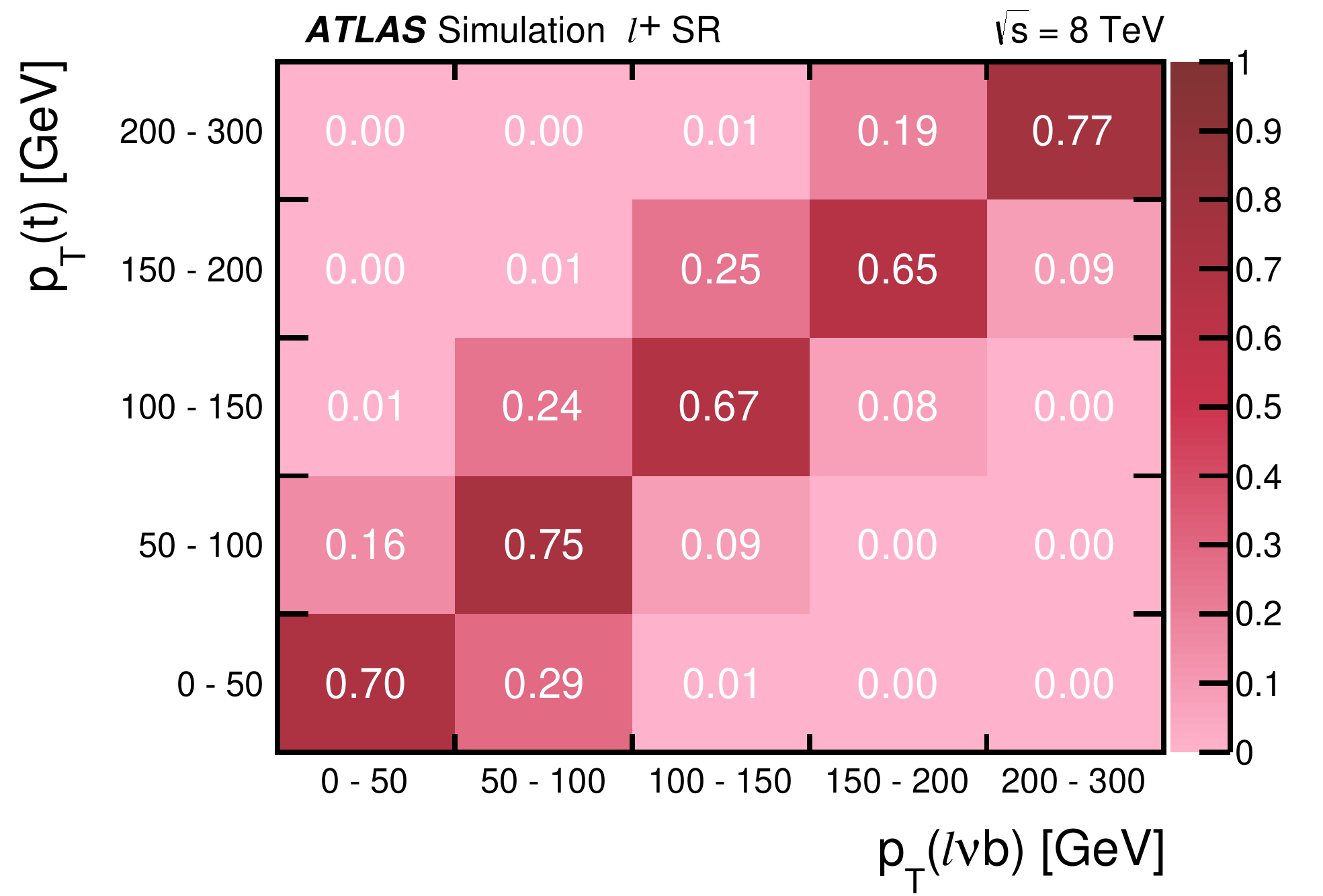}}\\
  \subfloat[][]{\includegraphics[width=0.46\textwidth]{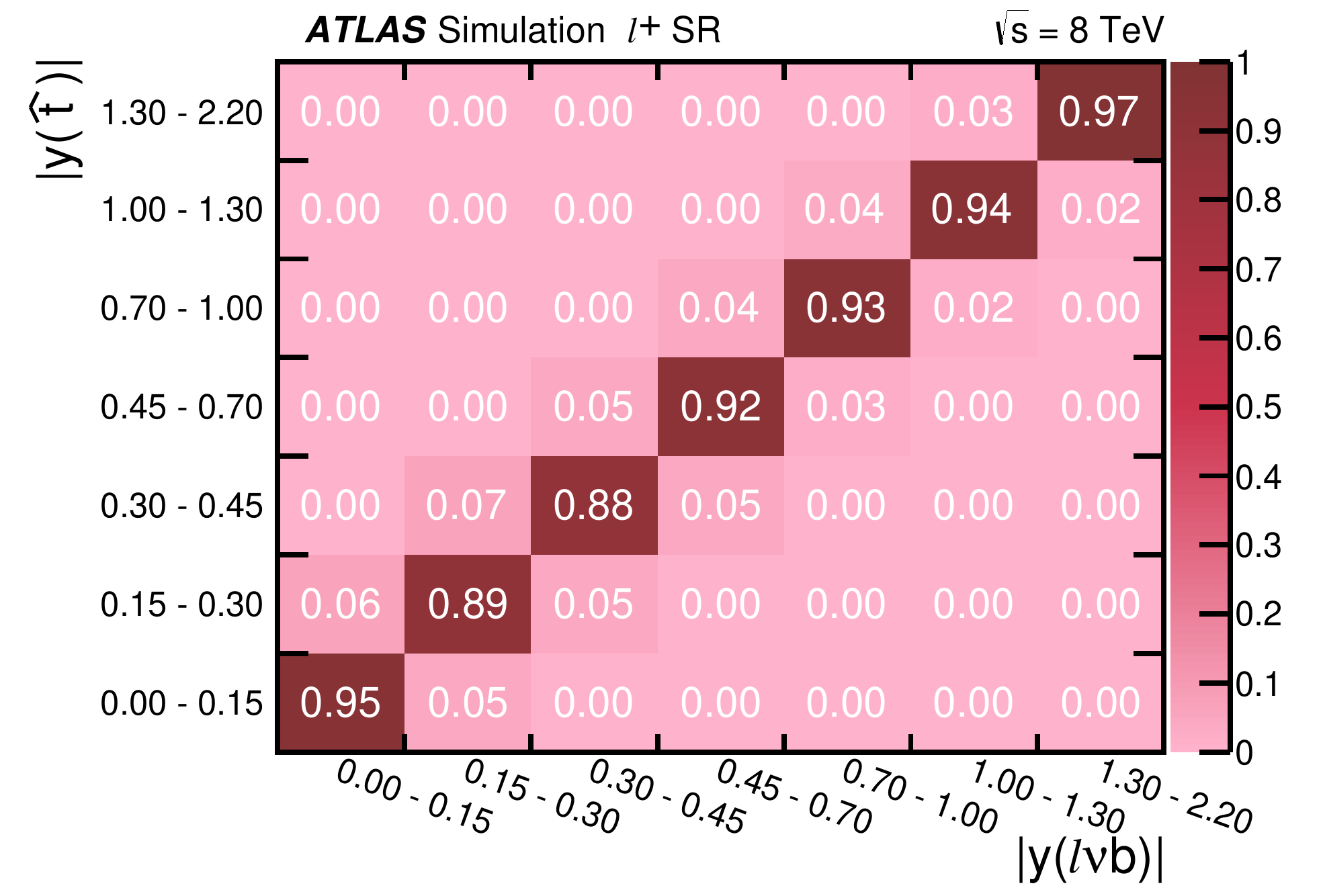}}\quad
  \subfloat[][]{\includegraphics[width=0.46\textwidth]{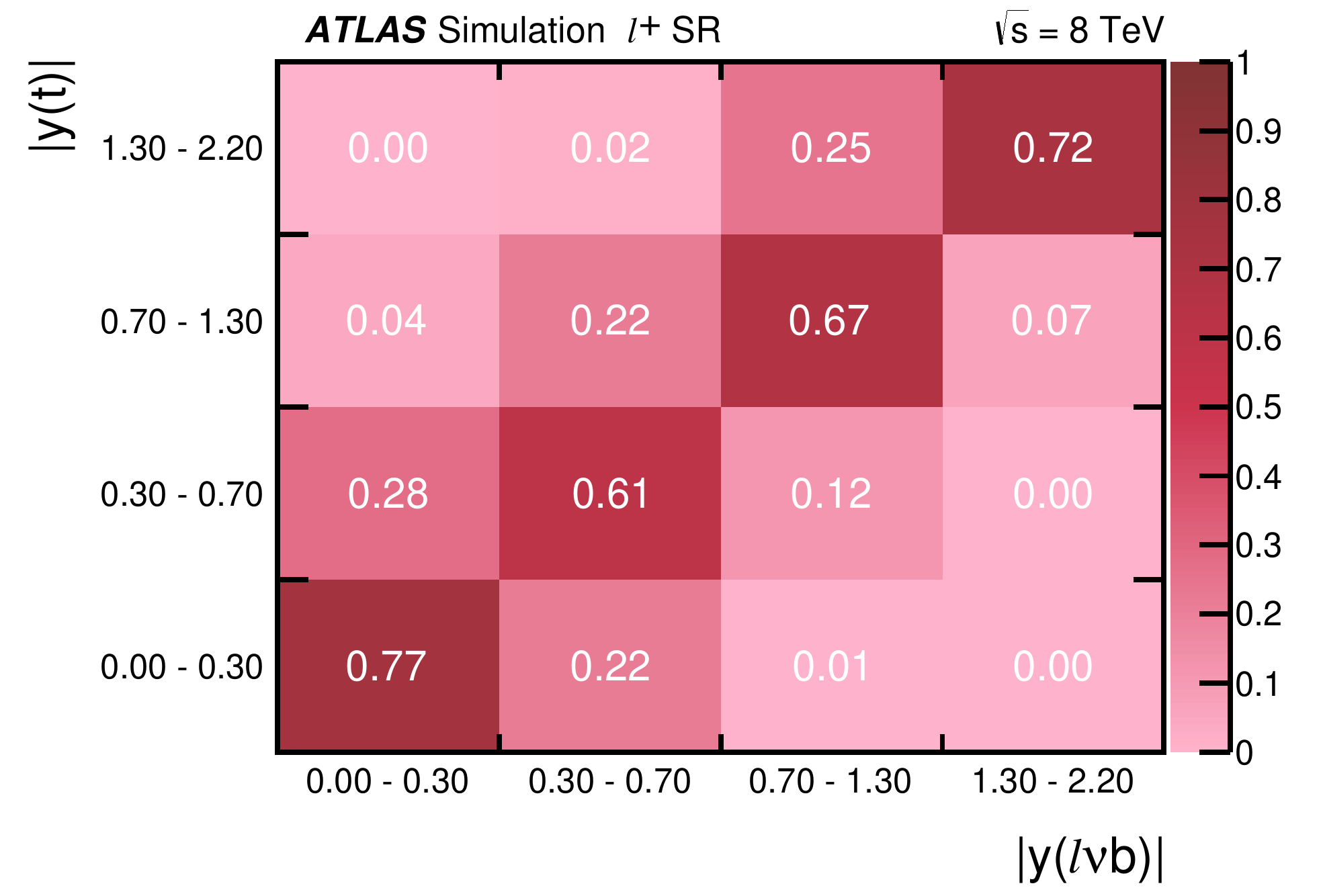}}
  \caption{Migration matrices for (a) \pTthat, (b) \pTt, (c) \absythat and (d) \absyt.
  (a) and (c) are for particle level, while (b) and (d) are for parton level.
  The pseudo top quark or parton-level quark is shown on the $y$-axis
  and the reconstructed variable is shown on the $x$-axis.}
  \label{fig:migration:pTt_yt}
\end{figure}

The migration matrix describes the probability of migration of generator-level events in bin $k$ to bin $j$ after detector reconstruction of the event.
Migration matrices, determined with the \POWHEGBOX + \PYTHIAV{6} MC sample,
for \pTthat and \absythat at particle level and \pTt and \absyt at parton level are shown in \Fig{\ref{fig:migration:pTt_yt}}.
The advantage of unfolding to particle level can clearly be seen;
the sizes of the off-diagonal elements in the particle-level migration matrices are much smaller,
which makes the unfolding less sensitive to the effect of systematic uncertainties.

The efficiency, $\epsilon_{k}$, includes signal acceptance, detector efficiencies
due to e.g.\ trigger and $b$-tagging, as well as the efficiency of the cut on the NN output:
\begin{equation}
\epsilon_k = \frac{S_{k}^{\text{sel,MC}}}{S_k^{\text{tot,MC}}}\,,
\label{eq:efficiency}
\end{equation}
where $S_k^{\text{tot,MC}}$ is the number of generated MC events in bin $k$
and $S_{k}^{\text{sel,MC}}$ is the number of selected MC events in bin $k$ after all cuts are applied.

$\hat{B}_j$ is calculated from the estimated number of background events, $\tilde{\nu}_j^b$,
resulting from the binned maximum-likelihood fit of the total cross-section
measurement:
\begin{equation}
\hat{B}_j = \sum_{b\in \text{all background}} \tilde{\nu}_j^b\,.
\end{equation}

\subsubsection{Unfolding to particle level}
\label{sec:xsect_diff:unfold:particle}

The reconstructed observables of both top quarks and untagged jets are unfolded to the particle level within the fiducial volume.
The detector efficiency and resolution effects are corrected using
\begin{equation}
\hat{\nu}_k^{\text{ptcl}} = 
\mathcal{C}_k^{\text{ptcl!reco}} 
\sum_j M^{-1}_{jk} \mathcal{C}_j^{\text{reco!ptcl}}  (N^{\text{data}}_j - \hat{B}_j)\,,
\label{eq:nuhat_particle}
\end{equation}
where $\hat{\nu}_k^{\text{ptcl}}$ is the measured expectation value for the number of signal events at particle level in bin $k$ of the fiducial volume,
$M^{-1}_{jk}$ represents the Bayesian unfolding procedure,
and $\mathcal{C}_j^{\text{reco!ptcl}}$ is a correction factor for signal events that pass the reconstruction-level selection but not the particle-level selection.
It is defined as
\begin{equation}
  \mathcal{C}_j^{\text{reco!ptcl}} =
  \frac{S^{\text{reco}}_j - S^{\text {reco!ptcl}}_j}{S^{\text{reco}}_j}\,, 
\end{equation}
where $S^{\text{reco}}_j$ is the number of reconstructed signal events in bin
$j$ and $S^{\text{reco!ptcl}}_j$ is the number of events that pass
the reconstruction-level selection but not the particle-level selection. 
$\mathcal{C}_k^{\text{ptcl!reco}}$ is a correction factor that accounts for signal events that pass the particle-level selection but not the reconstruction-level selection:
\begin{equation}
  \mathcal{C}_k^{\text{ptcl!reco}} = \frac{1}{\epsilon_{k}} =
  \frac{S^{\text{ptcl}}_k}{S^{\text{ptcl}}_k - S^{\text{ptcl!reco}}_k}\,, 
\end{equation}
where $S^{\text{ptcl}}_k$ is the number of signal events at particle level and
$S^{\text{ptcl!reco}}_j$ is the number of events that pass
the particle-level selection but not the reconstruction-level selection.
The cross-section in bin $k$ is evaluated from
\begin{equation}
  \dif\hat{\sigma}_k = \hat{\nu}_k^{\text{ptcl}} / L_{\text{int}}\,.
  \label{eq:xsect_particle}
\end{equation}
For following iterations,
the estimated number of events, $\hat{\nu}_k^{\text{ptcl}}$, is used as input.

\subsubsection{Unfolding to parton level}
\label{sec:xsect_diff:unfold:parton}

The differential cross-section at parton level is determined in a way similar 
to that for particle level using
\begin{equation}
\dif\hat{\sigma}_k = \frac{\sum_j M_{jk}^{-1}(N^{\text{data}}_j - \hat{B}_j)}{\epsilon_k L_{\text{int}}}\,,
\label{eq:xsect_parton}
\end{equation}
which can be obtained from
\Eqns{\eqref{eq:nuhat_particle}}{\eqref{eq:xsect_particle}} by replacing the
particle-level quantity $\mathcal{C}_k^{\text{ptcl!reco}}$ by $1/\epsilon_k$ and by omitting $\mathcal{C}_j^{\text{reco!ptcl}}$,
since the parton-level cross-section is fully inclusive and such a correction is not needed.

\subsection{Binning and convergence of unfolding}
\label{sec:xsect_diff:bin}

The migration matrices and efficiencies determined with the \POWHEGBOX + \PYTHIAV{6} MC sample are used
to extract the central values of the differential cross-sections.
A number of criteria are used to optimise the binning chosen for each differential cross-section.
These include the resolution of the measured quantity, the number of
events available in the bin and the size of the diagonal elements in
the migration matrix.
In general, the same binning is used for \tq and $\tbar q$ cross-sections,
except in a few cases when two bins are combined for $\tbar q$ cross-sections due to large statistical uncertainties.
The resolution of kinematic quantities of the pseudo top quark is better than
the resolution of the corresponding quantities at parton level.
Hence more bins are usually used for the particle-level cross-sections.

The number of iterations needed before the unfolding converges depends on both the shape of the distribution being measured and the resolution of the variable.
The cross-sections as a function of rapidity usually require fewer iterations before convergence,
while the cross-sections as a function of \pTthat need the largest number of
iterations, as the cross-section falls steeply and has a peak at low \pT. 
The criterion chosen for convergence is that the bias of the
unfolded cross-section, i.e.\ the difference between the unfolded result and the true distribution, should be less than \SI{1}{\%} in all bins.
The bias is determined from the difference between the unfolded result using the \MGMCatNLO + \HERWIG MC sample for unfolding and its generated distribution,
while using the nominal \POWHEGBOX + \PYTHIAV{6} MC sample for the migration matrix and efficiency.
Depending on the distribution being unfolded between three and nine iterations
are used.

\subsection{Uncertainties}
\label{sec:xsect_diff:uncert}

This section describes how the statistical and systematic uncertainties are propagated through the unfolding.
The uncertainty from each source is estimated individually and separately for signal and background,
taking correlations into account.
In addition, an uncertainty is assigned to the unfolding process.
All uncertainties are added in quadrature in each bin.

Systematic uncertainties enter the analysis in several places.
First, they affect the background yield and therefore the expected signal-to-background ratio.
The expected background is subtracted from data leading to a change in the input to the unfolding.
The migration matrix and differential efficiency measured using the signal MC sample are also affected by systematic uncertainties.

For uncertainties associated with the modelling of the $t$-channel process, the bias is taken as the uncertainty.
The bias is defined as the difference between the measured unfolded cross-section using a particular combination
of signal, migration matrix and efficiency, and the generator-level cross-section.

\subsubsection{Statistical uncertainties}
\label{sec:xsect_diff:uncert:stat}

The statistical uncertainty of the unfolded data result is determined by running
over an ensemble of pseudo-experiments, varying the content of each bin according to its expected statistical uncertainty.
Each pseudo-experiment is unfolded and the spread (RMS) of the result in each bin is taken as the measure of the statistical uncertainty.

For the statistical uncertainty due to the size of the signal MC sample,
the migration matrix and efficiency are fluctuated in pseudo-experiments with a
Gaussian function whose spread corresponds to the number of MC events in the
sample.
The unfolding is performed with each varied migration matrix and efficiency.
Again the RMS of the unfolded results in each bin is taken as the uncertainty.

\subsubsection{Systematic uncertainties}
\label{sec:xsect_diff:uncert:syst}

The list of systematic uncertainties considered and their definition is given in \Sect{\ref{sec:sys}}.
Different uncertainties need to be treated in different ways in the unfolding.
If an uncertainty is correlated between signal and background,
the effect is added linearly.
The methods used are described below.

\paragraph{Detector-related uncertainties affecting the signal:}
The effects of the detector-related uncertainties affecting the signal are evaluated by unfolding the varied MC signal distributions
using the nominal migration matrix and efficiency.
The difference from the unfolded distribution using the nominal signal MC sample as an input is taken as 
the uncertainty and propagated binwise to the measurement.
Thus, rate and shape uncertainties are taken into account simultaneously.

\paragraph{PDF uncertainties affecting the signal:}
The effect of the PDF uncertainty on the $t$-channel MC simulation is evaluated
by unfolding the MC signal distribution, using migration matrices and efficiencies created from different PDF MC signal sets: CT10 and the PDF4LHC15 combined PDF set.
The bias of each PDF is then calculated and the largest difference is taken as both the negative and positive PDF uncertainty bin by bin.
The difference between the bias of each eigenvector of the PDF4LHC15 and the bias of the central PDF4LHC15 is taken as an additional uncertainty.

\paragraph{Signal modelling uncertainties:}
To evaluate the effect of different MC generators for the $t$-channel production,
the MC signal distribution is unfolded using a migration matrix and efficiency created using either the MC signal of
\MGMCatNLO + \HERWIG or the MC signal of \POWHEGBOX + \HERWIG.
The full difference between the bias of \MGMCatNLO + \HERWIG and the bias of \POWHEGBOX + \HERWIG is assigned as
systematic uncertainty.
For the uncertainty associated with the parton-shower model, the full
difference between the bias of \POWHEGBOX + \PYTHIA6 and the bias of \POWHEGBOX + \HERWIG is assigned as the final uncertainty.
The bias of the up/down scale choice with \POWHEGBOX + \PYTHIAV{6} is used to estimate the uncertainty due to the scale variations.

\paragraph{Uncertainties in background rates:}
The normalisation uncertainties of all backgrounds are taken from the total
cross-section measurements.
These uncertainties are listed in Table~\ref{tab:backgroundunc}.
The uncertainty in the sum of backgrounds is estimated using pseudo-experiments,
and thus takes correlations into account.
The rate uncertainty of the background sum is applied by varying the background
sum up and down by the amount estimated in the total fiducial cross-section measurements.
The modified background-subtracted data is unfolded with the nominal migration matrix and efficiency.
The difference from the default unfolded distribution is taken as the rate uncertainty. 

\begin{table}[htbp]
  \centering
  \begin{tabular}{lS}
    \toprule
    Process        & \multicolumn{1}{c}{$\Delta N / N$ [\%]}    \\
    \midrule
    $\ttbar, Wt, t\bar{b}$     & 7.5 \\
    \wpjets        			   & 7.1 \\
    \wmjets   			       & 7.3 \\
    $Z, VV$\,+\,jets 		   & 20\\
    Multijets  & 16 \\
    \bottomrule
  \end{tabular}%
  \caption{Uncertainties in the normalisations of the different backgrounds 
    for all processes, as derived from the total cross-section measurement.
  \label{tab:backgroundunc}}
\end{table}

\paragraph{Uncertainties in shape of backgrounds:}
The uncertainty in the differential cross-sections due to 
the uncertainty in the shape of the background is determined
by evaluating the effect of the uncertainty in the NN output for each background
contribution.
Some of the systematic uncertainties have a very small effect on the analysis.
Hence, the shifts due to the variations reflecting the systematic uncertainties
are compared to the MC statistical error in each bin of each distribution, in
 order to avoid counting statistical fluctuations as a systematic uncertainty.
If the change in the bin content in at least two bins is larger than the MC statistical error in those bins,
the background shape uncertainty is taken into account.
The shifted backgrounds are subtracted from the data and the resulting distribution is unfolded using the nominal migration matrix and efficiency.
The difference from the measured unfolded distribution in each bin is assigned as the systematic uncertainty due to shape.
The main contribution to the shape uncertainty comes from the \ttbar modelling.

\paragraph{Unfolding uncertainty:}
In order to estimate the uncertainty due to the unfolding method, the \POWHEGBOX + \PYTHIAV{6} sample is divided into two.
One half is used to determine the migration matrix,
while the other half is used to unfold the cross-section.
The full difference between the unfolded MC $t$-channel distribution and the MC $t$-channel generator-level distribution
is taken as the uncertainty in the unfolding process.

As a cross-check, the results are compared with using a bin-by-bin correction factor and the single value decomposition (SVD) method~\cite{Hocker:1995kb},
which is an extension of a simple matrix inversion.
Consistent results are found and no extra uncertainty is assigned.

\subsection{Particle-level cross-sections}
\label{sec:xsect_diff:particle}

The absolute unfolded particle-level cross-sections for top quarks and top antiquarks as a function of
\pTthat are shown in \Fig{\ref{fig:xsectA:pTt:particle}},
while the cross-sections as a function of \absythat are shown in \Fig{\ref{fig:xsectA:yt:particle}}.
The numerical values of both the absolute and normalised unfolded cross-sections are given in \Tabrange{\ref{tab:resultTopPt_ptcl}}{\ref{tab:resultantiTopY_ptcl}}.
The measurements are compared to MC predictions using the \POWHEGBOX and \MGMCatNLO generators.
Good agreement between the measured differential cross-sections and the predictions is seen.
Separate predictions using \PYTHIA or \HERWIG interfaced to \POWHEGBOX are shown.
The ratio plots show that the hadronisation model has a very small effect on the predictions.

\newcommand*{\DiffXsectFigCaption}{The unfolded distributions are compared to various MC predictions.
    The vertical error bars on the data points denote the total uncertainty.
    The inner (yellow) band in the bottom part of each figure
    represents the statistical uncertainty of the measurement,
    and the outer (green) band the total uncertainty.
}
\begin{figure}[htbp]
  \centering
  \subfloat[][]{\includegraphics[width=0.46\textwidth]{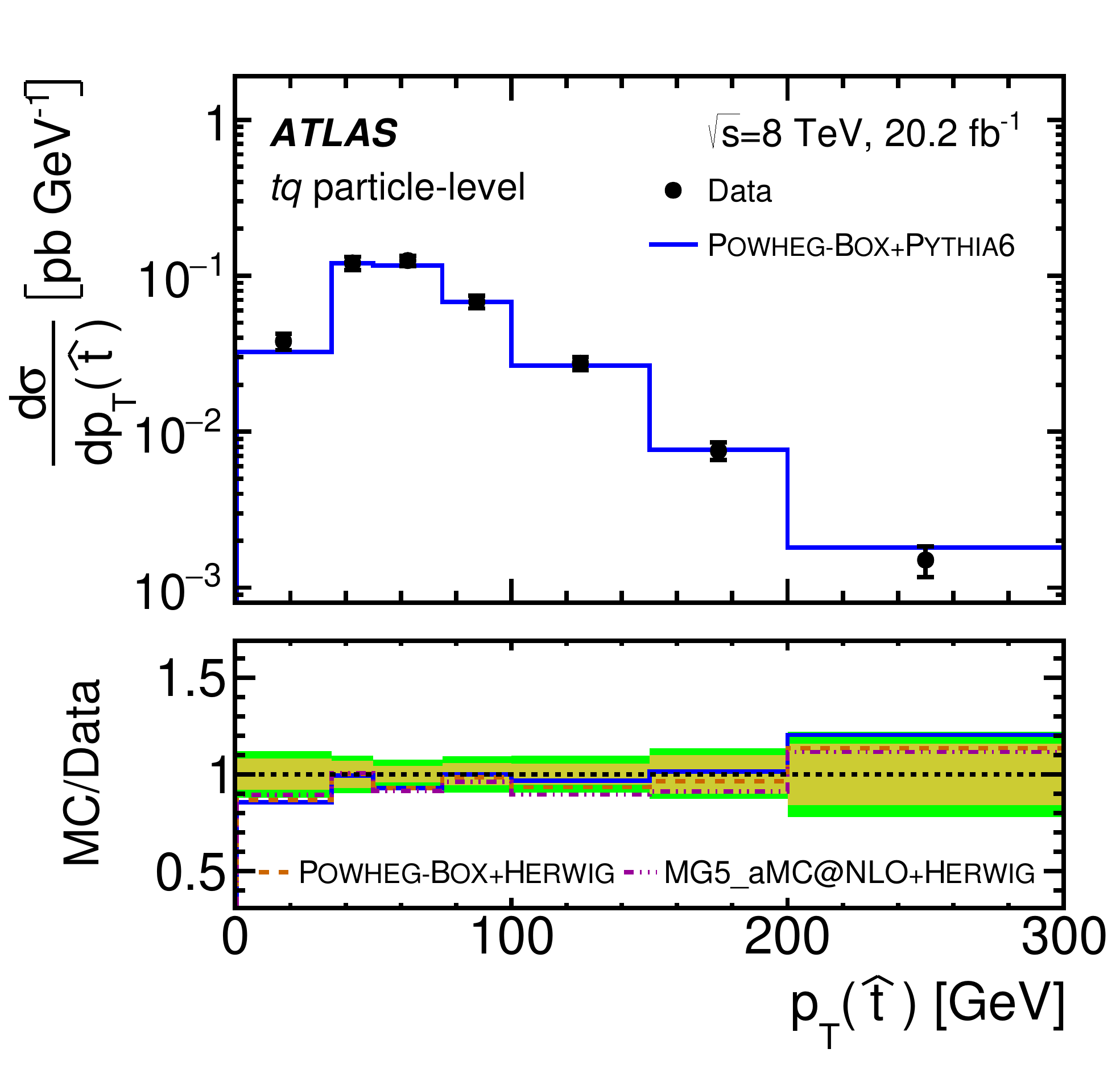}}\quad
  \subfloat[][]{\includegraphics[width=0.46\textwidth]{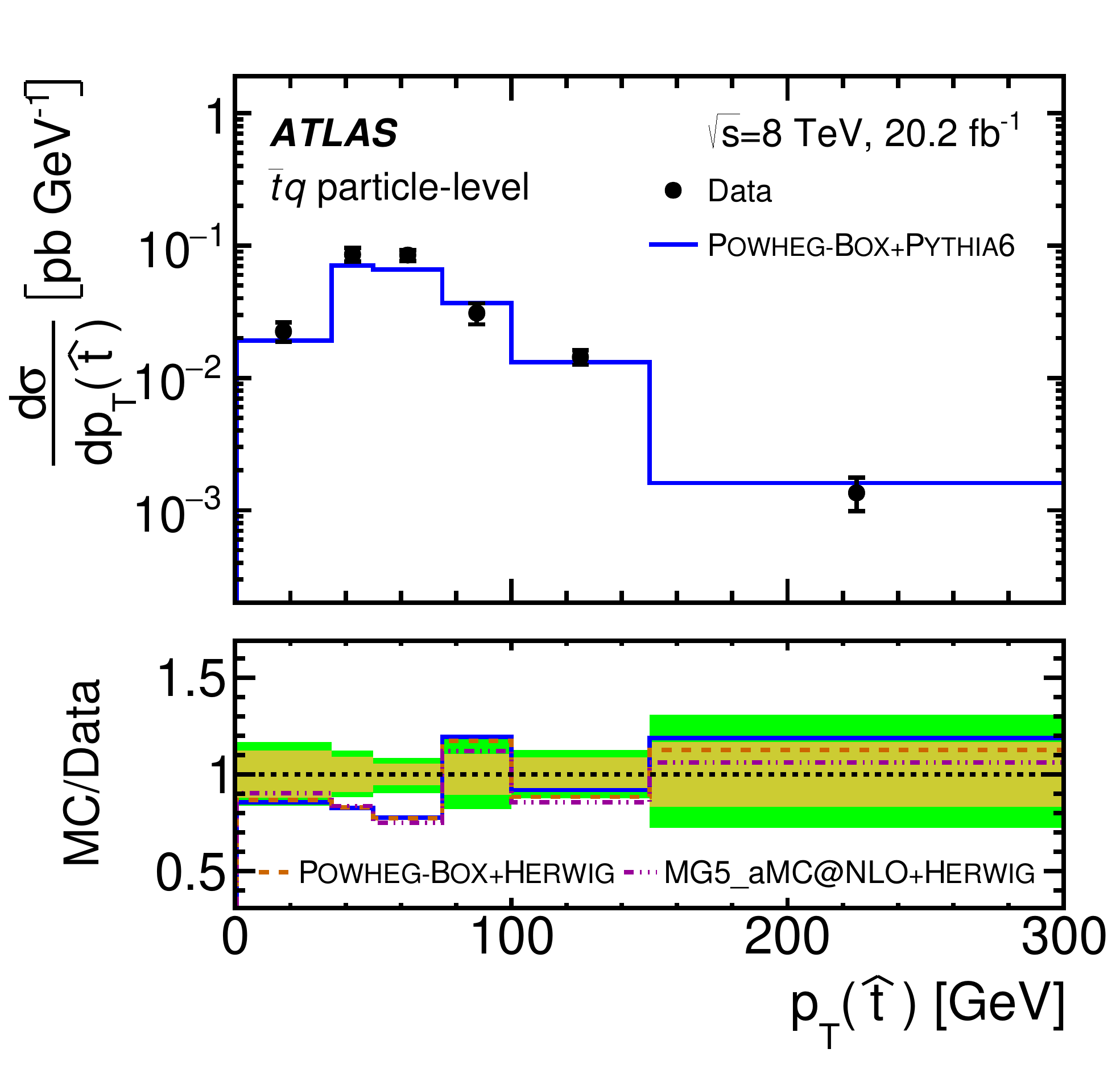}}
  \caption{Absolute  
    unfolded differential cross-sections as a function of \pTthat for
    (a) top quarks and (b) top antiquarks.
    \DiffXsectFigCaption
  }
  \label{fig:xsectA:pTt:particle}
\end{figure}

\begin{figure}[htbp]
  \centering
  \subfloat[][]{\includegraphics[width=0.46\textwidth]{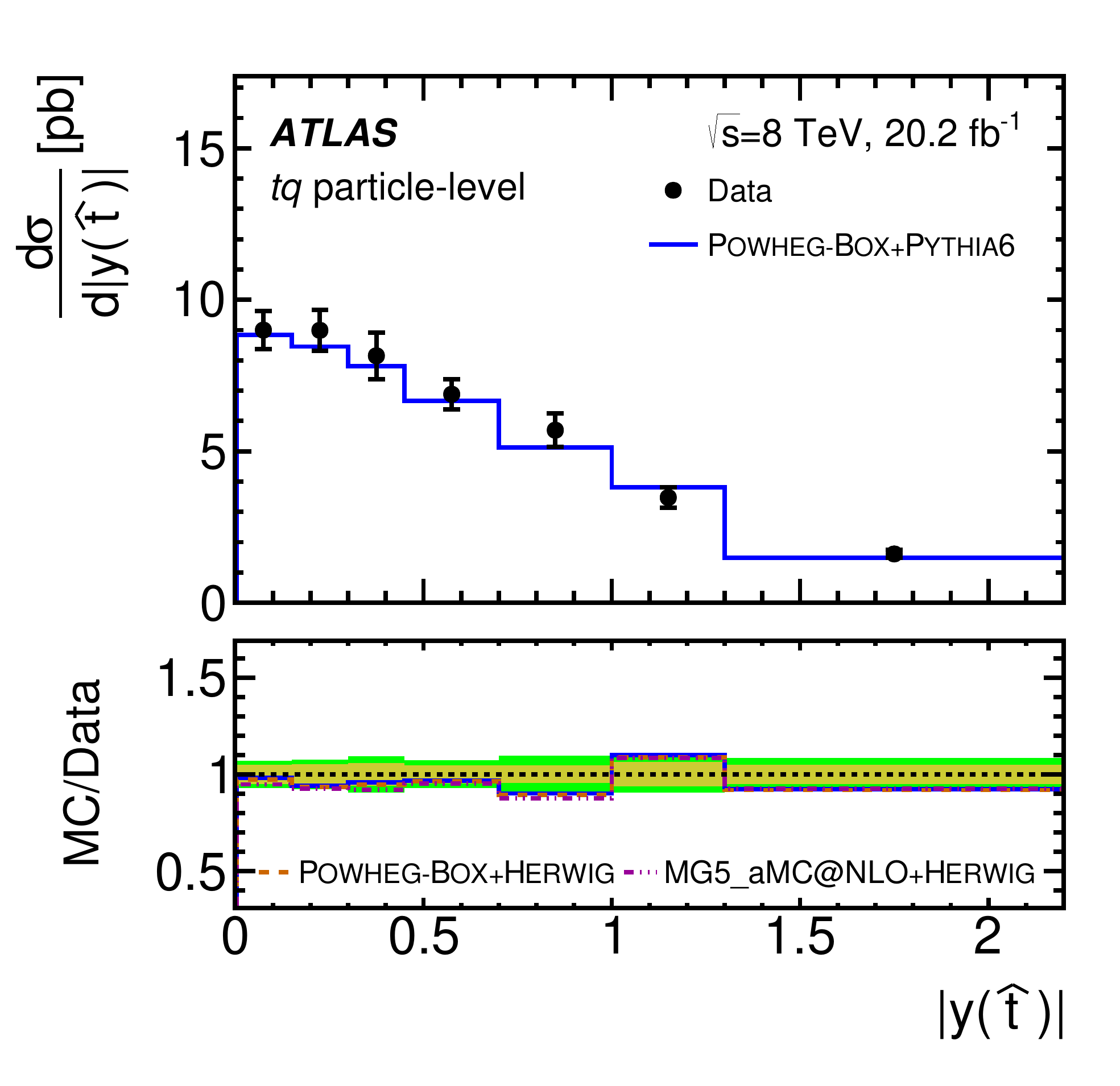}}\quad
  \subfloat[][]{\includegraphics[width=0.46\textwidth]{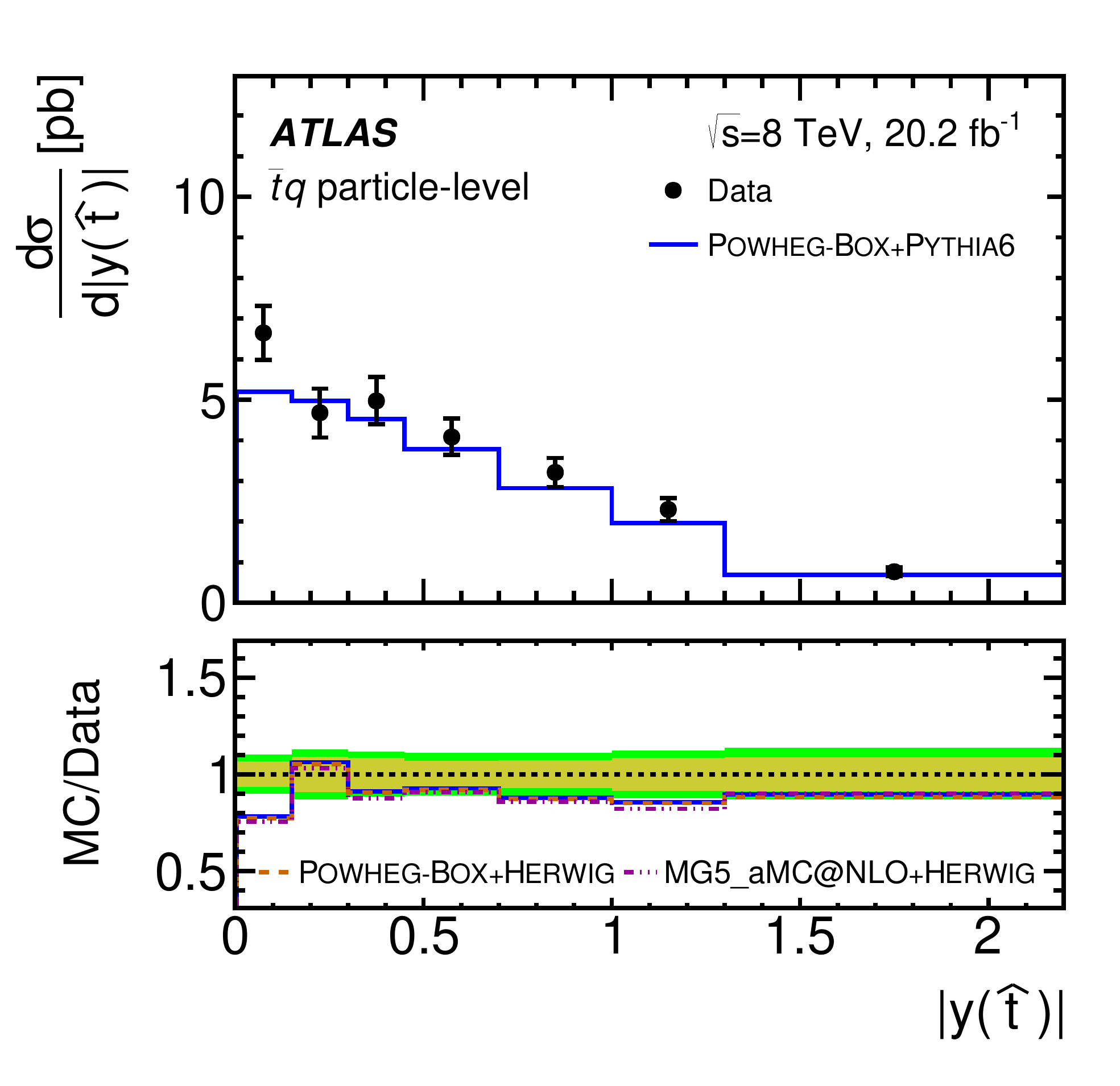}}
  \caption{Absolute 
    unfolded differential cross-sections as a function of \absythat for
    (a) top quarks and (b) top antiquarks.
    \DiffXsectFigCaption
  }
  \label{fig:xsectA:yt:particle}
\end{figure}

\newcommand*{\phoo}{\phantom{00}}
\newcommand*{\phd}{\phantom{.}}
\begin{table}[htbp]
	\centering
	\sisetup{round-mode=figures, retain-explicit-plus}
	\begin{tabular}{S[table-format=3.0]@{\,--\,}S[table-format=3.0]|
			S[round-precision=3, table-format=3.2]
			@{$\quad\pm$}S[round-precision=2, table-format=1.2]
			S[round-precision=2, table-format=-1.2]@{\,/\,}S[round-precision=2, table-format=-1.2]|
			S[round-precision=3, table-format=2.3]
			@{$\quad\pm$}S[round-precision=2, table-format=1.3]
			S[round-precision=2, table-format=-1.3]@{\,/\,}S[round-precision=2, table-format=-1.3]}
		\toprule
		\multicolumn{2}{c|}{$\pTthat$}
		& \multicolumn{4}{c|}{$\dif\sigma(tq)/\dif\pTthat$}
		& \multicolumn{4}{c}{$(1/\sigma) \dif\sigma(tq)/\dif\pTthat$}\\
		\multicolumn{2}{c|}{[\si{\GeV}]}
		& \multicolumn{4}{c|}{[\si{\fb\per\GeV}]}
		& \multicolumn{4}{c}{[\SI{E-3}{\per\GeV}]}\\
		\multicolumn{2}{c|}{\mbox{}}
		& \multicolumn{2}{r}{stat.\phantom{0}} & \multicolumn{2}{c|}{syst.}
		& \multicolumn{2}{r}{stat.\phantom{0}} & \multicolumn{2}{c}{syst.}\\
		\midrule
		0 & 35   &    37.9821   &   3.08236  & +3.28498 &-3.35333  &
		3.8457   &   0.293564 & +0.22017 &-0.218232  \\
		35 & 50   &  {\num[round-precision=4]{120.9}\pho}   &   8.44989  & +7.9536 &-8.19924  &
		{\num[round-precision=4]{12.2412}\pho}  &   0.821862  & +0.608325 &-0.591695  \\
		50  & 75  &  {\num[round-precision=4]{125.157}\pho}   &   5.28227  & +7.74687 &-7.90647  &
		{\num[round-precision=4]{12.6722}\pho}   &   0.491351   & +0.54279 &-0.537698  \\
		75 & 100  &   68.0584   &  3.89056 & +5.11061 &-5.01034  &
		6.89094   &  0.379785 & +0.359548 &-0.344677  \\
		100 & 150  &    27.5043   &  1.50612 & +2.13713 &-2.13797 &
		2.78482   &  0.146647 & +0.181069 &-0.183099  \\
		150& 200  &   7.55251   &  0.757755 & +0.673165 &-0.5634  &
		0.764694   &  0.0762591 & +0.0556805 &-0.0455267  \\
		200 & 300  &    1.50399   &  0.239733 & +0.228209 &-0.230379  &
		0.15228   &  0.0241607 & +0.0215418  &-0.0216462  \\
		\bottomrule
	\end{tabular}
  \caption{Absolute and normalised unfolded differential \tq production cross-section as a function of $\pTthat$ at particle level.
  }
  \label{tab:resultTopPt_ptcl} 
\end{table}

\begin{table}[htbp]
	\centering
	\sisetup{round-mode=figures, retain-explicit-plus}
	\begin{tabular}{S[table-format=3.0]@{\,--\,}S[table-format=3.0]|
			S[round-precision=3, table-format=2.2]
			@{$\quad\pm$}S[round-precision=2, table-format=1.2]
			S[round-precision=2, table-format=-1.2]@{\,/\,}S[round-precision=2, table-format=-1.2]|
			S[round-precision=3, table-format=2.3]
			@{$\quad\pm$}S[round-precision=2, table-format=1.3]
			S[round-precision=2, table-format=-1.3]@{\,/\,}S[round-precision=2, table-format=-1.3]}
		\toprule
		\multicolumn{2}{c|}{$\pTthat$}
		& \multicolumn{4}{c|}{$\dif\sigma(\bar{t}q)/\dif\pTthat$}
		& \multicolumn{4}{c}{$(1/\sigma) \dif\sigma(\bar{t}q)/\dif\pTthat$}\\
		\multicolumn{2}{c|}{[\si{\GeV}]}
		& \multicolumn{4}{c|}{[\si{\fb\per\GeV}]}
		& \multicolumn{4}{c}{[\SI{E-3}{\per\GeV}]}\\
		\multicolumn{2}{c|}{\mbox{}}
		& \multicolumn{2}{r}{stat.\phantom{0}} & \multicolumn{2}{c|}{syst.}
		& \multicolumn{2}{r}{stat.\phantom{0}} & \multicolumn{2}{c}{syst.}\\
		\midrule
		0 & 35   &    22.4548   &  2.74424 & +2.54973 &-2.41341 &
		3.81696   &  0.438451  & +0.265804 &-0.238147  \\
		35 & 50   &  85.6412   &  7.78107 & +7.1766 &-6.3348 &
		14.5577   &  1.26118 & {\num[round-precision=1]{+0.978357}\phoo} &{\num[round-precision=1]{-0.79127}\phoo}  \\
		50  & 75  &  84.7444   &  4.72938  & +5.40001 &-6.85511 &
		{\num[round-precision=4]{14.4052}\pho}   &  0.735222  & +0.513117 &-0.811282 \\
		75 & 100  &   30.9136  & 3.29293 & +4.6497 &-4.44412 &
		5.25482  & 0.539057  & +0.65496 &-0.616816 \\
		100 & 150  &    14.3567  & 1.27697 & +1.246 &-1.24813  &
		2.44041  & 0.210405  & +0.126553 &-0.13436 \\
		150& 300  &   1.35386   & 0.226079 & +0.348565 &-0.295932 &
		0.230135   & 0.0379425  & +0.0553175 &-0.0455645  \\
		\bottomrule
	\end{tabular}
  \caption{Absolute and normalised unfolded differential \tbarq production cross-section as a function of $\pTthat$ at particle level.
  }
  \label{tab:resultantiTopPt_ptcl} 
\end{table}

\begin{table}[htbp]
	\centering
	\sisetup{round-mode=places, retain-explicit-plus}
	\begin{tabular}{S[table-format=1.2]@{\,--\,}S[table-format=1.2]|
			S[round-precision=2, table-format=1.2]
			@{$\quad\pm$}S[round-precision=2, table-format=1.2]
			S[round-precision=2, table-format=-1.2]@{\,/\,}S[round-precision=2, table-format=-1.2]|
			S[round-precision=0, table-format=3.0]
			@{$\quad\pm$}S[round-precision=0, table-format=2.0]
			S[round-precision=0, table-format=-2.0]@{\,/\,}S[round-precision=0, table-format=-2.0]}
		\toprule
		\multicolumn{2}{c|}{$\absythat$}
		& \multicolumn{4}{c|}{$\dif\sigma(tq)/\dif\absythat$}
		& \multicolumn{4}{c}{$(1/\sigma) \dif\sigma(tq)/\dif\absythat$}\\
		\multicolumn{2}{c|}{}
		& \multicolumn{4}{c|}{[\si{\pb}]}
		& \multicolumn{4}{c}{[\SI{E-3}]}\\
		\multicolumn{2}{c|}{\mbox{}}
		& \multicolumn{2}{r}{stat.} & \multicolumn{2}{c|}{syst.}
		& \multicolumn{2}{r}{stat.} & \multicolumn{2}{c}{syst.}\\
		\midrule
       0.0 & 0.15  &  8.99845   & 0.445848 & +0.432062 &-0.433593  &
       914.213   & 42.8691  & +18.5626 &-18.0808 \\
       0.15 & 0.3  &  8.99116    & 0.47247 & +0.46788 &-0.486215  &
       913.473    & 45.8477  & +41.0742 &-43.2865 \\
       0.3 & 0.45 &   8.15121  & 0.481436 & +0.5919 &-0.598135  &
       828.136  & 46.4565  & +44.2233 &-46.4433  \\
       0.45 & 0.7   &   6.8815 & 0.320711 & +0.37837 &-0.374477  &
       699.138 & 30.1856 & +19.2242 &-17.3069 \\
       0.7 & 1.0  &  5.69545   & 0.261559 & +0.486858 &-0.484348 &
       578.639   & 24.4788 & +35.67 &-35.8154 \\
       1.0 & 1.3  &  3.47285    & 0.221394 & +0.262919 &-0.247179 &
       352.83    & 21.2572  & +12.7821 &-10.8588 \\
       1.3 & 2.2  &   1.61205  & 0.0812105  & +0.1066 &-0.107987 &
       163.779  & 7.59093 & +4.09496 &-4.21895  \\
		\bottomrule
	\end{tabular}
  \caption{Absolute and normalised unfolded differential \tq production cross-section as a function of $\absythat$ at particle level.
  }
  \label{tab:resultTopY_ptcl} 
\end{table}

\begin{table}[htbp]
	\centering
	\sisetup{round-mode=places, retain-explicit-plus}
	\begin{tabular}{S[table-format=1.2]@{\,--\,}S[table-format=1.2]|
			S[round-precision=2, table-format=1.2]
			@{$\quad\pm$}S[round-precision=2, table-format=1.2]
			S[round-precision=2, table-format=-1.2]@{\,/\,}S[round-precision=2, table-format=-1.2]|
			S[round-precision=0, table-format=4.0]
			@{$\quad\pm$}S[round-precision=0, table-format=2.0]
			S[round-precision=0, round-mode=places, table-format=-2.0]@{\,/\,}S[round-precision=0, round-mode=places, table-format=-2.0]}
		\toprule
		\multicolumn{2}{c|}{$\absythat$}
		& \multicolumn{4}{c|}{$\dif\sigma(\bar{t}q)/\dif\absythat$}
		& \multicolumn{4}{c}{$(1/\sigma) \dif\sigma(\bar{t}q)/\dif\absythat$}\\
		\multicolumn{2}{c|}{}
		& \multicolumn{4}{c|}{[\si{\pb}]}
		& \multicolumn{4}{c}{[\SI{E-3}]}\\
		\multicolumn{2}{c|}{\mbox{}}
		& \multicolumn{2}{r}{stat.} & \multicolumn{2}{c|}{syst.}
		& \multicolumn{2}{r}{stat.} & \multicolumn{2}{c}{syst.}\\
		\midrule
       0.0 & 0.15  &  6.64648   & 0.440524 & +0.501088 &-0.488067 &
       1144.68   & 70.2922 & +56.9735 &-54.8918 \\
       0.15 & 0.3  &  4.68012    & 0.428935 & +0.410988 &-0.430251 &
       806.027    & 70.6189 & +51.4831 &-57.4773 \\
       0.3 & 0.45 &   4.97217  & 0.422093  & +0.403697 &-0.390569 &
       856.326  & 69.1032 & +43.625 &-40.0602 \\
       0.45 & 0.7   &  4.08342  & 0.291041  & +0.342796 &-0.332635 &
       703.261  & 46.4174  & +37.7057 &-39.0871 \\
       0.7 & 1.0  &  3.21169   & 0.231241  & +0.26627 &-0.273135 &
       553.13   & 36.8309  & +28.367 &-30.067 \\
       1.0 & 1.3  &  2.29735    & 0.196409  & +0.204665 &-0.206689 &
       395.658   & 31.9915  & +17.4474 &-16.6249 \\
       1.3 & 2.2  &  0.764468  & 0.0668885 & +0.0798768 &-0.0735545 &
       131.66  & 10.7269  & +7.97459 &-6.63075 \\
		\bottomrule
	\end{tabular}
  \caption{Absolute and normalised unfolded differential \tbarq production cross-section as a function of $\absythat$ at particle level.
  }
  \label{tab:resultantiTopY_ptcl} 
\end{table}

\begin{figure}[htbp]
  \centering
  \subfloat[][]{\includegraphics[width=0.46\textwidth]{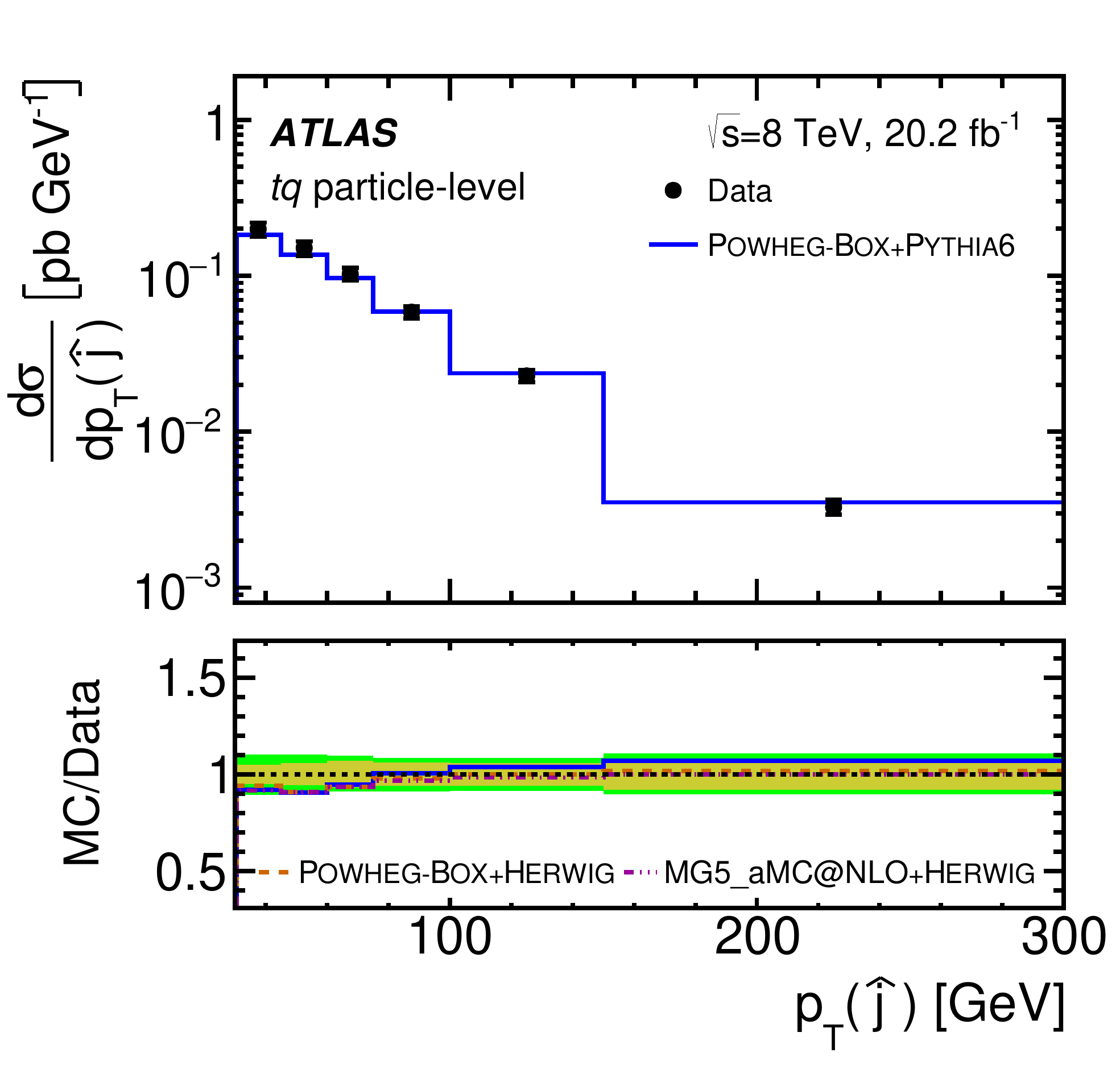}}\quad
  \subfloat[][]{\includegraphics[width=0.46\textwidth]{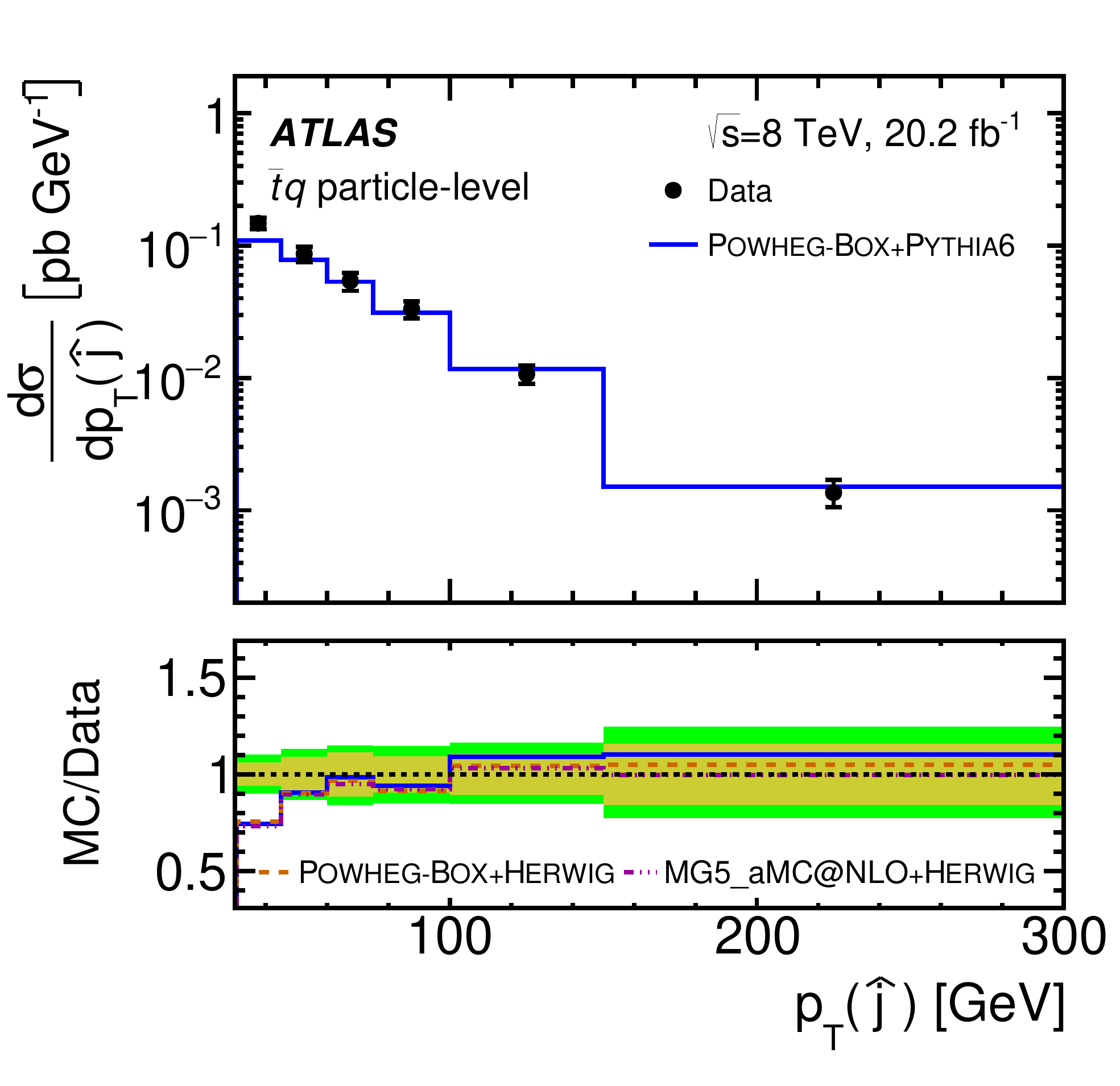}}
  \caption{Absolute  
  unfolded differential cross-sections as a function of \pTjhat for
  (a) top quarks (b) top antiquarks.
  \DiffXsectFigCaption
  }
  \label{fig:xsectA:pTjt:particle}
\end{figure}

\begin{figure}[htbp]
  \centering
  \subfloat[][]{\includegraphics[width=0.46\textwidth]{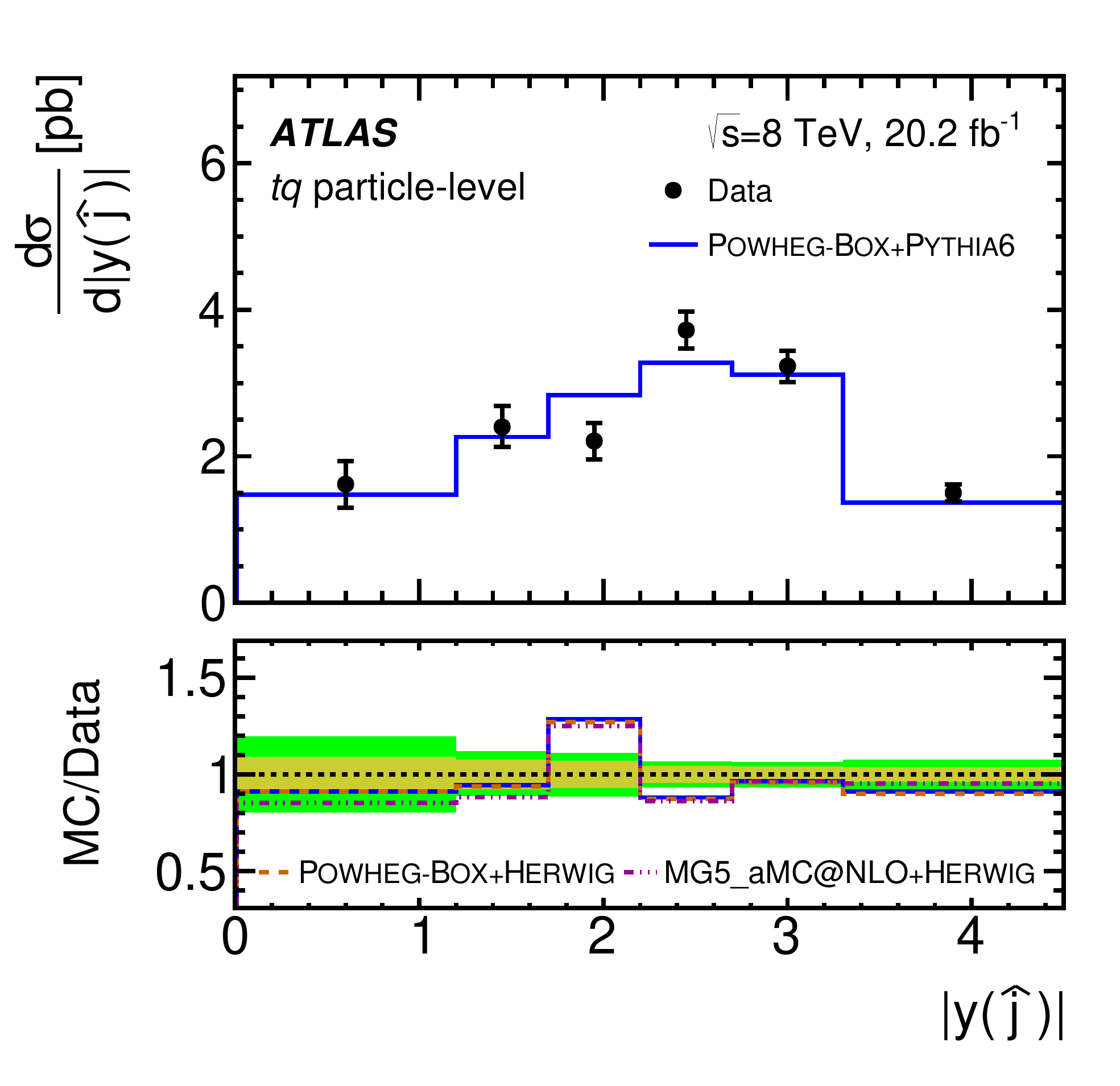}}\quad
  \subfloat[][]{\includegraphics[width=0.46\textwidth]{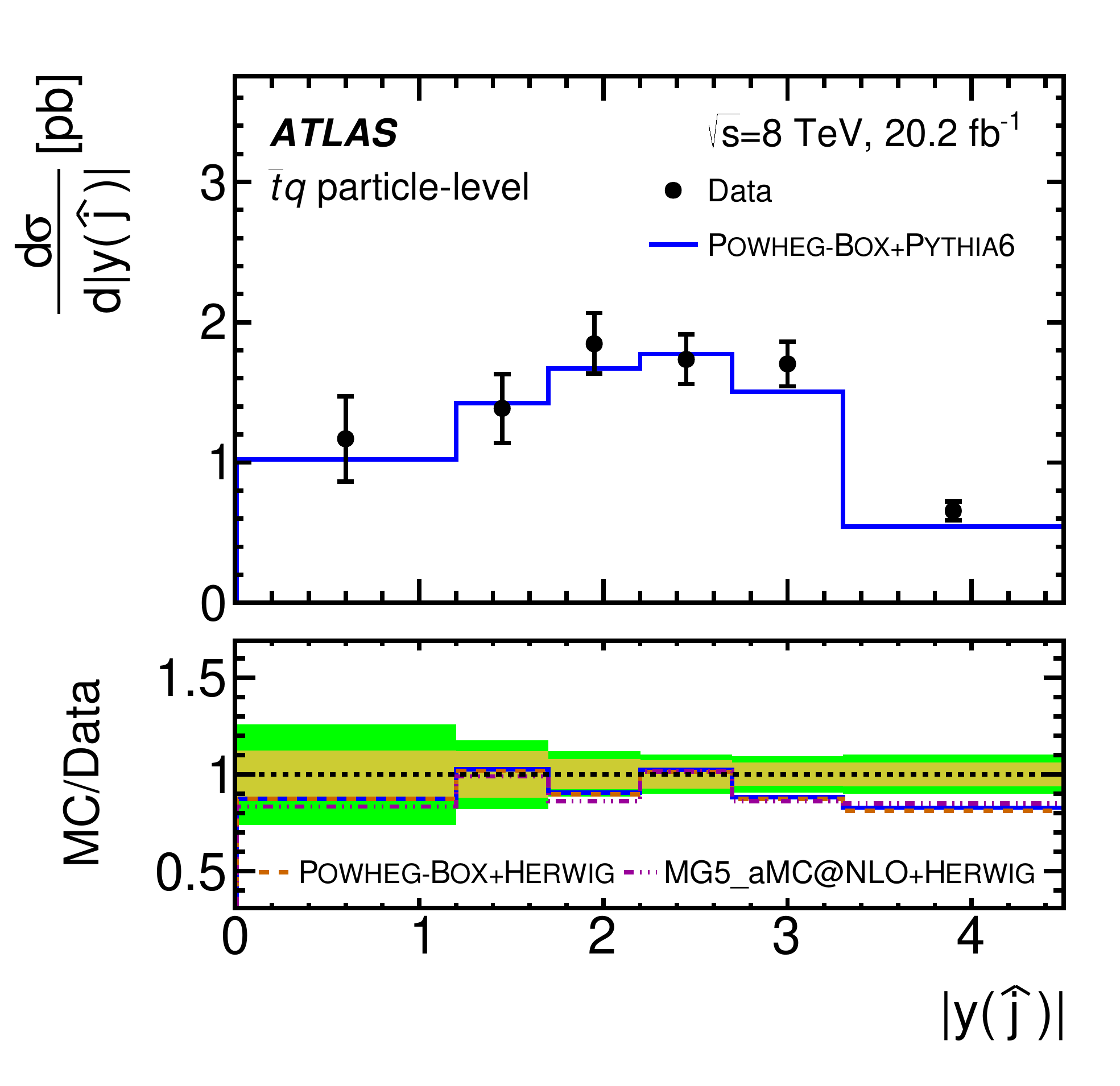}}
  \caption{Absolute  
  unfolded differential cross-sections as a function of \absyjhat for
  (a) top quarks and (b) top antiquarks.
  \DiffXsectFigCaption
  }
  \label{fig:xsectA:yjt:particle}
\end{figure}

The absolute cross-sections for the untagged jet as a function of the same variables 
are shown in \Figs{\ref{fig:xsectA:pTjt:particle}}{\ref{fig:xsectA:yjt:particle}}
and both the absolute and normalised cross-sections are tabulated in \Tabrange{\ref{tab:resultlightPtplus_ptcl}}{\ref{tab:resultlightYminus_ptcl}}.
The measurement as a function of \absyjhat uses the neural network without $|\eta(j)|$,
while all other measurements use the default network.
The measured cross-sections are again well described by the predictions,
although there is a tendency for the prediction to be somewhat harder than the data as a function of \pTjhat.

\begin{table}[htbp]
	\centering
	\sisetup{round-mode=figures, retain-explicit-plus}
	\begin{tabular}{S[table-format=3.0, round-mode=off]@{\,--\,}S[table-format=3.0, round-mode=off]|
			S[round-precision=3, table-format=3.2]
			@{$\quad\pm$}S[round-precision=2, table-format=2.2]
			S[round-precision=2, table-format=-2.2]@{\,/\,}S[round-precision=2, table-format=-2.2]|
			S[round-precision=3, table-format=2.3]
			@{$\quad\pm$}S[round-precision=2, table-format=1.3]
			S[round-precision=2, table-format=-1.3]@{\,/\,}S[round-precision=2, table-format=-1.3]}
		\toprule
		\multicolumn{2}{c|}{$\pTjhat$}
		& \multicolumn{4}{c|}{$\dif\sigma(tq)/\dif\pTjhat$}
		& \multicolumn{4}{c}{$(1/\sigma) \dif\sigma(tq)/\dif\pTjhat$}\\
		\multicolumn{2}{c|}{[\si{\GeV}]}
		& \multicolumn{4}{c|}{[\si{\fb\per\GeV}]}
		& \multicolumn{4}{c}{[\SI{E-3}{\per\GeV}]}\\
		\multicolumn{2}{c|}{\mbox{}}
		& \multicolumn{2}{r}{stat.\phantom{0}} & \multicolumn{2}{c|}{syst.}
		& \multicolumn{2}{r}{stat.\phantom{0}} & \multicolumn{2}{c}{syst.}\\
		\midrule
		30 & 45   &    198.953   &  {\num[round-precision=1]{9.82721}\phoo\phd} & +17.7972 & -18.6604 &
		20.1471   &  {\num[round-precision=1]{0.838262}\phoo}  & +1.15364 & -1.23738  \\
		45 & 60   &  150.677   &  {\num[round-precision=1]{8.58713}\pho\phd}  & +12.8457 & -13.6733 &
		{\num[round-precision=4]{15.2585}\pho}  &  0.847513 & +0.912555 & -0.943079  \\
		60  & 75  &  {\num[round-precision=4]{102.327}}   &  7.0052   & +6.7987 & -5.80996 &
		{\num[round-precision=4]{10.3622}\pho}   &  0.689861   & +0.459282 & -0.346571  \\
		75 & 100  &   58.467   & 3.53369  & +3.48226 & -3.86931 &
		5.9207   & 0.348034  & +0.237818 & -0.266168  \\
		100 & 150  &    22.8096   & 1.32256  & +1.42166 & -1.43312 &
		2.30983   & 0.130327 & +0.108393 & -0.108532  \\
		150 & 300  &    3.28992   & 0.259695  & +0.244589 & -0.21728 &
		0.333157   & 0.0259176 & +0.0193141 & -0.0154939  \\
		\bottomrule
	\end{tabular}
  \caption{Absolute and normalised unfolded differential \tq production cross-section as a function of $\pTjhat$ at particle level.
  }
  \label{tab:resultlightPtplus_ptcl} 
\end{table}

\begin{table}[htbp]
	\centering
	\sisetup{round-mode=figures, retain-explicit-plus}
	\begin{tabular}{S[table-format=3.0, round-mode=off]@{\,--\,}S[table-format=3.0, round-mode=off]|
			S[round-precision=3, table-format=3.2]
			@{$\quad\pm$}S[round-precision=2, table-format=1.2]
			S[round-precision=2, table-format=-2.2]@{\,/\,}S[round-precision=2, table-format=-2.2]|
			S[round-precision=3, table-format=2.3]
			@{$\quad\pm$}S[round-precision=2, table-format=1.3]
			S[round-precision=2, table-format=-1.3]@{\,/\,}S[round-precision=2, table-format=-1.3]}
		\toprule
		\multicolumn{2}{c|}{$\pTjhat$}
		& \multicolumn{4}{c|}{$\dif\sigma(\bar{t}q)/\dif\pTjhat$}
		& \multicolumn{4}{c}{$(1/\sigma) \dif\sigma(\bar{t}q)/\dif\pTjhat$}\\
		\multicolumn{2}{c|}{[\si{\GeV}]}
		& \multicolumn{4}{c|}{[\si{\fb\per\GeV}]}
		& \multicolumn{4}{c}{[\SI{E-3}{\per\GeV}]}\\
		\multicolumn{2}{c|}{\mbox{}}
		& \multicolumn{2}{r}{stat.\phantom{0}} & \multicolumn{2}{c|}{syst.}
		& \multicolumn{2}{r}{stat.\phantom{0}} & \multicolumn{2}{c}{syst.}\\
		\midrule
		30 & 45   &    147.269   &  {\num[round-precision=1]{8.89656}\phoo\phd}  & +11.9188 & -11.5565 &
		25.033   &  1.24226 & +1.09542 & -1.01387 \\
		45 & 60   &  86.4151   &  7.82594 & +8.30239 & -8.48865 &
		14.6889   &  1.28925  & {\num[round-precision=1]{+0.964446}\phoo} & {\num[round-precision=1]{-0.987447}\phoo}  \\
		60  & 75  &  54.1745   & 6.22486   & +5.06756 & -5.98638 &
		9.20864   &  {\num[round-precision=3]{1.03421}\pho}  & +0.682496 & -0.880624  \\
		75 & 100  &   33.0354   & 3.096  & +3.67527 & -3.86201 &
		5.61539   & 0.511432 & +0.355965 & -0.408109  \\
		100 & 150  &    10.703   & 1.13019  & +1.33326 & -1.20099 &
		1.81932   & 0.185858 & +0.144845 & -0.108771  \\
		150 & 300  &    1.3606   & 0.216429  & +0.255334 & -0.218634 &
		0.231276   & 0.0362549 & +0.044357 & -0.0380769  \\
		\bottomrule
	\end{tabular}
  \caption{Absolute and normalised unfolded differential \tbarq production cross-section as a function of $\pTjhat$ at particle level.
  }
  \label{tab:resultlightPtminus_ptcl} 
\end{table}

\begin{table}[htbp]
	\centering
	\sisetup{round-mode=figures, retain-explicit-plus}
	\begin{tabular}{S[table-format=1.1, round-mode=off]@{\,--\,}S[table-format=1.1, round-mode=off]|
			S[round-precision=3, table-format=1.3]
			@{$\quad\pm$}S[round-precision=2, table-format=1.3]
			S[round-precision=2, table-format=-1.3]@{\,/\,}S[round-precision=2, table-format=-1.3]|
			S[round-precision=3, table-format=3.1]
			@{$\quad\pm$}S[round-precision=2, table-format=2.1]
			S[round-precision=2, table-format=-2.1]@{\,/\,}S[round-precision=2, table-format=-2.1]}
		\toprule
		\multicolumn{2}{c|}{$\absyjhat$}
		& \multicolumn{4}{c|}{$\dif\sigma(tq)/\dif\absyjhat$}
		& \multicolumn{4}{c}{$(1/\sigma) \dif\sigma(tq)/\dif\absyjhat$}\\
		\multicolumn{2}{c|}{}
		& \multicolumn{4}{c|}{[\si{\pb}]}
		& \multicolumn{4}{c}{[\SI{E-3}]}\\
		\multicolumn{2}{c|}{\mbox{}}
		& \multicolumn{2}{r}{stat.\phantom{0}} & \multicolumn{2}{c|}{syst.}
		& \multicolumn{2}{r}{stat.\phantom{0}} & \multicolumn{2}{c}{syst.}\\
		\midrule
       0.0 & 1.2  &  1.61852   & 0.144376 & +0.28301 & -0.28335 &
       164.376   &  12.1599 & +21.7738 & -21.666 \\
       1.2 & 1.7  &  2.39945   & 0.182407  & +0.221638 & -0.198547 &
       243.687   & 17.2387 & +14.6894 & -10.8829 \\
       1.7 & 2.2  &   2.20784  & 0.154842 & +0.190588 & -0.197112 &
       224.228  & 14.9543 & {\num[round-precision=1]{+9.7645}\pho} & -11.1333  \\
       2.2 & 2.7   &   3.72113  & 0.164255  & +0.187314 & -0.187103 &
       377.917  & 16.3132 & +15.5878 & -15.5938 \\
       2.7 & 3.3   &   3.23115  & 0.130699  & +0.157415 & -0.171822 &
       328.155  & 13.4039 & +15.0746 & -15.3603  \\
       3.3 & 4.5   &   1.50108  & {\num[round-precision=1]{0.0564995}\pho}  & {\num[round-precision=1]{+0.0964218}\pho} & -0.101423 &
       {\num[round-precision=4]{152.45}}  & 5.99187 & +9.15866 & -9.33458 \\
		\bottomrule
	\end{tabular}
  \caption{Absolute and normalised unfolded differential \tq production cross-section as a function of $\absyjhat$ at particle level.
  }
  \label{tab:resultlightYplus_ptcl} 
\end{table}

\begin{table}[htbp]
	\centering
	\sisetup{round-mode=figures, retain-explicit-plus}
	\begin{tabular}{S[table-format=1.1, round-mode=off]@{\,--\,}S[table-format=1.1, round-mode=off]|
			S[round-precision=3, table-format=1.3]
			@{$\quad\pm$}S[round-precision=2, table-format=1.3]
			S[round-precision=2, table-format=-1.3]@{\,/\,}S[round-precision=2, table-format=-1.3]|
			S[round-precision=3, table-format=3.0]
			@{$\quad\pm$}S[round-precision=0, round-mode=places, table-format=2.0]
			S[round-precision=0, round-mode=places, table-format=-2.0]@{\,/\,}S[round-precision=0, round-mode=places, table-format=-2.0]}
		\toprule
		\multicolumn{2}{c|}{$\absyjhat$}
		& \multicolumn{4}{c|}{$\dif\sigma(\bar{t}q)/\dif\absyjhat$}
		& \multicolumn{4}{c}{$(1/\sigma) \dif\sigma(\bar{t}q)/\dif\absyjhat$}\\
		\multicolumn{2}{c|}{}
		& \multicolumn{4}{c|}{[\si{\pb}]}
		& \multicolumn{4}{c}{[\SI{E-3}]}\\
		\multicolumn{2}{c|}{\mbox{}}
		& \multicolumn{2}{r}{stat.\phantom{0}} & \multicolumn{2}{c|}{syst.}
		& \multicolumn{2}{r}{stat.} & \multicolumn{2}{c}{syst.}\\
		\midrule
       0.0 & 1.2  &  1.16904  & 0.14309  & +0.265635 & -0.26927 &
       205.304  & 19.8056 & +30.5707 &-31.4172 \\
       1.2 & 1.7  &  1.38581    & 0.166154  & +0.177278 & -0.184406 &
       243.372    & 27.3808  & +13.8654 &-16.2968 \\
       1.7 & 2.2  &   1.84628  & 0.144834  & +0.164061 & -0.155034 &
       324.24  & 24.7764 & +19.7248 &-16.7799  \\
       2.2 & 2.7   &   1.73466  & 0.127123  & +0.122707 & -0.120751 &
       304.637  & 22.1042 & +19.9441 &-19.3901  \\
       2.7 & 3.3   &   1.70304  & 0.101722  & +0.121385 & -0.122585 &
       299.084  & 18.6226 & +26.3968 &-26.3275 \\
       3.3 & 4.5   &   0.655119  & 0.0400426 & +0.0526757 & -0.0511378 &
       115.05  & 7.53798 & +11.2206 &-11.0947 \\ 
		\bottomrule
	\end{tabular}
  \caption{Absolute and normalised unfolded differential \tbarq production cross-section as a function of $\absyjhat$ at particle level.
  }
  \label{tab:resultlightYminus_ptcl} 
\end{table}

In general, the main sources of uncertainty in the differential cross-sections
are similar to those for the fiducial cross-section measurements:
the JES calibration and uncertainties associated with the modelling of both the signal and the \ttbar\ background.
The background normalisation uncertainty is typically about half of the total systematic uncertainty,
while the statistical uncertainty in each bin is similar to the total systematic uncertainty for the 
absolute cross-section measurements.
For the normalised cross-sections, the luminosity and $b/\bbar$ efficiency uncertainties cancel and
the size of many other systematic uncertainty contributions is reduced.
Uncertainties due to the unfolding are small compared to the total uncertainty.

\subsection{Parton-level cross-sections}
\label{sec:xsect_diff:parton}

Differential cross-sections for the top quark and antiquark at parton level are
measured as a function of \pTt and \yt.
The absolute cross-sections are shown in \Figs{\ref{fig:xsectA:pTt:parton}}{\ref{fig:xsectA:yt:parton}}
and the numerical values for both the absolute and normalised cross-sections 
are given in \Tabrange{\ref{tab:resultTopPt_parton}}{\ref{tab:resultantiTopY_parton}}.
The measured cross-sections are compared to both NLO QCD predictions
as well as the same MC predictions used for the comparison of the particle-level cross-sections.
A calculation at NLO+NNLL QCD is available for the top-quark \pT~\cite{Kidonakis:2013yoa}.
This is compared to the data in \Fig{\ref{fig:xsectA:pTt:parton}}.
All predictions agree well with the data,
with the same tendency for almost all MC predictions to be somewhat harder than the data as a function of \pTt.
The NLO+NNLL prediction describes the data better than the MC predictions as a function of \pTt.

\newcommand*{\DiffXsectpTFigCaption}{%
    The vertical error bars on the data points denote the total uncertainty.
    The dashed (red) line in the central distribution shows the NLO prediction calculated using MCFM.
    The dash-dot (blue) line is the NLO+NNLL prediction~\cite{Kidonakis:2015nna}.
    The bottom distribution compares the data with the MC predictions from 
    \POWHEGBOX (orange dashed line) and \MGMCatNLO (purple dash-dotted line).
    The inner (yellow) band  in the bottom part of each figure represents the
    statistical uncertainty of the measurement, and the outer (green) band the total uncertainty.
}
\newcommand*{\DiffXsectEtaFigCaption}{%
    The vertical error bars on the data points denote the total uncertainty.
    The dashed (red) line in the central distribution shows the NLO prediction calculated using MCFM.
    The bottom distribution compares the data with the MC predictions from 
    \POWHEGBOX (orange dashed line) and \MGMCatNLO (purple dash-dotted line).
    The inner (yellow) band  in the bottom part of each figure represents the
    statistical uncertainty of the measurement, and the outer (green) band the total uncertainty.
}
\begin{figure}[htbp]
  \centering
  \subfloat[][]{\includegraphics[width=0.46\textwidth]{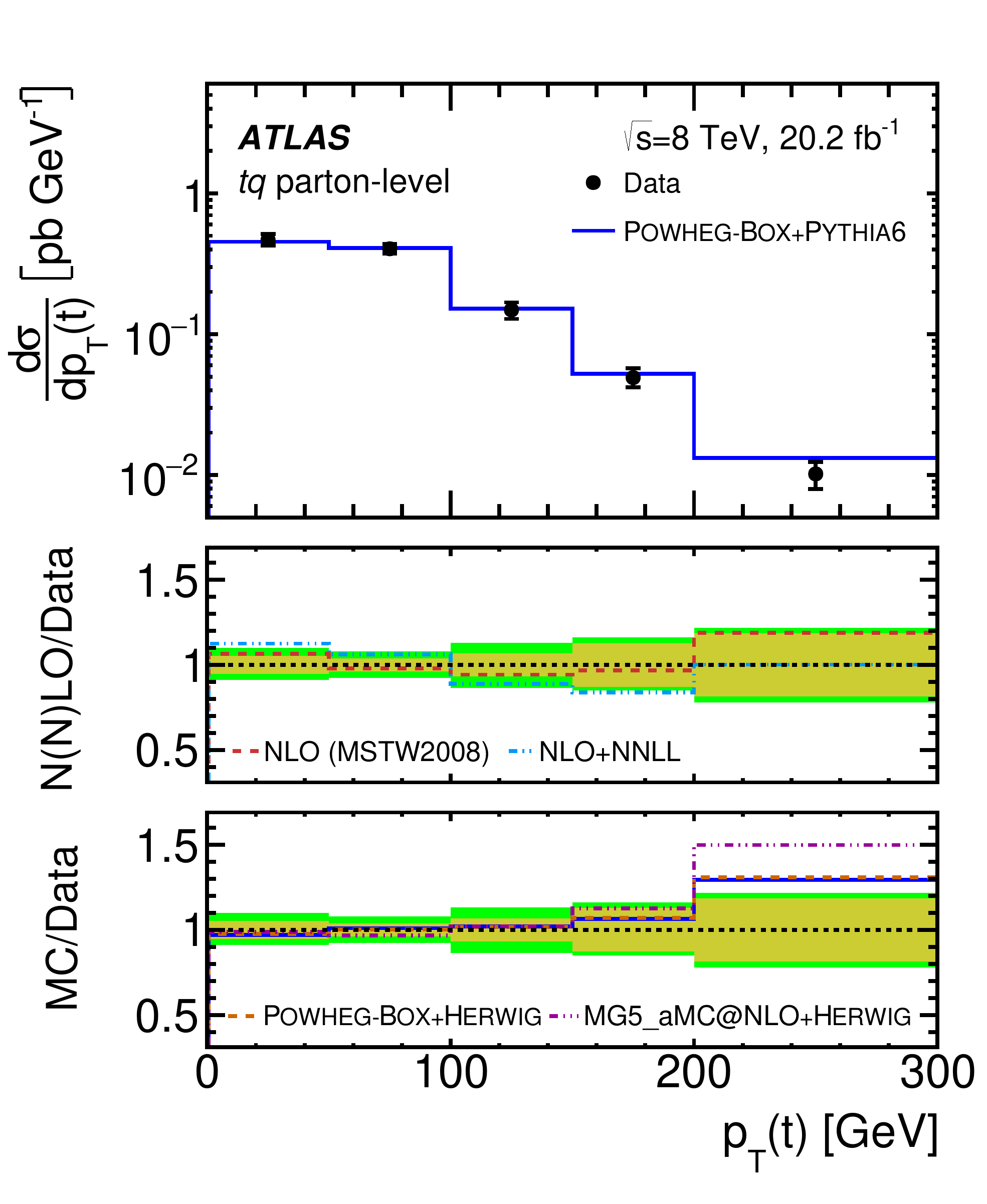}}\quad
  \subfloat[][]{\includegraphics[width=0.46\textwidth]{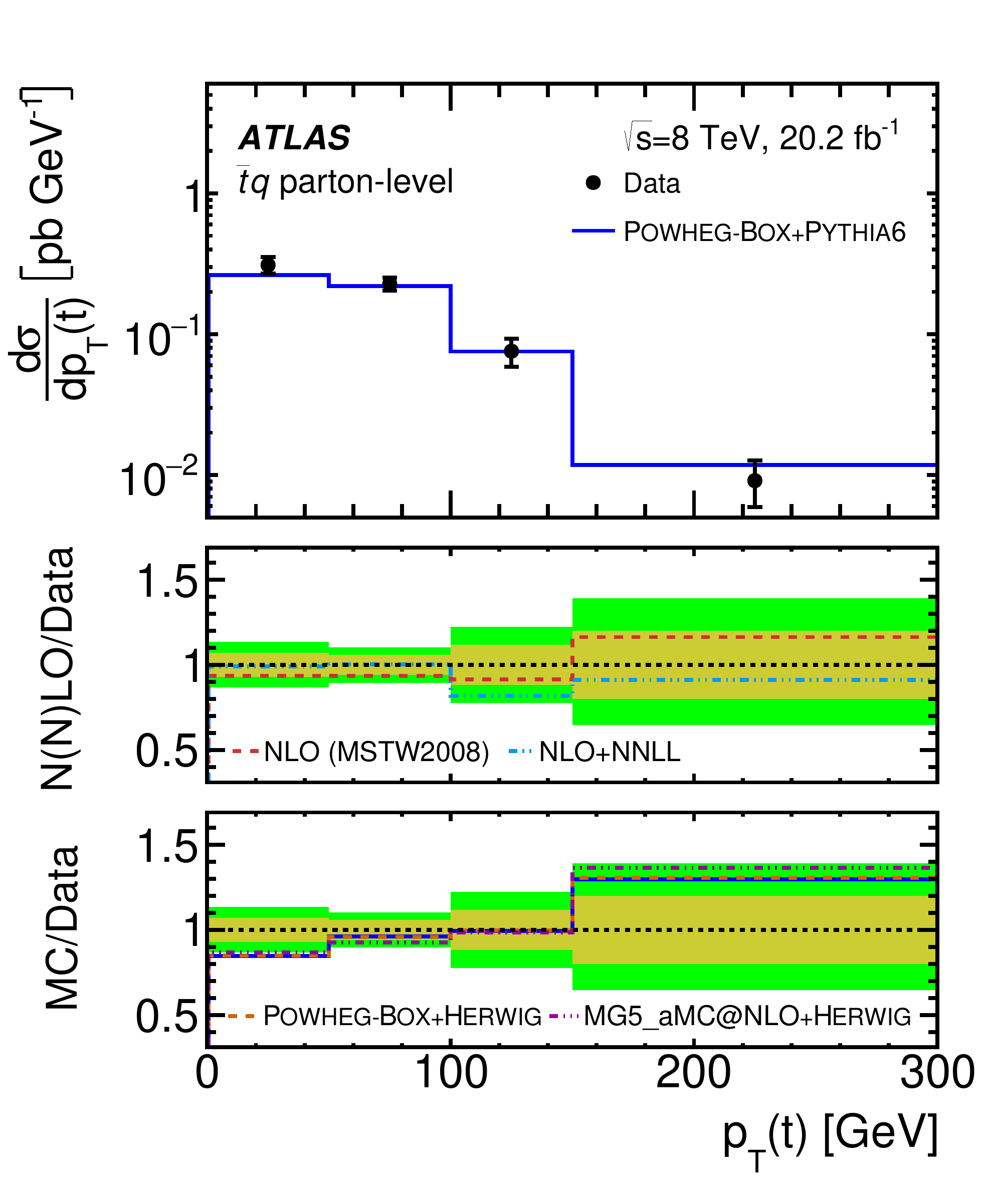}}
  \caption{Absolute  
    unfolded differential cross-sections as a function of \pTt for
    (a) top quarks and (b) top antiquarks.
    The unfolded distributions are compared to QCD NLO and NLO+NNLL calculations 
    as well as various MC predictions. 
    \DiffXsectpTFigCaption
  }
  \label{fig:xsectA:pTt:parton}
\end{figure}

\begin{figure}[htbp]
  \centering
  \subfloat[][]{\includegraphics[width=0.46\textwidth]{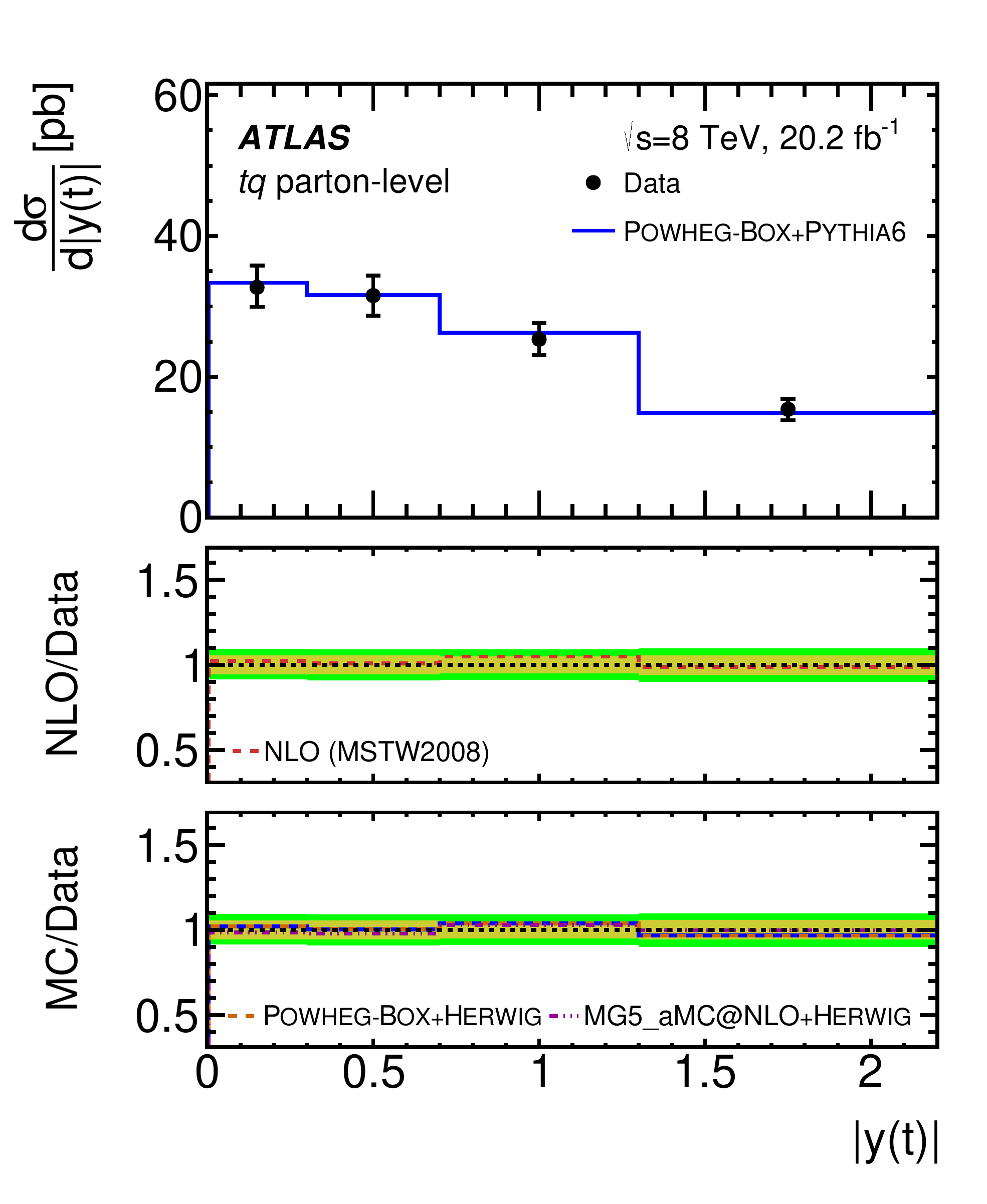}}\quad  
  \subfloat[][]{\includegraphics[width=0.46\textwidth]{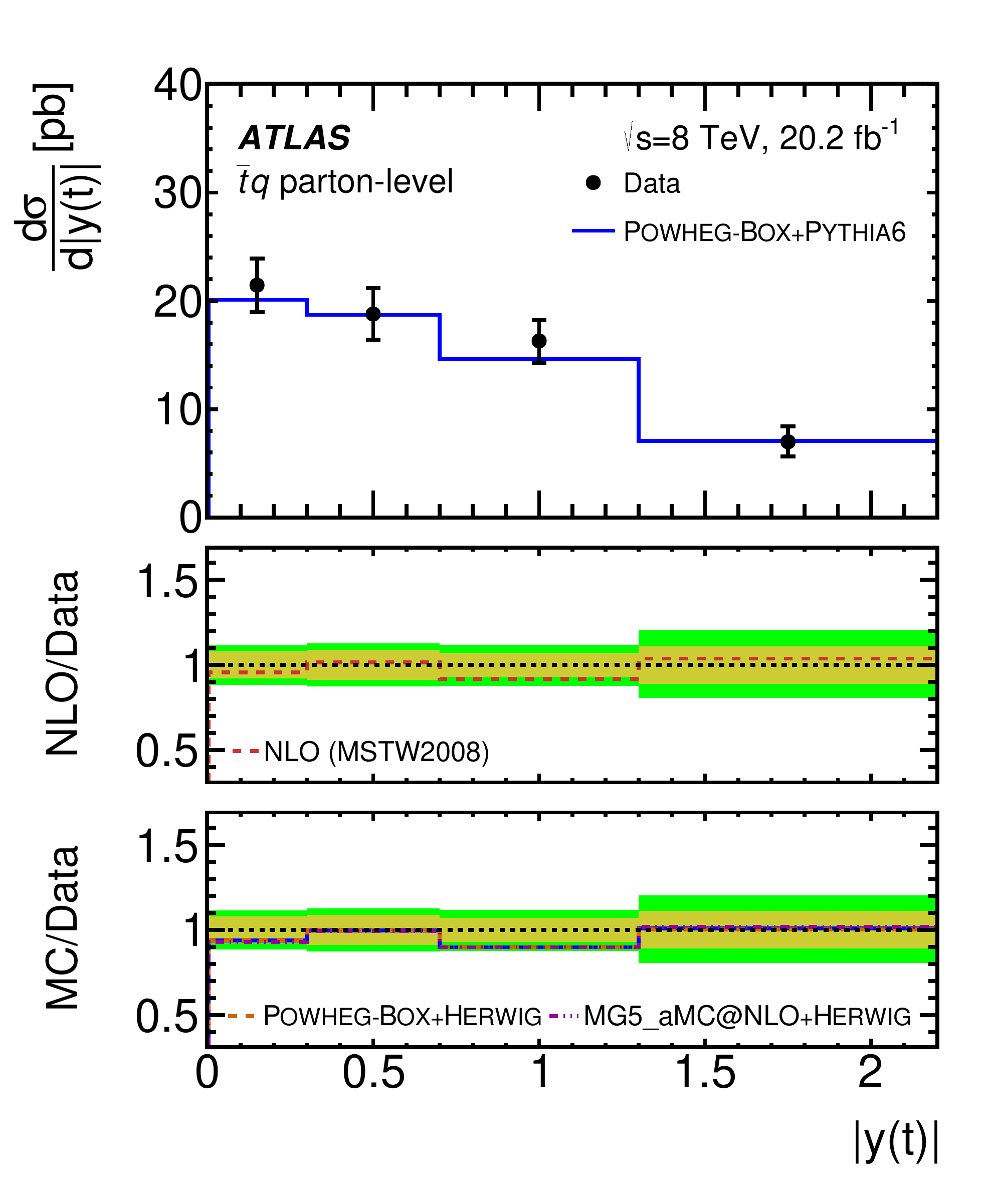}}
  \caption{Absolute  
    unfolded differential cross-sections as a function of \absyt for
    (a) top quarks and (b) top antiquarks.
    The unfolded distributions are compared to a QCD NLO calculation and various MC predictions. 
    \DiffXsectEtaFigCaption
  }
  \label{fig:xsectA:yt:parton}
\end{figure}

\begin{table}[htbp]
  \centering
  \sisetup{round-mode=figures, retain-explicit-plus}
  \begin{tabular}{S[table-format=3.0, round-mode=off]@{\,--\,}S[table-format=3.0, round-mode=off]|
			S[round-precision=3, table-format=3.1]
			@{$\quad\pm$}S[round-precision=2, table-format=2.1]
			S[round-precision=2, table-format=-2.1]@{\,/\,}S[round-precision=2, table-format=-2.1]|
			S[round-precision=3, table-format=1.3]
			@{$\quad\pm$}S[round-precision=2, table-format=1.3]
			S[round-precision=2, table-format=-1.3]@{\,/\,}S[round-precision=2, table-format=-1.3]}
    \toprule
    \multicolumn{2}{c|}{$\pT(t)$}
    & \multicolumn{4}{c|}{$\dif\sigma(tq)/\dif\pT(t)$}
    & \multicolumn{4}{c}{$(1/\sigma) \dif\sigma(tq)/\dif\pT(t)$}\\
    \multicolumn{2}{c|}{[\si{\GeV}]}
    & \multicolumn{4}{c|}{[\si{\fb\per\GeV}]}
    & \multicolumn{4}{c}{[\SI{E-3}{\per\GeV}]}\\
    \multicolumn{2}{c|}{\mbox{}}
    & \multicolumn{2}{r}{stat.\phantom{0}} & \multicolumn{2}{c|}{syst.}
    & \multicolumn{2}{r}{stat.\phantom{0}} & \multicolumn{2}{c}{syst.}\\
    \midrule
      0 & 50   &    466.532 & 24.9818   & +33.59030   & -39.18869  &
     8.56815 & 0.334556   & +0.31702   & -0.42841  \\
     50 & 100   &  404.214   & 15.3967   & +27.89077   & -27.08234  &
     7.42363   & 0.320675 & +0.46769  &-0.40088 \\
    100  & 150  &  148.695   & 10.1706   & +16.65384  & -17.54601 &
    2.73087   & 0.183076  & +0.27036  & -0.29220 \\
    150 & 200  &   49.1889 & 6.2896 & +4.96808 & -4.13187 &
    {\num[round-precision=2]{0.903384}\pho}   & 0.115238 & {\num[round-precision=1]{+0.07950}\pho} &{\num[round-precision=1]{-0.06956}\pho} \\
    200 & 300  &    10.1813 & 1.89892 & +1.17085 & -1.32357   &
    0.186985   & 0.0347674 & +0.01907  &-0.02206 \\
    \bottomrule
  \end{tabular}
  \caption{Absolute and normalised unfolded differential \tq production cross-section as a function of \pTt at parton level.
  }
  \label{tab:resultTopPt_parton} 
\end{table}

\begin{table}[htbp]
  \centering
  \sisetup{round-mode=figures, retain-explicit-plus}
  \begin{tabular}{S[table-format=3.0, round-mode=off]@{\,--\,}S[table-format=3.0, round-mode=off]|
      S[round-precision=3, table-format=3.1]
      @{$\quad\pm$}S[round-precision=2, table-format=2.1]
      S[round-precision=2, table-format=-2.1]@{\,/\,}S[round-precision=2, table-format=-2.1]|
      S[round-precision=3, table-format=1.3]
      @{$\quad\pm$}S[round-precision=2, table-format=1.3]
      S[round-precision=2, table-format=-1.3]@{\,/\,}S[round-precision=2, table-format=-1.3]}
    \toprule
    \multicolumn{2}{c|}{$\pT(t)$}
    & \multicolumn{4}{c|}{$\dif\sigma(\bar{t}q)/\dif\pT(t)$}
    & \multicolumn{4}{c}{$(1/\sigma) \dif\sigma(\bar{t}q)/\dif\pT(t)$}\\
    \multicolumn{2}{c|}{[\si{\GeV}]}
    & \multicolumn{4}{c|}{[\si{\fb\per\GeV}]}
    & \multicolumn{4}{c}{[\SI{E-3}{\per\GeV}]}\\
    \multicolumn{2}{c|}{\mbox{}}
    & \multicolumn{2}{r}{stat.\phantom{0}} & \multicolumn{2}{c|}{syst.}
    & \multicolumn{2}{r}{stat.\phantom{0}} & \multicolumn{2}{c}{syst.}\\
    \midrule
    0 & 50   &   309.832 & 21.3898   & +36.25034   & -35.32085  &
    9.66805 & 0.481771   & +0.77344   & -0.76378  \\
    50 & 100   &   228.011   & 13.1474   & +19.38094   & -20.29298 &
    7.11491   & 0.474081 & +0.49093 &-0.51227  \\
    100  & 150  &   {\num[round-precision=2]{75.7445}\phd}   & {\num[round-precision=1]{8.84054}\phd}   & +14.39146  & -14.39146 &
    2.36355   & 0.274309  & +0.45144 & -0.45617  \\
    150 & 300  & {\phoo\num[round-precision=2]{9.11728}} & 1.7539 & +3.06341 & -2.63489 &
    0.284497   & 0.0565622 & +0.08933 &-0.07596 \\
    \bottomrule
  \end{tabular}
  \caption{Absolute and normalised unfolded differential \tbarq production cross-section as a function of \pTt at parton level.
  }
  \label{tab:resultantiTopPt_parton} 
\end{table}

\begin{table}[htbp]
	\centering
	\sisetup{round-mode=figures, retain-explicit-plus}
	\begin{tabular}{S[round-mode=off]@{\,--\,}S[round-mode=off]|
      S[round-precision=3, table-format=2.1]
      @{$\quad\pm$}S[round-precision=1, round-mode=places, table-format=1.1]
      S[round-precision=2, table-format=-1.1]@{\,/\,}S[round-precision=2, table-format=-1.3]|
      S[round-precision=3, table-format=3.0]
      @{$\quad\pm$}S[round-precision=2, table-format=2.0]
      S[round-precision=2, table-format=-2.0]@{\,/\,}S[round-precision=2, table-format=-2.0]}
    \toprule
    \multicolumn{2}{c|}{$\absyt$}
    & \multicolumn{4}{c|}{$\dif\sigma(tq)/\dif\absyt$}
    & \multicolumn{4}{c}{$(1/\sigma) \dif\sigma(tq)/\dif\absyt$}\\
    \multicolumn{2}{c|}{}
    & \multicolumn{4}{c|}{[\si{\pb}]}
    & \multicolumn{4}{c}{[\SI{E-3}]}\\
    \multicolumn{2}{c|}{\mbox{}}
    & \multicolumn{2}{r}{stat.} & \multicolumn{2}{c|}{syst.}
    & \multicolumn{2}{r}{stat.} & \multicolumn{2}{c}{syst.}\\
		\midrule
		0.0 & 0.3  &  32.7059 & 1.8375   & +2.45294   & -2.12588  &
		635.799 & 34.7527   & +47.04913   & -38.78374  \\
		0.3 & 0.7  &  31.5304   & 1.77459   & +2.23866   & -2.36478  &
		612.947   & 33.8529 & +30.64735  &-33.09914 \\
		0.7 & 1.3  &   25.3165   & 1.27364   & +1.92405  & -1.92405 &
		492.148   & 23.7603  & +25.59170  & -26.57599 \\
		1.3 & 2.2   &   15.3632 & 0.881928 & +1.21369 & -1.24442 &
		298.659   & 14.426 & +14.33563 &-14.63429  \\
		\bottomrule
	\end{tabular}
  \caption{Absolute and normalised unfolded differential \tq production cross-sections as a function of \absyt at parton level.
  }
  \label{tab:resultTopY_parton} 
\end{table}

\begin{table}[htbp]
	\centering
	\sisetup{round-mode=figures, retain-explicit-plus}
	\begin{tabular}{S[round-mode=off]@{\,--\,}S[round-mode=off]|
      S[round-precision=1, round-mode=places, table-format=2.1]
      @{$\quad\pm$}S[round-precision=1, round-mode=places, table-format=1.1]
      S[round-precision=2, table-format=-1.1]@{\,/\,}S[round-precision=2, table-format=-1.3]|
      S[round-precision=3, table-format=3.0]
      @{$\quad\pm$}S[round-precision=2, table-format=2.0]
      S[round-precision=2, table-format=-2.0]@{\,/\,}S[round-precision=2, table-format=-2.0]}
    \toprule
    \multicolumn{2}{c|}{$\absyt$}
    & \multicolumn{4}{c|}{$\dif\sigma(\bar{t}q)/\dif\absyt$}
    & \multicolumn{4}{c}{$(1/\sigma) \dif\sigma(\bar{t}q)/\dif\absyt$}\\
    \multicolumn{2}{c|}{}
    & \multicolumn{4}{c|}{[\si{\pb}]}
    & \multicolumn{4}{c}{[\SI{E-3}]}\\
    \multicolumn{2}{c|}{\mbox{}}
    & \multicolumn{2}{r}{stat.} & \multicolumn{2}{c|}{syst.}
    & \multicolumn{2}{r}{stat.} & \multicolumn{2}{c}{syst.}\\
		\midrule
		0.0 & 0.3   &    21.4538 & 1.69738   & +1.80212   & -1.88793  &
		714.499 & 54.7555   & +41.44094   & -45.72794 \\
		0.3 & 0.7  &  18.7902   & 1.64303   & +1.74749   & -1.70991  &
		625.791   & 53.273 & +46.30853  &-46.30853 \\
		0.7 & 1.3  &  16.3086   & 1.17094   & +1.61455  & -1.58193 &
		543.141   & 37.4549  & +43.99442  & -42.90814 \\
		1.3 & 2.2  &   6.98779  & 0.772933 & +1.19491 & -1.12503 &
		232.722   & 22.6026 & +30.48658  &-28.62481 \\
    \bottomrule
  \end{tabular}
  \caption{Absolute and normalised unfolded differential \tbarq production cross-sections as a function of \absyt at parton level.
  }
  \label{tab:resultantiTopY_parton} 
\end{table}

%% file: conclusion.tex
\section{Conclusion}
\label{sec:conclusion}

Measurements of $t$-channel single top-quark production 
using data collected by the ATLAS experiment in $pp$ collisions at \SI{8}{\TeV}
at the LHC are presented. The data set corresponds to an integrated
luminosity of \SI{20.2}{\per\fb}. 
An artificial neural network is used to separate signal from background.
Total and fiducial cross-sections are measured for both top quark and top
antiquark production.
The fiducial cross-section is measured with a precision of \SI{5.8}{\%}
(top quark) and \SI{7.8}{\%} (top antiquark), respectively.
In addition, the cross-section ratio of top-quark to top-antiquark production
is measured,
resulting in a precise value to compare with predictions,
$\rt = 1.72 \pm 0.09$.
The total cross-section is used to extract
the $Wtb$ coupling: $\flvtb = 1.029\pm 0.048$,
which corresponds to $|V_{tb}|>0.92$ at the \SI{95}{\%} confidence level,
when assuming $\fl=1$ and restricting the range of $|V_{tb}|$ to the
interval $[0, 1]$.

Requiring a high value of the neural-network discriminant leads to relatively
pure $t$-channel samples, which are used to measure differential cross-sections
for both \tq and \tbarq production.
Differential cross-sections as a function of the transverse momentum and
absolute value of the rapidity of the top quark, the top antiquark,
as well as the accompanying jet from the $t$-channel scattering are
measured at particle level.
The measurements of cross-sections as a function of the accompanying-jet
transverse momentum and absolute value of the rapidity extend previous results,
which only measured top-quark and top-antiquark distributions.
Differential cross-sections as a function of the transverse momentum and
rapidity of the top quark and top antiquark are also measured at parton
level.
All measurements are compared to different Monte Carlo predictions as well as
to fixed-order QCD calculations where these are available.
The SM predictions provide good descriptions of the data.

%% file: Acknowledgements.tex

We thank CERN for the very successful operation of the LHC, as well as the
support staff from our institutions without whom ATLAS could not be
operated efficiently.

We acknowledge the support of ANPCyT, Argentina; YerPhI, Armenia; ARC, Australia; BMWFW and FWF, Austria; ANAS, Azerbaijan; SSTC, Belarus; CNPq and FAPESP, Brazil; NSERC, NRC and CFI, Canada; CERN; CONICYT, Chile; CAS, MOST and NSFC, China; COLCIENCIAS, Colombia; MSMT CR, MPO CR and VSC CR, Czech Republic; DNRF and DNSRC, Denmark; IN2P3-CNRS, CEA-DSM/IRFU, France; SRNSF, Georgia; BMBF, HGF, and MPG, Germany; GSRT, Greece; RGC, Hong Kong SAR, China; ISF, I-CORE and Benoziyo Center, Israel; INFN, Italy; MEXT and JSPS, Japan; CNRST, Morocco; NWO, Netherlands; RCN, Norway; MNiSW and NCN, Poland; FCT, Portugal; MNE/IFA, Romania; MES of Russia and NRC KI, Russian Federation; JINR; MESTD, Serbia; MSSR, Slovakia; ARRS and MIZ\v{S}, Slovenia; DST/NRF, South Africa; MINECO, Spain; SRC and Wallenberg Foundation, Sweden; SERI, SNSF and Cantons of Bern and Geneva, Switzerland; MOST, Taiwan; TAEK, Turkey; STFC, United Kingdom; DOE and NSF, United States of America. In addition, individual groups and members have received support from BCKDF, the Canada Council, CANARIE, CRC, Compute Canada, FQRNT, and the Ontario Innovation Trust, Canada; EPLANET, ERC, ERDF, FP7, Horizon 2020 and Marie Sk{\l}odowska-Curie Actions, European Union; Investissements d'Avenir Labex and Idex, ANR, R{\'e}gion Auvergne and Fondation Partager le Savoir, France; DFG and AvH Foundation, Germany; Herakleitos, Thales and Aristeia programmes co-financed by EU-ESF and the Greek NSRF; BSF, GIF and Minerva, Israel; BRF, Norway; CERCA Programme Generalitat de Catalunya, Generalitat Valenciana, Spain; the Royal Society and Leverhulme Trust, United Kingdom.

The crucial computing support from all WLCG partners is acknowledged gratefully, in particular from CERN, the ATLAS Tier-1 facilities at TRIUMF (Canada), NDGF (Denmark, Norway, Sweden), CC-IN2P3 (France), KIT/GridKA (Germany), INFN-CNAF (Italy), NL-T1 (Netherlands), PIC (Spain), ASGC (Taiwan), RAL (UK) and BNL (USA), the Tier-2 facilities worldwide and large non-WLCG resource providers. Major contributors of computing resources are listed in Ref.~\cite{ATL-GEN-PUB-2016-002}.

%% file: atlas_authlist.tex
\begin{flushleft}
{\Large The ATLAS Collaboration}

\bigskip

M.~Aaboud$^\textrm{\scriptsize 137d}$,
G.~Aad$^\textrm{\scriptsize 88}$,
B.~Abbott$^\textrm{\scriptsize 115}$,
J.~Abdallah$^\textrm{\scriptsize 8}$,
O.~Abdinov$^\textrm{\scriptsize 12}$,
B.~Abeloos$^\textrm{\scriptsize 119}$,
O.S.~AbouZeid$^\textrm{\scriptsize 139}$,
N.L.~Abraham$^\textrm{\scriptsize 151}$,
H.~Abramowicz$^\textrm{\scriptsize 155}$,
H.~Abreu$^\textrm{\scriptsize 154}$
% ...
\bigskip
\\
% ...
$^{8}$ Department of Physics, The University of Texas at Arlington, Arlington TX, United States of America\\
% ...
$^{12}$ Institute of Physics, Azerbaijan Academy of Sciences, Baku, Azerbaijan\\
% ...
$^{88}$ CPPM, Aix-Marseille Universit{\'e} and CNRS/IN2P3, Marseille, France\\
% ...
$^{115}$ Homer L. Dodge Department of Physics and Astronomy, University of Oklahoma, Norman OK, United States of America\\
% ...
$^{119}$ LAL, Univ. Paris-Sud, CNRS/IN2P3, Universit{\'e} Paris-Saclay, Orsay, France\\
% ...
$^{137}$ $^{(a)}$ Facult{\'e} des Sciences Ain Chock, R{\'e}seau Universitaire de Physique des Hautes Energies - Universit{\'e} Hassan II, Casablanca; $^{(b)}$ Centre National de l'Energie des Sciences Techniques Nucleaires, Rabat; $^{(c)}$ Facult{\'e} des Sciences Semlalia, Universit{\'e} Cadi Ayyad, LPHEA-Marrakech; $^{(d)}$ Facult{\'e} des Sciences, Universit{\'e} Mohamed Premier and LPTPM, Oujda; $^{(e)}$ Facult{\'e} des sciences, Universit{\'e} Mohammed V, Rabat, Morocco\\
% ...
$^{139}$ Santa Cruz Institute for Particle Physics, University of California Santa Cruz, Santa Cruz CA, United States of America\\
% ...
$^{151}$ Department of Physics and Astronomy, University of Sussex, Brighton, United Kingdom\\
% ...
$^{154}$ Department of Physics, Technion: Israel Institute of Technology, Haifa, Israel\\
$^{155}$ Raymond and Beverly Sackler School of Physics and Astronomy, Tel Aviv University, Tel Aviv, Israel\\
% ...
$^{a}$ Also at Department of Physics, King's College London, London, United Kingdom\\
$^{b}$ Also at Institute of Physics, Azerbaijan Academy of Sciences, Baku, Azerbaijan\\
$^{c}$ Also at Novosibirsk State University, Novosibirsk, Russia\\
$^{d}$ Also at TRIUMF, Vancouver BC, Canada\\
$^{e}$ Also at Department of Physics {\&} Astronomy, University of Louisville, Louisville, KY, United States of America\\
% ...
\end{flushleft}
